\documentclass[a4paper]{article}
\usepackage{etex}
\usepackage{a4wide}
\usepackage{graphicx}
\usepackage{epstopdf} %
\usepackage{amsmath,amsbsy,amssymb,amsgen,amsfonts}
\usepackage{color,latexsym}
\usepackage[breaklinks=true,bookmarks=true,colorlinks=true,linkcolor=blue,citecolor=blue,allcolors=blue]{hyperref}
\usepackage[toc,title]{appendix}
\usepackage{natbib}

\usepackage{doi}
\usepackage{relsize}

\def\solidthick{\protect\rule[2pt]{10.pt}{1pt}}
\def\solidshort{\protect\rule[2pt]{3.pt}{1pt}}
\def\dashed{\solidshort$\,$\solidshort$\,$\solidshort}
\def\chndot{\solidshort$\,\cdot\,$\solidshort}

\def\solidthickbullet{\solidthick$\hspace{-1ex}
                      \bullet\hspace{-1ex}$\solidthick}

\newcommand{\soliddiamond}{$\mathsmaller{\blacklozenge}$}      
\newcommand{\solidsquare}{$\mathsmaller{\blacksquare}$}

\usepackage[section]{placeins} %
\definecolor{darkgreen}{RGB}{0,102,0}
\definecolor{forestgreen}{rgb}{0.2344,0.6992,0.4414}
\definecolor{niceorange}{rgb}{0.9216,0.5059,0.1059}
\usepackage{sectsty}
\allsectionsfont{\sffamily}
\let\LaTeXmaketitle\maketitle
\renewcommand{\maketitle}{{\sf\LaTeXmaketitle}}

\begin{document}
\epstopdfsetup{suffix=} %
\title{%
  Clustering and preferential concentration of 
  finite-size particles in forced homogeneous-isotropic turbulence 
} 
\author{%
  Markus Uhlmann\footnote{\href{mailto:markus.uhlmann@kit.edu}{markus.uhlmann@kit.edu}}
  \hspace*{1ex} and 
  Agathe Chouippe\footnote{\href{mailto:agathe.chouippe@kit.edu}{agathe.chouippe@kit.edu}}
  \\[1ex]
  {\small 
    Institute for Hydromechanics, Karlsruhe Institute of
    Technology}\\ 
  {\small 
    76131 Karlsruhe, Germany
  }
}
\date{\tiny (Nov.\ 30, 2016, manuscript accepted for publication in {\it J.\ Fluid Mech.})} 
\maketitle
\begin{abstract}
  We have performed interface-resolved direct numerical simulations of
  forced homogeneous-isotropic turbulence in a dilute suspension of
  spherical particles in the Reynolds number range
  $Re_\lambda=115-140$. 
  The solid-fluid density ratio was set to $1.5$, 
  gravity was set to zero,  
  and two particle diameters were investigated corresponding to
  approximately $5$ and $11$ Kolmogorov lengths.  
    Note that these particle sizes are clearly outside the range of
    validity of the point-particle approximation, as has been shown by
    \cite{homann:10}. 
    At the present parameter points the global effect of the
    particles upon the fluid flow is weak. 
    We observe that the dispersed phase exhibits clustering with
    moderate intensity.
  The tendency to cluster, which was quantified in terms of the standard
  deviation of Vorono\"i cell volumes, decreases with the particle
  diameter. 
  We have analyzed the relation between particle locations and the
  location of intense vortical flow structures. The results do not
  reveal any significant statistical correlation. 
  Contrarily, we have detected a small but statistically significant
  preferential location of particles with respect to the `sticky
  points' proposed by \cite{goto:08},  
  i.e.\ points where 
  the fluid acceleration field is acting such as to increase the local
  particle concentration in one-way coupled point-particle models
  under Stokes drag. 
  The presently found statistical correlation between the `sticky
  points' and the particle locations further increases when focusing
  on regions with high local concentration. 
  Our results suggest that small finite-size particles can be brought
  together along the expansive directions of the fluid acceleration
  field, as previously observed only for the simplest model for
  sub-Kolmogorov particles. 
  We further discuss the effect of density ratio and collective
  particle motion upon the basic Eulerian and Lagrangian statistics. 
\end{abstract}
\section{Introduction}
\label{sec-intro}
It is well known that solid particles suspended in a turbulent flow
may exhibit a non-trivial spatial distribution as a consequence of
their hydrodynamic interaction with coherent flow structures 
\citep{eaton:94}.
The resulting spatial structure of the dispersed phase, in turn, is
key to understanding many technologically relevant aspects of these
systems, such as particle collision statistics, dispersion/mixing
behavior as well as the particles' impact upon the carrier phase. 
It is therefore not a surprise that much effort has been devoted in
the past to the understanding of the mechanisms responsible for
the formation of agglomerations of particles in turbulent flows. 
In the following review we will for the sake of brevity focus our
attention on the idealized flow configuration of 
(approximately) homogeneous-isotropic turbulence.

Let us first consider the case of small inertial particles (i.e.\ of 
sub-Kolmogorov size). 
The formation of clusters has indeed been observed in many
laboratory experiments in the presence of gravity 
\citep[e.g.][]{aliseda:02,wood:05}, 
and under conditions of micro-gravity \citep{fallon:02}. 
It has also been often reproduced in direct numerical simulations with 
point-particles under the assumption of Stokes drag
\citep[e.g][]{squires:91,bec:06,calzavarini:08}.   
  A number of studies suggest that clustering of sub-Kolmogorov
  particles is most intense when the Stokes number $St$ (defined as
  the ratio between the relevant particle and fluid time scales) takes
  values of order unity with respect to the dissipative scales of the
  turbulent carrier flow \citep{hogan:01,balachandar:10,monchaux:12}. 
  This conclusion, however, somewhat depends upon the measure which is
  chosen for quantifying the intensity of clustering, and 
  \cite{yoshimoto:07} have observed in their point-particle
  simulations that clustering occurs over a substantial range of
  values of the Stokes number, with different flow scales contributing
  accordingly.  
  In a recent experimental study on polydisperse particles in
  grid-generated wind-tunnel turbulence, \cite{sumbekova:16} conclude
  that the intensity of particle clustering (quantified by means of 
  Vorono\"i tesselation analysis) is practically insensitive to the
  particles' dissipation-scale Stokes number. 

  When investigating the statistical relation between particle
  positions and the local topology of the flow field, it is often
  found that sub-Kolmogorov particles are preferentially located in 
  low-vorticity, strain-dominated regions 
  \citep[e.g.][]{squires:91,sundaram:97,bec:06,balachandar:10}. 
  The mechanism for this type of preferential particle accumulation
  has traditionally been linked to the action of coherent vortices. 
  Based on an asymptotic expansion argument \cite{maxey:87} has proposed
  that small particles should centrifuge out of vortices due to their
  inertia, i.e.\ that particle clustering should indeed occur in
  low-vorticity regions. The aforementioned evidence can then be taken
  as an indication of the effectiveness of the centrifugal effect. 
  However, since the analysis of \cite{maxey:87} strictly only applies
  to the simplest point-particle model in the limit of small Stokes
  numbers, it is questionable how preferential accumulation could be
  caused by this mechanism at larger values of $St$. 

    Various authors have investigated the structure of the disperse
    phase from the Lagrangian point of view. 
    Starting from a description of the relative motion of a pair of
    particles, both \cite{zaichik:03} and \cite{chun:05} have
    constructed models which are able to predict the radial
    distribution function (RDF) based upon prior knowledge about
    statistical features of the carrier flow. The former approach
    has later been refined \citep{zaichik:07}, and it is a priori
    applicable to the entire range of Stokes numbers. 
    Note that the clustering mechanism inherent in the description of
    \cite{zaichik:07} is non-local due to the fact that the relative
    motion of two particles is a function of the history of the fluid
    velocity gradient tensor along the particle paths \cite[cf.\ also
    the recent review by][]{gustavsson:16}.  
    \cite{bragg:14a} have shown that for Stokes numbers $St\geq{\cal
      O}(1)$ this non-local-in-time effect starts to outweigh the local 
    centrifugal mechanism. 
    \cite{bragg:15a} further argue that the non-local clustering
    mechanism is additionally biased by preferential sampling of
    strain-dominated regions along the particle paths, leading to the
    observed statistical correlation between particle locations and such
    fluid regions even in the Stokes number regime where the centrifugal
    mechanism alone is not expected to be efficient. 
Another point of view on clustering of point-particles in
homogeneous-isotropic turbulence has been proposed by
\cite*{chen:06} and \cite{goto:06} in two-dimensions, and
later by \cite{goto:08} and \cite{coleman:09} in the
three-dimensional setting. 
These authors have established a link between the accumulation of
point-particles and the local properties of the fluid acceleration
field, which is based upon the observation by \cite{maxey:87} that in
the limit of small particle inertia (i.e.\ small Stokes number)  
the particle velocity deviates from the local fluid velocity
proportionally to the fluid acceleration. This effect has been
termed ``sweep-stick'' mechanism, since it relates to the sweeping of
particles by the fluid flow field and their ``sticking'' at
zero-acceleration points.  
  A recent paper by \cite{bragg:15b} has challenged the effectiveness 
  of the sweep-stick mechanism at length scales $r$ when the
  scale-dependent Stokes number is of the order of (or exceeds) unity, 
  $St_r\geq{\cal O}(1)$.
  These authors argue that, while for 
  $St_r\ll1$
  it is the preferential sampling of low-vorticity regions of the
  fluid velocity field ``coarse-grained'' at the length $r$ which
  dominates, 
  for $St_r\geq{\cal O}(1)$ 
  the essential mechanism is rather a combination of the preferential
  sampling 
  and of the 
  above-mentioned non-local effect. 
Much less is known about the dynamics of particles with diameters equal
to or larger than the Kolmogorov scale. 
Systematic studies of finite-size effects have been carried out in von
K\'arm\'an flows (which feature a central region of approximately
homogeneous-isotropic turbulence) by \cite{voth:02}, \cite{brown:09} and
\cite{volk:11}, while \cite{qureshi:07} have performed a complementary
analysis in grid turbulence in a wind-tunnel. In all these studies the
particles were approximately neutrally-buoyant, and the concentration
was sufficiently low such that collective effects can be excluded. 
It was found that the normalized probability density function (p.d.f.)
of particle acceleration roughly collapses on a functional form which
is consistent with the log-normal scaling of the acceleration vector
magnitude first suggested by \cite{mordant:04} for fluid particles. 
Finite-size effects can then conveniently be 
described by the acceleration variance alone, which is found to be
reasonably well represented by Kolmogorov inertial-range scaling,
supposing that the scales which determine the particle acceleration
are of the same order as the particle diameter $D$ itself, i.e.\
implying a $D^{-2/3}$ power-law. 
Note that \cite{qureshi:08} have extended the analysis to heavy
particles with densities up to 66 times the fluid density, suggesting
a non-trivial influence of the density ratio upon the particle
acceleration variance. 
The more recent measurements of \cite{fiabane:12} are to our knowledge
the only available source of experimental data on clustering of
finite-size particles suspended in sustained, approximately
homogeneous-isotropic turbulence. They investigated two particle
types (either neutrally-buoyant or with a density  of 2.5 times
that of the fluid) in a chamber with random large-scale turbulence
forcing. The authors observe that in the same range of the Stokes
number (roughly from 0.25 to 1.25) the density-matched particles do
not form clusters, while the heavy particles clearly do.  
This result demonstrates that the Stokes number alone is not a
sufficient parameter for determining the spatial distribution of
finite-size particles. 
Unfortunately it is not possible to rigorously separate the effect of
gravity from the effect of enhanced particle inertia in the laboratory
measurements of \cite{qureshi:08} and \cite{fiabane:12}, which is
where numerical simulations can contribute to fill the knowledge
gaps. 

Turning to numerical work,  
the first simulations of finite-size particles suspended in forced,
homogeneous-isotropic turbulence were performed by \cite{tencate:04}
in moderately dense systems where the solid/fluid density ratio was
above unity and gravity set to zero. 
\cite{YeoClimentMaxey:10} have considered similar systems 
\citep[using the force-coupling approach of][]{lomholt:03}, while 
varying the density ratio over a certain range.  
On the other hand, \cite{homann:10} 
have carried out a systematic study of
finite-size effects upon the motion of a single, neutrally-buoyant
particle in forced, homogeneous-isotropic turbulence. The parameter
space was later extended by \cite{cisse:13} and \cite{cisse:15} to
larger turbulent Reynolds numbers and particle sizes. 
Together, these numerical studies provide important information on the
scaling of a number of particle-related properties (cf.\ discussion in
\S~\ref{sec-results-basic-stats}). 
However, the aspect
of preferential particle location with respect to coherent structures
has not been addressed by these authors.  
In the present contribution we have performed DNS describing the 
motion of a dilute suspension of particles with diameter larger than  
the Kolmogorov length (up to 11 times), while fully resolving the flow
around each particle. 
We consider inertial particles with a mass density of 1.5
times the fluid density. This parameter point has been chosen in order
to be able to address the effects of gravity and turbulence upon the
motion of a finite-size particle collective separately: the former
effect has been singled out in the study of \cite{uhlmann:14a}, while
the latter is the subject of the present work. Ultimately we are
interested in analyzing the combined effects of gravity and turbulence
upon the motion of heavy particles; this next step will be considered  
in a forthcoming contribution.  
In the present work a turbulent flow is maintained statistically
stationary with the aid of large-scale random forcing. This allows us
to obtain information about the fully-developed state of the
particulate phase in response to a well-defined turbulent carrier
flow, which is difficult to achieve when decaying turbulence is
considered \citep[as done e.g.\ in the studies
of][]{doychev:10a,lucci:10,lucci:11}.  
The main objective of the present work is to contribute to the
clarification of the tendency of finite-size particles to concentrate
locally in a turbulent flow. Here we perform -- for the first time to
our knowledge -- a detailed analysis of the particles' preferential
location with respect to coherent structures of the turbulent flow
field.  

This paper is organized as follows. The chosen methodology is 
described in terms of the numerical method as well as the flow
configuration and parameter set in section~\ref{sec-setup}. 
Section~\ref{sec-results} is devoted to the presentation of the
results, where we set out to quantify the feedback of the dispersed
phase upon the carrier phase, and where we document the basic Eulerian
and Lagrangian statistics in relation to previous results from the
literature. 
Next we address more specifically the question of particle clustering  
by quantifying the deviation from a random arrangement. 
  In the final part we then explore the relation
  between coherent vortices and particle locations, before turning to
  the analysis of the role of the fluid acceleration field in the
  present context. 
The paper closes with a summary and discussion in
section~\ref{sec-conclusion}.   

\section{Computational set-up}
\label{sec-setup}
\subsection{Numerical method}
\label{sec-setup-numa}
The basic numerical method employed in the present simulations 
has been described in detail by \cite{uhlmann:04}; it has recently
been extended by including a random turbulence forcing scheme
\citep{chouippe:15a}. 
The incompressible Navier-Stokes equations are solved by a fractional
step approach with an implicit treatment of the viscous terms
(Crank-Nicolson) and a low-storage, three-step Runge-Kutta scheme for the
non-linear terms. The spatial discretization employs second-order
central finite-differences on a staggered mesh which is uniform and
isotropic (i.e.\ $\Delta x=\Delta y=\Delta z=cst.$).  
The no-slip condition at the surface of moving solid particles is
imposed by means of a specifically designed immersed boundary technique
\citep{uhlmann:04}. This gives rise to an additional volume force term
in the momentum equation which we denote as $\mathbf{f}^{(ibm)}$.  
In this setting the equations which describe the motion of an
incompressible fluid with constant density $\rho_f$ and constant
kinematic viscosity $\nu$ can be written as follows: 
\begin{subequations}\label{equ-navier-stokes-general}
  \begin{eqnarray}\label{equ-navier-stokes-general-mom}
    \frac{\partial\mathbf{u}_f}{\partial t}
    +(\mathbf{u}_f\cdot\nabla)\mathbf{u}_f 
    +\frac{1}{\rho_f}\nabla p 
    &=&
    \nu\nabla^2\mathbf{u}_f
    +\mathbf{f}^{(t)}
    +\mathbf{f}^{(ibm)}
    \,,
    \\\label{equ-navier-stokes-general-conti}
    \nabla\cdot\mathbf{u}_f&=&0
    \,,
  \end{eqnarray}
\end{subequations}
where $\mathbf{u}_f$ denotes the fluid velocity vector, $p$ the
hydrodynamic pressure, and $\mathbf{f}^{(t)}$ is a volume force
introduced in order to generate and maintain turbulent motion. 
\cite{chouippe:15a} provide the details of the presently chosen temporal
integration of the Navier-Stokes equations in the presence of solid
particles and artificial turbulence forcing. 

On the other hand, the motion of the particles is computed from the
Newton equations for linear and angular motion of rigid bodies, driven
by buoyancy, hydrodynamic force/torque and contact forces (in case of
particle-particle collisions). 
Since the particle suspensions under consideration here are dilute,
collisions are occurring infrequently. 
In fact it is found that the average collision-free period is of the
order of one large-eddy time scale (or, equivalently, of the order of
50 Kolmogorov time scales) for the systems simulated herein. 
Therefore, in the present work particle contact is treated by a
simple repulsive force mechanism 
\citep{glowinski:99} formulated such as to keep colliding particles
from overlapping non-physically. In the case of dense particle
arrangements, the discrete element method of
\cite{kidanemariam:14a} can be employed instead. 

The computational code has been extensively validated in unbounded
flows as well as in wall-bounded shear flows 
\citep[][]{uhlmann:04,uhlmann:05a,uhlmann:06c,uhlmann:08a,uhlmann:13a,kidanemariam:14a,chouippe:15a}. 

\subsection{Flow configuration and parameters}
\label{sec-setup-config}
Before turning to the description of the flow configuration, let us 
first fix some notational details. 
The usual Reynolds decomposition of the fluid phase velocity is
introduced, where the average $\langle\cdot\rangle_{\Omega_f}$ is
computed over the region $\Omega_f$ filled with fluid, viz. 
\begin{equation}\label{equ-def-decomp-fluid-only}
  \mathbf{u}_f(\mathbf{x},t)
  =
  \langle\mathbf{u}_f\rangle_{\Omega_f}(t)
  +
  \mathbf{u}^\prime_f(\mathbf{x},t)
  \,.
\end{equation}
The corresponding kinetic energy of the fluctuations is then defined
as  
\begin{equation}\label{equ-ek-fluct-fluid-only}
  k(t)
  =
  \frac{1}{2}
  \langle \mathbf{u}^\prime\cdot \mathbf{u}^\prime \rangle_{\Omega_f} 
  =
  \frac{3}{2}
  \,u_{rms}^2(t)
  \,,
\end{equation}
where the characteristic velocity scale $u_{rms}(t)$ has been defined
simultaneously. 
We also define a dissipation rate averaged over the fluid phase, viz. 
\begin{equation}\label{equ-def-eps-box-avg-fluid-only}
  \varepsilon(t)=2\nu\langle S_{ij}^\prime S_{ij}^\prime\rangle_{\Omega_f}
  \,, 
\end{equation}
with $S_{ij}=(u_{i,j}+u_{j,i})/2$. 
After further averaging $k(t)$, $\varepsilon(t)$ and $u_{rms}(t)$ over
time in the statistically stationary interval (here indicated by
simply omitting the argument ``$t$''), we can define the following
global quantities:   
the Kolmogorov length scale 
$\eta=(\nu^3/\varepsilon)^{1/4}$, 
the Taylor micro-scale 
$\lambda=(15\nu u_{rms}^2/\varepsilon)^{1/2}$ 
and the associated Reynolds number 
$Re_\lambda=\lambda u_{rms}/\nu$, 
the large-eddy length-scale 
$L=k^{3/2}/\varepsilon$, 
the large-eddy turn-over time 
$T=u_{rms}^2/\varepsilon$, 
and the Kolmogorov time 
$\tau_\eta=(\nu/\varepsilon)^{1/2}$. 
The particle response time (based upon Stokes drag)
is given by $\tau_p=D^2\rho_p/(18\nu\rho_f)$; 
the Stokes number based upon Kolmogorov scales is then defined as
$St_\eta=\tau_p/\tau_\eta$, while a large-eddy Stokes number can be
written as $St_T=\tau_p/T$. 
The particle velocity vector of the $i$th particle is denoted by
$\mathbf{u}_p^{(i)}(t)$, and analogously for the position vector 
$\mathbf{x}_p^{(i)}(t)$. The instantaneous average over the set of particles is
denoted by the operator $\langle \cdot\rangle_p$; 
as a shorthand we use the angular brackets without subscript
``$\langle\cdot\rangle$'' for the combined average over the respective
phase and over time (in the statistically stationary interval). 

In the present work we consider cubic boxes with side-lengths ${\cal
  L}_x={\cal L}_y={\cal L}_z$ along the 
three Cartesian coordinate directions $(x,y,z)$. The flow field and
the particle motion are assumed triply-periodic in space. 
Two different particulate flow cases are simulated, with global
physical parameters as shown in table~\ref{tab-phys-params}. 
The solid-to-fluid density ratio of $1.5$ as well as the global solid
volume fraction of $0.005$ are kept the same in these two cases. 
The principal difference is the ratio of the particle diameter and the 
Kolmogorov scale, which measures $5.5$ and $11$, respectively.   
Henceforth, these cases will be denoted as ``D5'' and ``D11''. 
Homogeneous-isotropic turbulence is generated and maintained
with the aid of the random forcing scheme of \cite{eswaran:88}, as
described in detail by \cite{chouippe:15a}. 
Two slightly different turbulence-forcing parameter sets have been
used for the two cases D5 and D11 (cf.\
table~\ref{tab-turbforce-params}) which lead to mildly  different
Reynolds numbers of $Re_\lambda=115$ and $142$, respectively.
In the following we will also use data from two single-phase
simulations (denoted as ``S5'' and ``S11'') which have been performed
with identical turbulence-forcing parameters as the respective
particulate cases \citep[note that the present case S11 is identical
to case B of][]{chouippe:15a}.

\begin{figure}%
  \begin{minipage}{2ex}
    \rotatebox{90}
    {budget}
  \end{minipage}
  \begin{minipage}{0.45\linewidth}
    \centerline{$(a)$}
    \includegraphics[width=\linewidth]
    {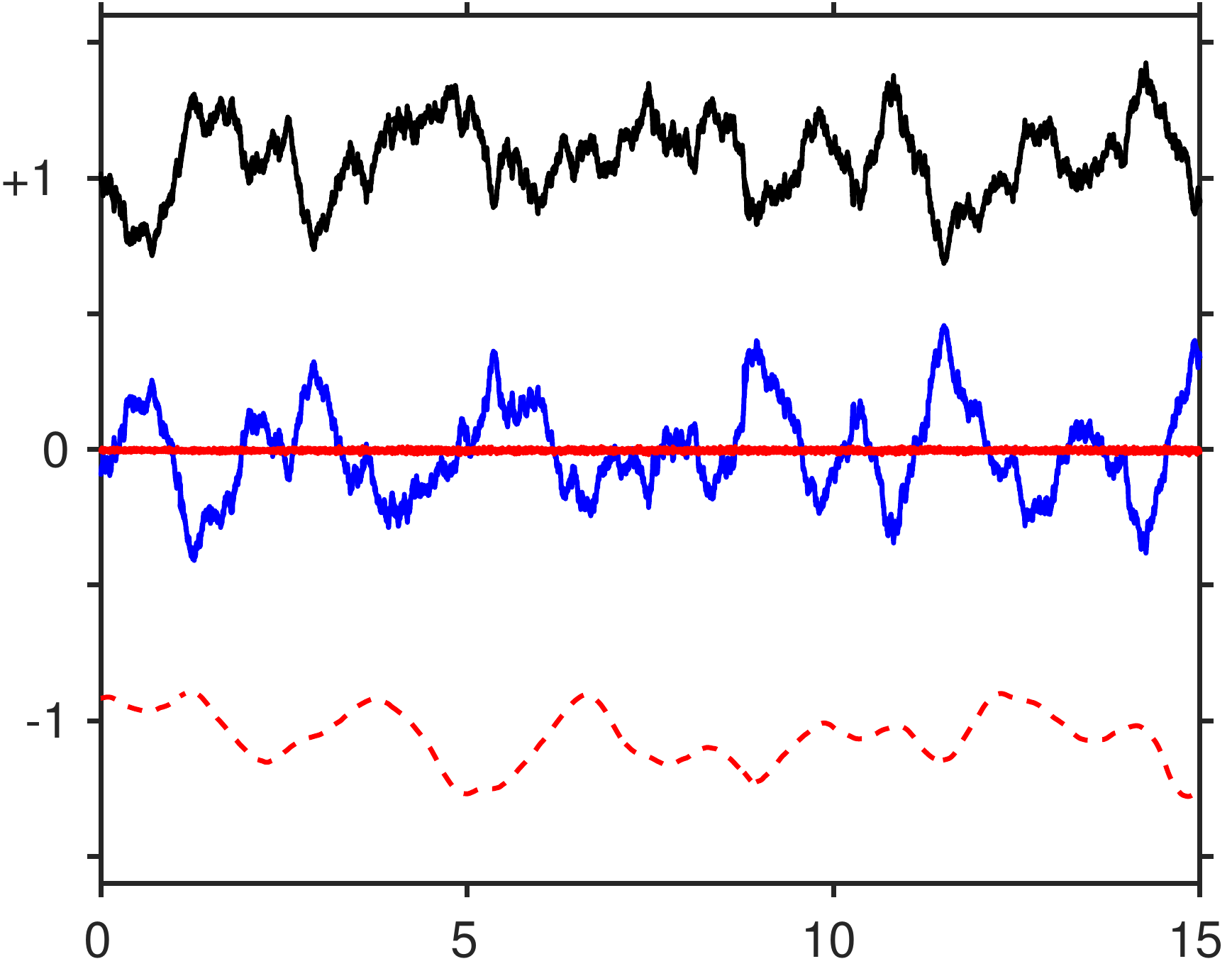}
    \\
    \centerline{$t/T$}
  \end{minipage}
  \hfill
  \begin{minipage}{2ex}
    \rotatebox{90}
    {budget}
  \end{minipage}
  \begin{minipage}{0.45\linewidth}
    \centerline{$(b)$}
    \includegraphics[width=\linewidth]
    {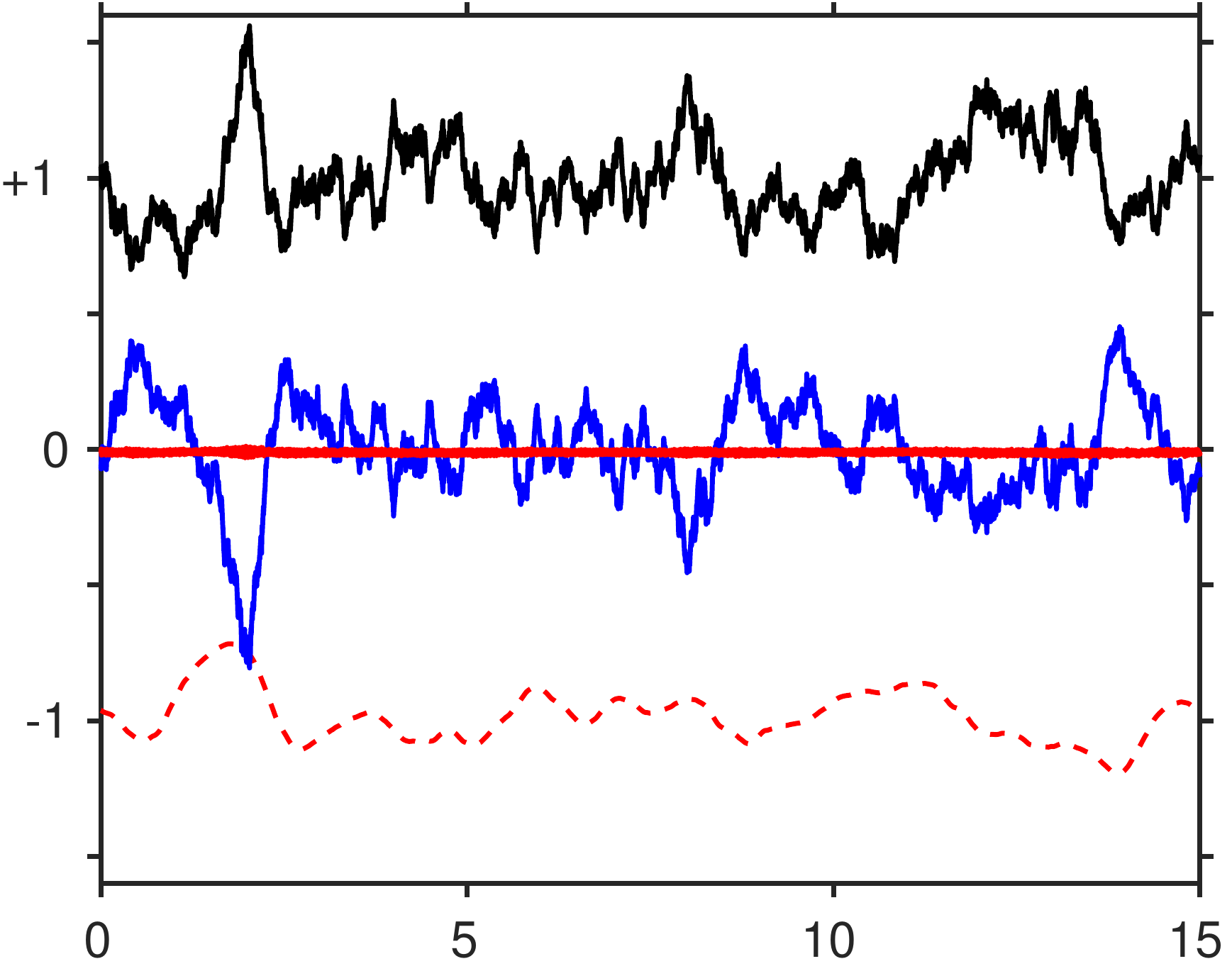}
    \\
    \centerline{$t/T$}
  \end{minipage}
  \caption{%
    Time evolution of the terms in the equation for kinetic energy
    $E_k=\mathbf{u}\cdot\mathbf{u}/2$, averaged over the computational domain: 
    $(a)$ case D5; 
    $(b)$ case D11.
    As shown by \cite{chouippe:15a} the evolution equation reads as
    follows:  
    $0=
    -\mbox{d}\langle E_k\rangle_\Omega/\mbox{d}t
    -\varepsilon_\Omega
    +\Psi^{(t)}
    +\Psi^{(p)}$, 
    where $\langle\cdot\rangle_\Omega$ indicates
    averaging over the spatial domain, 
    $\varepsilon_\Omega$ is the box-averaged instantaneous dissipation
    rate (in the graph: {\color{red}\dashed}), 
    $\Psi^{(t)}$ the power-input due to turbulence forcing
    ({\color{black}\solidthick}), and  
    $\Psi^{(p)}$ the two-way coupling term
    ({\color{red}\solidthick});  
    the time-rate-of-change term is plotted in blue
    ({\color{blue}\solidthick}).  
  }
  \label{fig-ek-budget-vs-time}
\end{figure}

Table~\ref{tab-num-params} shows the further numerical parameters
which characterize the present particulate simulations. It can be seen
that the computational domain is fairly large with respect to the
large-eddy size $L$, which is also manifest through a reasonable  
decay of the energy spectrum at small wavenumbers (not shown). 
The length of one edge of the domain corresponds to $128$ ($64$)
particle diameters in case D5 (D11). 
The small-scale resolution can be considered as excellent with $2.9$
($1.4$) grid widths per Kolmogorov length in case D5 (D11), and $16$
grid widths covering one particle diameter. 
The domain accomodates a comparatively large number of roughly $20000$
and $2500$ finite-size particles in case D5 and D11, respectively. 
Finally, it is important to note that we have run the simulations for
sufficiently long observation intervals $T_{obs}$ after discarding
initial transients (cf.\ table~\ref{tab-num-params}), both in terms of
the large-eddy time-scale $T$ and in terms of the particle response
time $\tau_p$.  
The Stokes number values of the two simulations are as follows: with
respect to the Kolmogorov scales we have 
  $St_\eta=2.5$ 
($10.7$) in
case D5 (D11); 
with respect to the energetic eddies the values are 
  $St_T=0.06$ 
and $0.29$, respectively. 
  Note that some authors define the particle time scale in a slightly
  different manner in the case of moderate density-ratio particles,
  viz.\ $\tilde{\tau}_p=D^2(1+2\rho_p/\rho_f)/(36\nu)$; this
  alternative definition then leads to Stokes numbers which are 
  larger by a factor $4/3$ than with the presently chosen definition
  $\tilde{\tau}_p$.   
The fact that the systems have indeed reached a statistically
stationary state has been determined by monitoring various statistical
quantities of both phases, some of which will be discussed below
(e.g.\ the kinetic energy budget shown in
figure~\ref{fig-ek-budget-vs-time} and the Vorono\"i tesselation of
the disperse phase in figure~\ref{fig-voronoi-cell-vols}), others not
being shown here (e.g.\ the longitudinal velocity derivative skewness). 
\section{Results}  
\label{sec-results}
\subsection{Basic Eulerian and Lagrangian statistics}
\label{sec-results-basic-stats}
Figure~\ref{fig-ek-budget-vs-time} shows the time-evolution of the
terms in the budget of the kinetic energy when averaging over the
entire computational domain \citep[cf.][]{chouippe:15a}.  
In both cases the turbulence-forcing power-input is on time-average
balanced by the dissipation rate, while the two-way coupling term does
not play a significant role 
(its time-average value corresponds to $0.004$ and $0.016$ times
that of the dissipation rate in case D5 and D11, respectively). 
The instantaneous imbalance between power input and dissipation leads
to temporal fluctuations of the turbulent kinetic energy, whose
temporal variance, however, is smaller than 8.5\% in both cases.
Even though the direct effect of the presence of solid particles
through the two-way coupling term is practically negligible, the
presence of the particles might still have an important impact upon
the flow. In the present configurations the comparison between
particle-laden and unladen flows shows that this is not the case. 
In particular, the time-averaged turbulent kinetic energy is only
slightly reduced to $k=0.99\,k_{SP}$ (where the sub-script ``SP''
refers to the  corresponding single-phase simulation) in both cases,
while the value of the time-averaged dissipation rate either mildly 
increases  
    ($\varepsilon=1.03\,\varepsilon_{SP}$ in case D5) 
or slightly decreases ($\varepsilon=0.99\,\varepsilon_{SP}$ in case
D11). As a result, the obtained Taylor-scale Reynolds numbers in the
present particulate flow simulations are mildly different from the
unladen counterparts: 
  $Re_\lambda/Re_\lambda^{(SP)}=0.98$ in case D5,
and $Re_\lambda/Re_\lambda^{(SP)}=0.99$ in case D11.
As a conclusion of this short analysis of the global flow quantities
we can state that the addition of particles at the current parameter
points does not induce a significant modification of the carrier flow,
except possibly locally in the vicinity of the particles. This
statement will be further corroborated by the following analysis. 
Note that an investigation of the mechanisms of particle-induced
turbulence attenuation is outside the present scope, and the reader is
referred to \cite{lucci:10,lucci:11} for a detailed discussion in the
context of finite-size particles suspended in decaying
homogeneous-isotropic turbulence.  

The velocity field in homogeneous-isotropic turbulence is
well-known to be nearly Gaussian distributed, with a slightly
sub-Gaussian flatness \citep{jimenez:98b,wilczek:11}. 
We observe the same distribution for the fluid velocity in the
particulate case, the flatness measuring $2.89$ at both presently
investigated parameter points (figure omitted). 
The one-component p.d.f.\ of particle velocity (figure omitted) is
also found to be close to a Gaussian distribution, with a flatness
measuring $2.84$ ($2.82$) in case D5 (D11). This result is consistent
with the observations of previous authors in similar flow
configurations at roughly comparable parameter points
\citep{homann:10,YeoClimentMaxey:10}.  

\begin{figure}%
  \centering 
  \begin{minipage}{2ex}
    \rotatebox{90}
    {$(\langle|\mathbf{u}_f|^2\rangle-\langle|\mathbf{u}_p|^2\rangle)/
      \langle|\mathbf{u}_f|^2\rangle$}
  \end{minipage}
  \begin{minipage}{0.45\linewidth}
      \includegraphics[width=\linewidth]
      {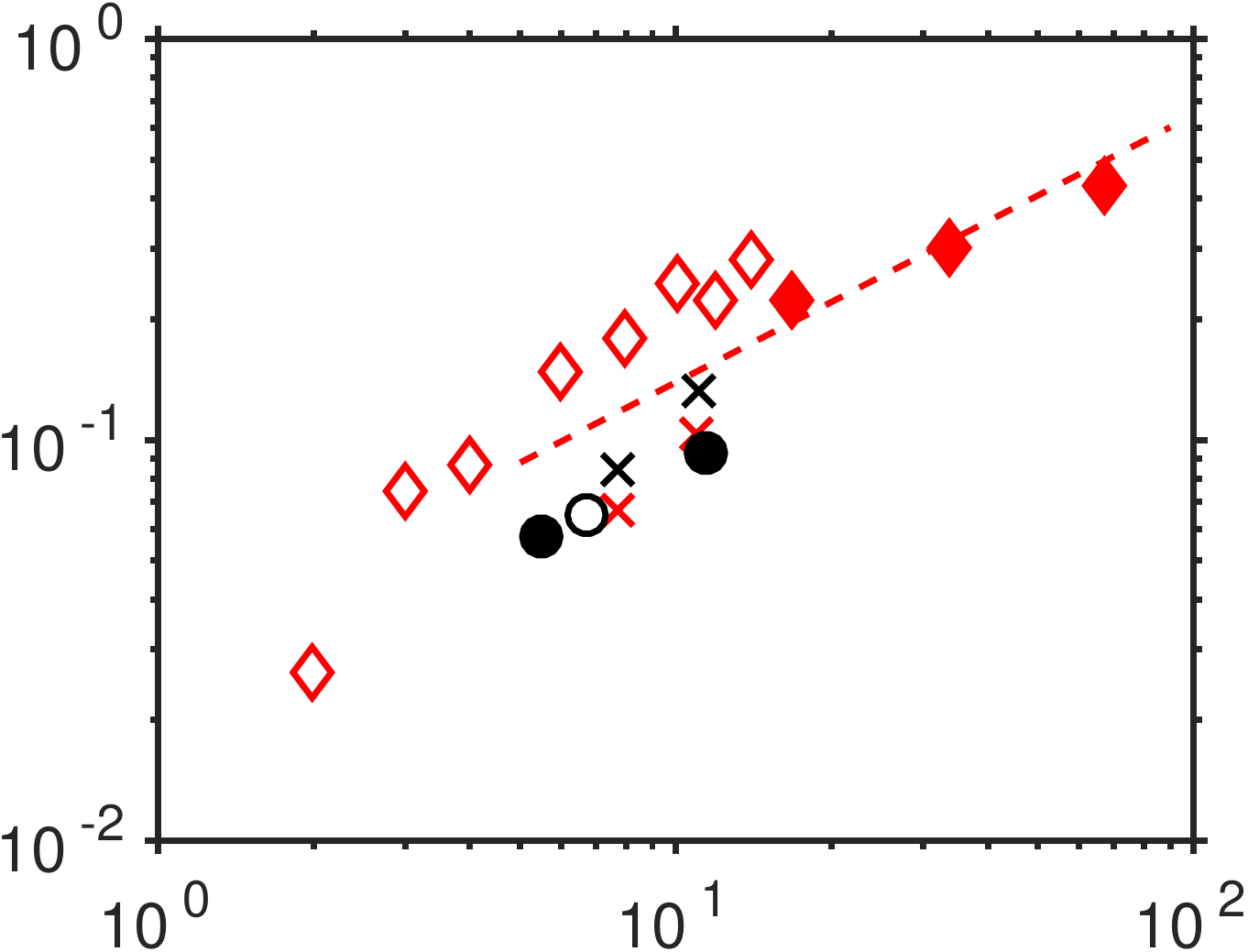}
    \\
    \centerline{$D/\eta$}
  \end{minipage}
  \caption{%
      Normalized difference between the variance of the fluid velocity
      and that of the particle velocity. 
    The symbols correspond to:
    {\color{red}$\mathsmaller{\lozenge}$}, \cite{homann:10}; 
    {\color{red}\soliddiamond}, \cite{cisse:15}; 
    {\color{red}$\times$}, density-matched particles (cases ``N1'',
    ``N2'') of \cite{YeoClimentMaxey:10}; 
    {\color{black}$\times$}, particles with density ratio
    $\rho_p/\rho_f=1.4$ (cases ``S1'', ``S2'') of
    \cite{YeoClimentMaxey:10};  
    {\color{black}$\bullet$}, present simulations; 
    {\color{black}$\circ$}, case ``A-G0'' from \cite{chouippe:15a}. 
    The dashed line ({\color{red}\dashed}) indicates a power-law
    proportional to $(D/\eta)^{2/3}$. 
  }
  \label{fig-part-vel-variance-diff}
\end{figure}

Let us turn to the variance of the velocities, which can be
conveniently discussed in terms of the relative difference between the
value of the fluid phase and the one of the particle phase, viz.\  
$\langle|\mathbf{u}_f|^2\rangle
-\langle|\mathbf{u}_p|^2\rangle)/
\langle|\mathbf{u}_f|^2\rangle$. 
It is clear that the tracer limit corresponds to a vanishing
difference, and it can be expected that the value increases with
particle size. 
\cite{homann:10} have shown that the difference of the velocity
variance roughly follows a $D^{2/3}$ power-law for sufficiently large
particle diameters, roughly for $D/\eta\gtrsim5$. 
As remarked by these authors, dimensional arguments in the
inertial-range indeed suggest such a dependency. 
Note that this scaling implies the following assumptions:  
(i) an inertial-range exists (i.e.\ sufficiently large Reynolds number
$Re_\lambda$);
(ii) supposing that the particle velocity is determined by flow scales
of size equal to or larger than the particle diameter. 
Figure~\ref{fig-part-vel-variance-diff} shows that although
assumption~(i) is not completely fulfilled by the numerical
simulations, such a power-law is found to be consistent with the data
of \cite{homann:10} and the more recent data of \cite{cisse:15}
(performed at larger Reynolds number). 
The $2/3$ power-law also roughly matches the variation of the
difference of velocity variance measured in the simulations of
\cite{YeoClimentMaxey:10}, which feature an ensemble of particles, in
both density-matched and non-density-matched cases (with a
density ratio of $\rho_p/\rho_f=1.4$). In the latter case it was
observed that the mild excess of particle density has the tendency to
increase the difference of velocity variances somewhat -- as might be
expected from the enhanced particle inertia. 
The present data-set, together with case A-G0 of \cite{chouippe:15a}, 
adds three data-points to figure~\ref{fig-part-vel-variance-diff}. 
\%
  It can be seen that our present results are consistent with the
  power-law proposed by \cite{homann:10} for particles with diameter
  $D/\eta\gtrsim3$.  
  It should be noted that the difference of velocity variances at
  small particle diameters (particularly in case D5) is a very small
  quantity compared to the velocity variances themselves, which makes
  it sensitive to statistical sampling errors. 
  This underlines the necessity to consider observation intervals
  which are large multiples of the large-eddy time-scales. 

A quantity of considerable interest for the purpose of modeling the 
particle motion (e.g.\ in the context of a point-particle approach)
is the relative velocity between the two phases. 
Note that even the definition of a relevant fluid velocity in the
vicinity of the particles (or equivalently a fluid velocity ``seen''
by the particles) is a matter of debate, and various alternative
expressions have been proposed
\citep{bagchi:03,lucci:10,klein:12,kidanemariam:13,cisse:13,uhlmann:14a}.  
Here we proceed as in \cite{uhlmann:14a}, and define the fluid
velocity seen by the $i$th particle, $\mathbf{u}_{f}^{(i)}$,  by
computing an average of the fluid velocity over the surface of a
sphere with diameter $D_S$, centered at the $i$th particle's center
location. 
  Recall that the averaging diameter should neither be too
  small (in which case the fluid velocity will tend to the particle
  velocity as the no-slip boundary is approached) nor too large (in
  which case the resulting fluid velocity will no longer be
  relevant to the motion of the particle under consideration). 
  Based upon our previous work \citep{kidanemariam:13} 
  we have chosen the averaging diameter as $D_S=3D$ throughout the 
  present study. 
We can then compute a relative velocity 
$\mathbf{u}_{pr}^{(i)}=\mathbf{u}_{p}^{(i)}-\mathbf{u}_{f}^{(i)}$  
from which we define an instantaneous particle Reynolds
number $Re_p^{(i)}(t)=|\mathbf{u}_{pr}^{(i)}|D/\nu$. 
Note that the probability distributions of the components of the
relative velocity $\mathbf{u}_{pr}^{(i)}$ (figure omitted) feature
exponential tails.  
The p.d.f.\ of the particle Reynolds number $Re_p$ (normalized to 
variance unity) is then found to be close to a Gamma distribution with
shape parameter $k=3$, as shown in figure~\ref{fig-rep-shell-avg}.
The match is particularly good in case D5, except for extremely large 
values of $Re_p$, where the p.d.f.\ is somewhat wider than the tail of
the Gamma distribution 
(the flatness measures $5.9$, while the
flatness is equal to $5$ for the Gamma distribution). 
Case D11, on the other hand, exhibits a slightly narrower shape with a
flatness of $4.69$.
The mean and variance of the particle Reynolds number recorded in the
two cases are also listed in figure~\ref{fig-rep-shell-avg}. 
They demonstrate that appreciable values of $Re_p$ can be attained 
instantaneously, although the particle sizes are still relatively
small and the Stokes numbers are not enormously high. In fact, the
particle Reynolds numbers are sufficient for the formation of attached
vortical structures in the vicinity of the particles, as will be
further discussed in \S~\ref{sec-results-preferential}. 
  These findings suggest that a quasi-steady linear drag assumption
  would not be appropriate when modelling the hydrodynamic force in
  the framework of a point-particle model in the present parameter
  range. 
  We have checked the sensitivity of these results with respect to the
  choice of the averaging diameter $D_S$: 
  going from $D_S=2.75D$ to $D_S=3.25D$ increases
  the mean particle Reynolds number from $6.8$ to $9.0$ ($29.8$ to
  $34.4$) in case D5 (D11). 

\begin{figure}%
  \centering
  \begin{minipage}{2ex}
    \rotatebox{90}
    {p.d.f.}
  \end{minipage}
  \begin{minipage}{0.45\linewidth}
    \includegraphics[width=\linewidth]
    {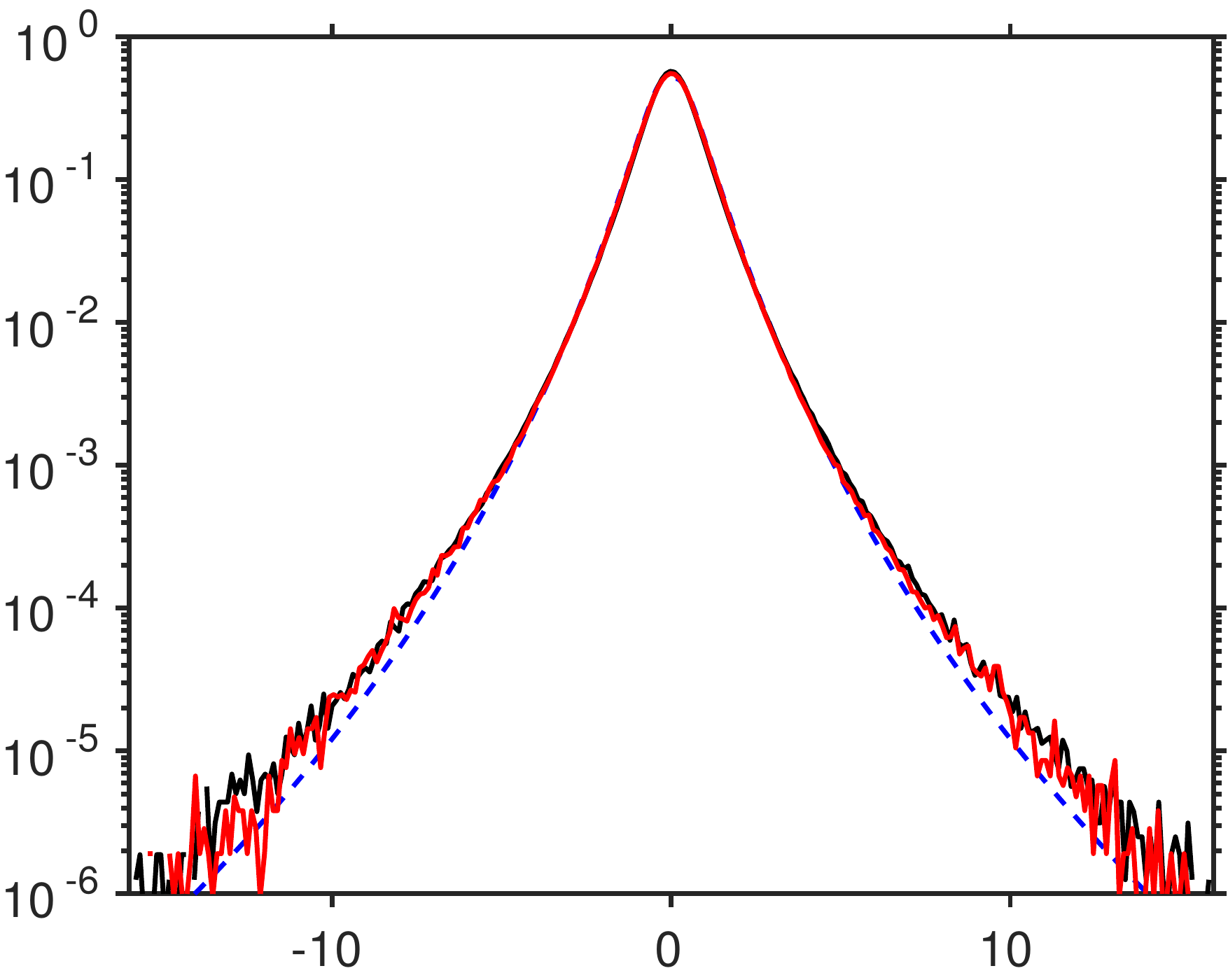}
    \\
    \centerline{$a^{(i)}_{p,\alpha}/\sigma(a^{(i)}_{p,\alpha})$}
  \end{minipage}
  \caption{%
    Normalized p.d.f.s of the linear particle acceleration. Samples
    corresponding to particles in contact with other particles have
    been eliminated. 
    Line styles correspond to: 
    {\color{black}\solidthick},~case~D5; 
    {\color{red}\solidthick},~case~D11; 
    {\color{blue}\dashed}, fit of experimental data proposed by
    \cite{qureshi:07}.  
  }
  \label{fig-part-accel-pdf}
\end{figure}

The statistics of particle acceleration are still a matter of debate
in the literature. 
It is now well established that the normalized p.d.f.\ of the linear
particle acceleration has a nearly universal shape as proposed by
\cite{mordant:04} and fitted to an experimental data-set with a wide
range of parameters by \cite{qureshi:07}. 
The data corresponding to the two present simulations in 
figure~\ref{fig-part-accel-pdf} nearly overlap, and they are indeed
very well represented by the empirical formula of these latter
authors,  
although some deviation in the extreme tails leads to a somewhat
larger flatness of $10.58$ ($9.86$) in case D5 (D11) as compared to
the original empirically fitted value of $8.37$.
Turning now to the variance of particle acceleration, the appropriate
scaling with particle size is not clear to date. 
The data from von K\'arm\'an flow experiments by \cite{voth:02}, 
\cite{brown:09} and \cite{volk:11}, 
as well as the wind-tunnel data of \cite{qureshi:07}
suggest a $(D/\eta)^{-2/3}$ power-law for particle diameters exceeding
a few multiples of the Kolmogorov length. Such a power-law can be
inferred from classical Kolmogorov inertial-range arguments, supposing
that the particle acceleration will be principally determined by flow
scales of the order of the particle diameter. 
\cite{volk:11} have refined the analysis and suggested an
intermittency correction which accounts for the slightly steeper slope
(of $D^{-0.81}$) in their measurements. 
Figure~\ref{fig-part-accel-variance} shows the particle acceleration
variance normalized by the fluid counterpart, which 
allows the comparison of data measured at different Reynolds numbers
\citep[cf.\ ][for the scaling of the fluid acceleration variance with
Reynolds number]{vedula:99,voth:02}.  
On the other hand, the normalized particle acceleration variance from
the DNS of \cite{homann:10} (performed at $Re_\lambda=32$) and
\cite{cisse:15} ($Re_\lambda=160$) approximately follows a
$(D/\eta)^{-4/3}$ trend. 
\cite{homann:10} argue that at low Reynolds numbers the predominant
mechanism affecting the particle motion is sweeping by the large
scales, which might explain the different scaling. 
Now let us consider the particle acceleration variance obtained
in the present simulations. 
Recall that the principal difference with respect to the previously
mentioned experiments and simulations is the fact that the present
particles are not density-matched. 
It can be seen from figure~\ref{fig-part-accel-variance} that the
results for case D5 and D11, as well as the one for case A-G0 of
\cite{chouippe:15a} are consistent with the results of \cite{homann:10}
and \cite{cisse:15} to within the scatter of the available
reference data-points. 
This observation suggests that the effect of a mild excess particle
density ($\rho_p/\rho_f=1.5$) as well as the collective effects in the
present dilute regime ($\Phi_s=0.005$) 
do not have a significant impact upon the particle acceleration
  variance in the absence of gravity. 
The data-set of \cite{YeoClimentMaxey:10}, which features parameters
comparable to our present simulations, allows us to discuss the
specific impact of the particle-to-fluid density ratio upon the
particle acceleration variance. It should be noted that the authors 
have estimated the fluid acceleration variance from the value provided
by \cite{yeung:89} in a separate single-phase simulation at the same
Reynolds number. These data points are included in the present
figure~\ref{fig-part-accel-variance}; however, they should probably be
treated with some caution. 
It can be seen that increasing the particle density from the
density-matched value by a factor of $1.4$ does not have a very large
effect in the work of \cite{YeoClimentMaxey:10}. 
Both data points (for $D/\eta=7.7$ and $11$) are shifted by
roughly the same fraction downward, i.e.\ increasing density decreases
the acceleration variance, as can be expected. 
Note that the ratio between the value for $\rho_p/\rho_f=1.4$ and for the
density-matched counterparts amounts to approximately $0.8$ (for both
particle diameters), which is substantially larger than what might be
expected when directly applying the factor due to the density ratio,
i.e.\ $(\rho_p/\rho_f)^{-2}\approx0.51$, which enters through the
Newton-Euler equation for particle motion. 
As a conclusion of this short discussion of density effects, it can be 
stated that according to the work of \cite{YeoClimentMaxey:10} the
variance of particle acceleration varies more slowly than inversely
proportional to the solid-to-fluid density ratio. 
It should also be pointed out that \cite{qureshi:08} have revealed
intricate effects of particle density upon the acceleration variance;
however, their laboratory measurements were performed in the presence
of non-zero gravitational acceleration, and, therefore, a direct
comparison with the present data does not seem pertinent. 

\begin{figure}%
  \begin{minipage}{2ex}
    \rotatebox{90}
    {auto-correlation coefficient}
  \end{minipage}
  \begin{minipage}{0.45\linewidth}
    \centerline{$(a)$}
    \includegraphics[width=\linewidth]
    {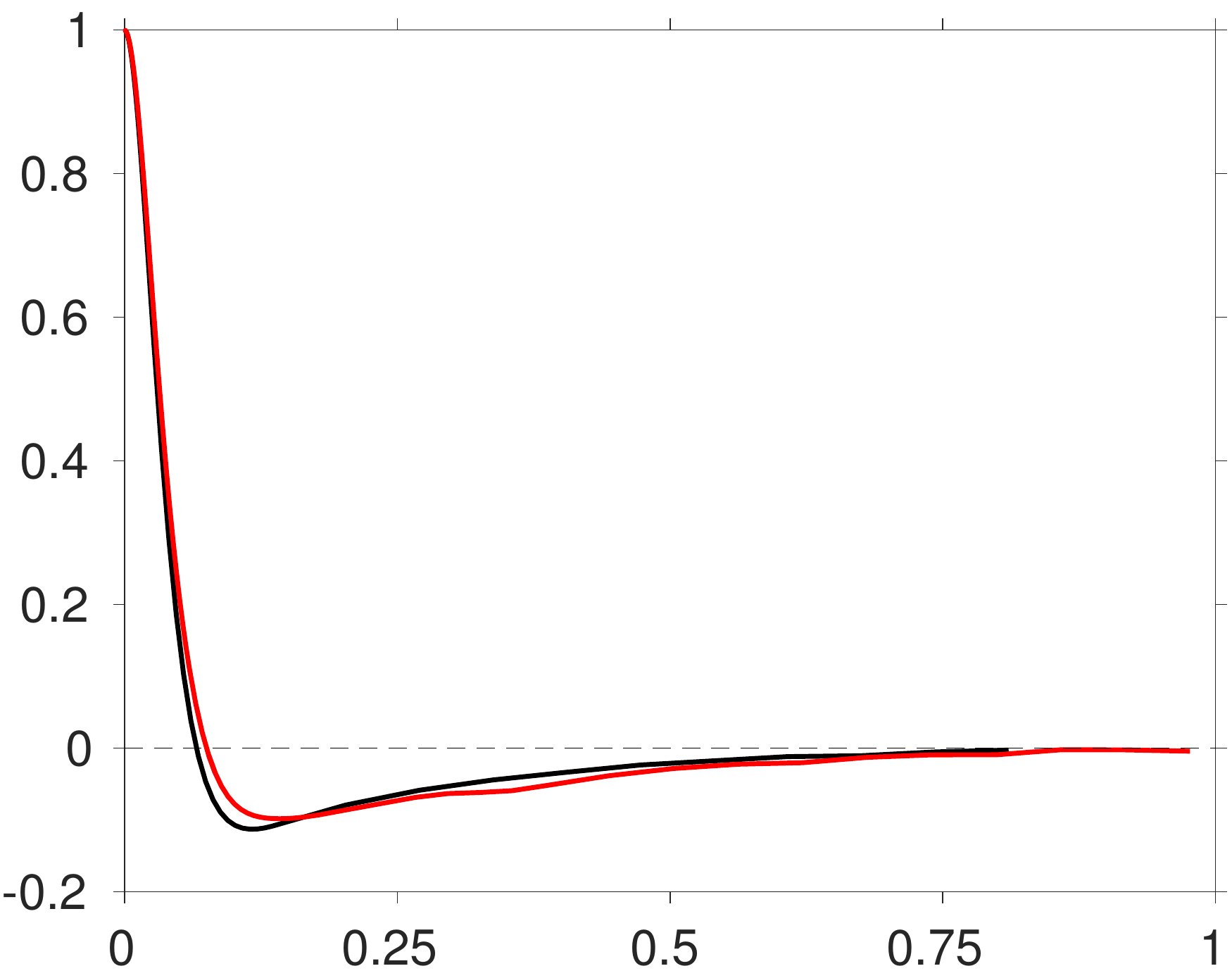}
    \hspace*{-.75\linewidth}
    \raisebox{.48\linewidth}{
      \begin{minipage}{0.55\linewidth}
        \includegraphics[width=\linewidth]
        {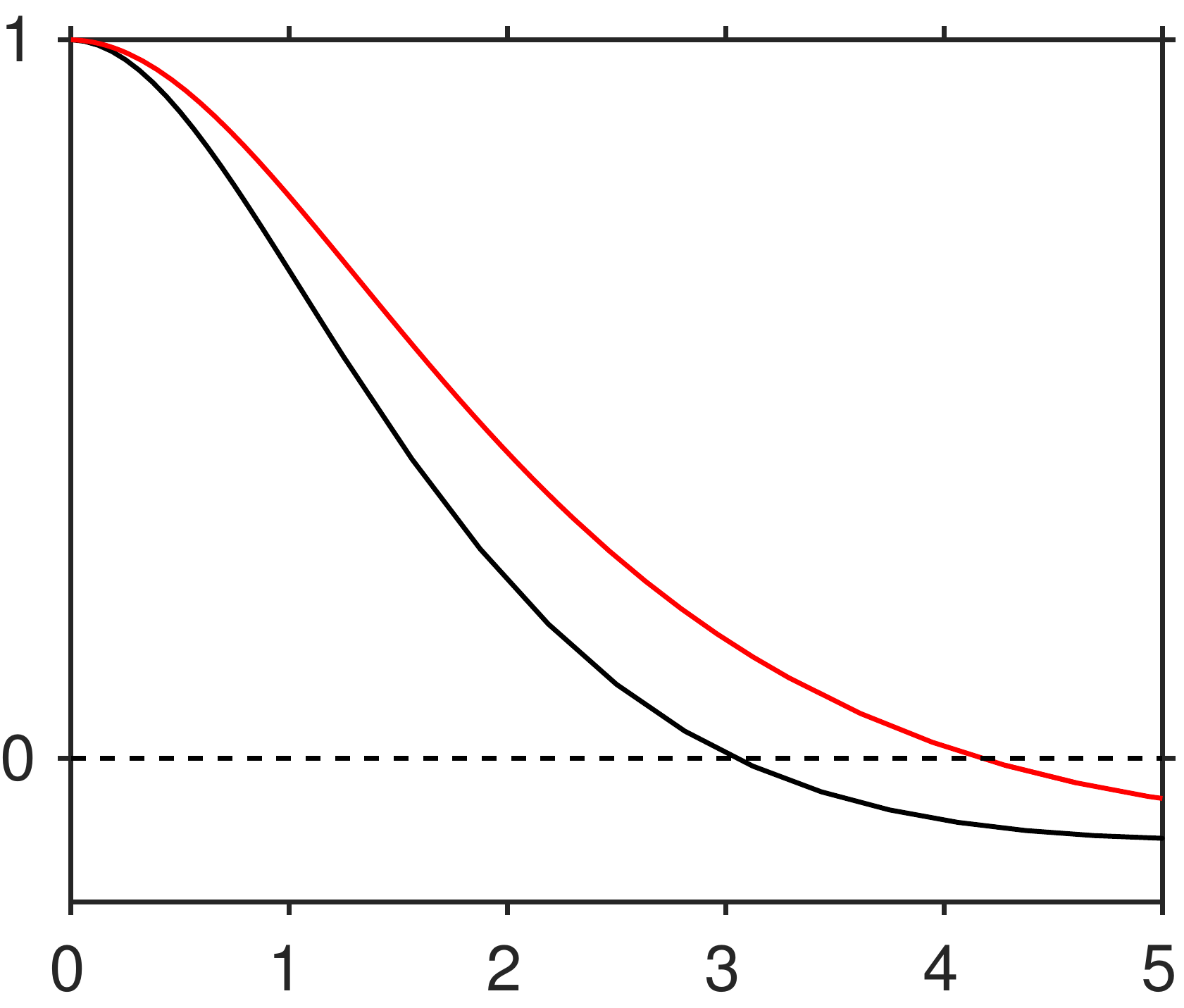}
        \\
        \centerline{$\tau/\tau_\eta$}
      \end{minipage}
    }
    \\
    \centerline{$\tau/T$}
  \end{minipage}
  \hfill
  \begin{minipage}{2ex}
    \rotatebox{90}{$\tau_{int}/\tau_\eta$}
  \end{minipage}
  \begin{minipage}{0.45\linewidth}
    \centerline{$(b)$}
    \includegraphics[width=\linewidth]
    {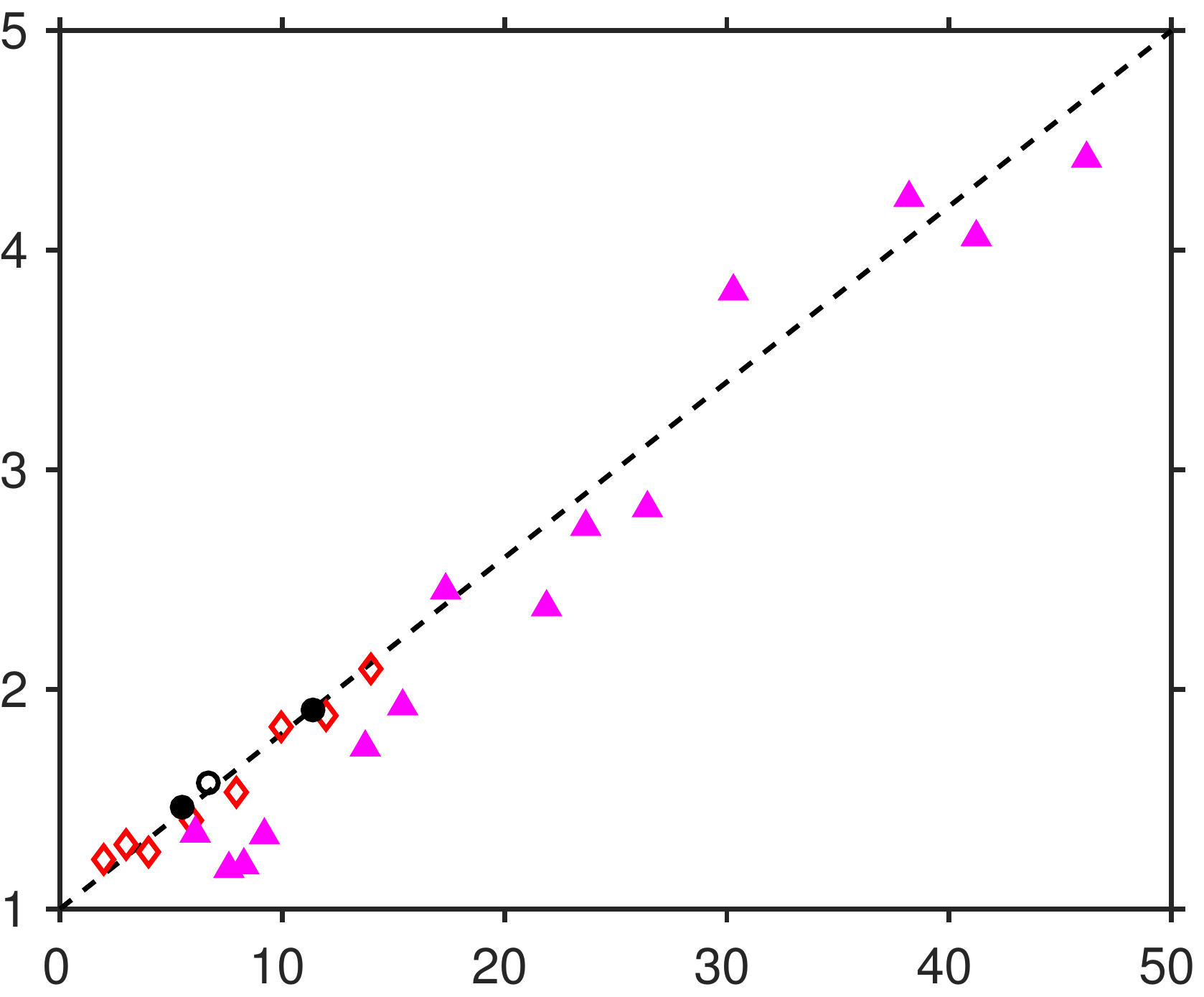}
    \\
    \centerline{$D/\eta$}
  \end{minipage}
  \caption{%
    $(a)$ 
    Lagrangian auto-correlation coefficient of one-component, linear
    particle acceleration, normalized with the large-eddy turnover
    time $T$. 
    The correlations have been computed during collision-free
    intervals only. 
    The inset shows the same data for smaller separation times $\tau$,
    and normalized in Kolmogorov time units $\tau_\eta$.
    Line styles correspond to: 
    {\color{black}\solidthick},~case~D5; 
    {\color{red}\solidthick},~case~D11,  
    $(b)$ The integral scale $\tau_{int}$ of the auto-correlation
    (from integration up to the first zero-crossing) shown in $(a)$,
    plotted as a function of the particle size. 
    The present data is indicated by solid circles ($\bullet$); 
    the black open circle ({\color{black}$\circ$}) indicates the data
    of case ``A-G0'' from \cite{chouippe:15a}. 
    The magenta-colored triangles 
    ({\color{magenta}$\mathsmaller{\blacktriangle}$}) correspond to
    data from the von K\'arm\'an flow experiments of \cite{volk:11}; 
    the red diamonds ({\color{red}$\mathsmaller{\lozenge}$}) represent
    the simulation data of \cite{homann:10}. 
    The dashed line (\dashed) indicates a linear increase 
    $\tau_{int}/\tau_\eta=1+0.08(D/\eta)$
  }
  \label{fig-part-accel-auto-corr}
\end{figure}

As a final point in this section we wish to discuss the Lagrangian 
auto-correlation of particle acceleration. This quantity provides the
information on the time scales over which particles are accelerated, 
which was so far missing from the above discussion. 
Figure~\ref{fig-part-accel-auto-corr}$(a)$ shows the data in both
large-eddy scaling and Kolmogorov scaling. It can be seen that the
long-time behavior of both present cases nearly overlaps, with a
negative loop and a gradual approach to zero from below on the order
of one large-eddy turnover time. The short-time behavior, i.e.\ the
first zero-crossing time-scale, on the contrary, increases with
particle diameter. 
Let us define an integral time-scale $\tau_{int}$ by integration of
the auto-correlation coefficient up to the time of the first
zero-crossing. Figure~\ref{fig-part-accel-auto-corr}$(b)$ shows the
variation of $\tau_{int}$ as a function of the particle diameter, both
axes being normalized with Kolmogorov scales. 
The data for cases D5, D11 and case A-G0 of \cite{chouippe:15a} is
obviously well represented by a linear variation
$\tau_{int}/\tau_\eta=1+0.08(D/\eta)$. 
The same is true for the DNS results of \cite{homann:10}, and -- with
some scatter -- for the experimental data of \cite{volk:11}. 
However, as remarked by \cite{volk:11}, over the limited range of
particle sizes one cannot definitely exclude an alternative scaling
with $(D/\eta)^{2/3}$, as implied by classical Kolmogorov
inertial-range arguments.  
The main conclusion from the present data is the insensitivity of the
Lagrangian time scales which characterize the particle acceleration,
i.e.\ the hydrodynamic forces, with respect to mild variations of the
density ratio.

\subsection{Particle clustering}
\label{sec-results-clustering}
\begin{figure}%
   \begin{minipage}{2.5ex}
     \rotatebox{90}{$\sigma(V^{(i)}/\langle V^{(i)}\rangle)$}
   \end{minipage}
   \begin{minipage}{0.45\linewidth}
     \centerline{$(a)$}
      \includegraphics[width=\linewidth]
      {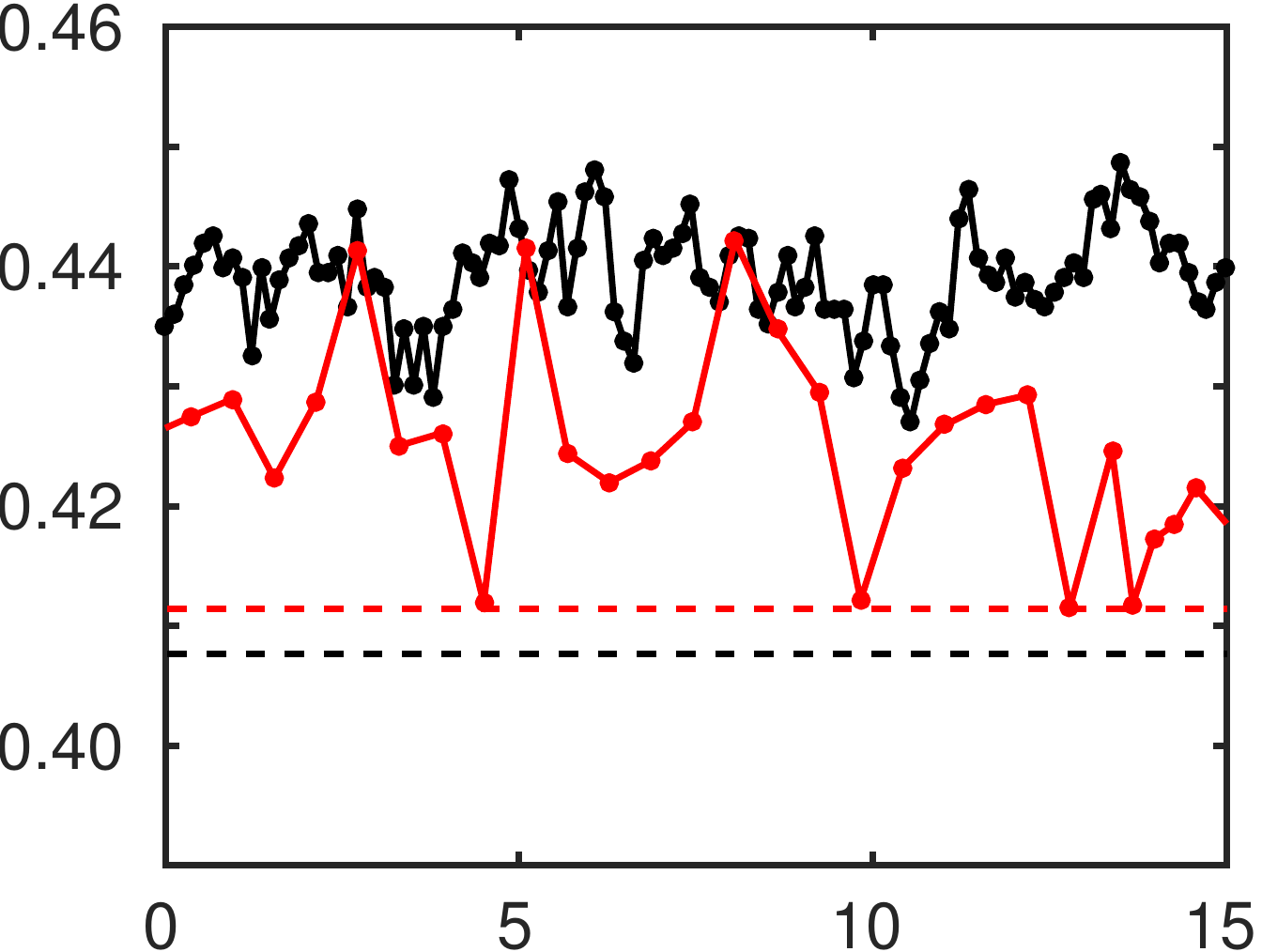}
      \\
      \centerline{$t/T$}
   \end{minipage}
   \hfill
   \begin{minipage}{2.5ex}
     \rotatebox{90}{$\sigma/\sigma_{RAND}$}
   \end{minipage}
   \begin{minipage}{0.45\linewidth}
     \centerline{$(b)$}
      \includegraphics[width=\linewidth]
      {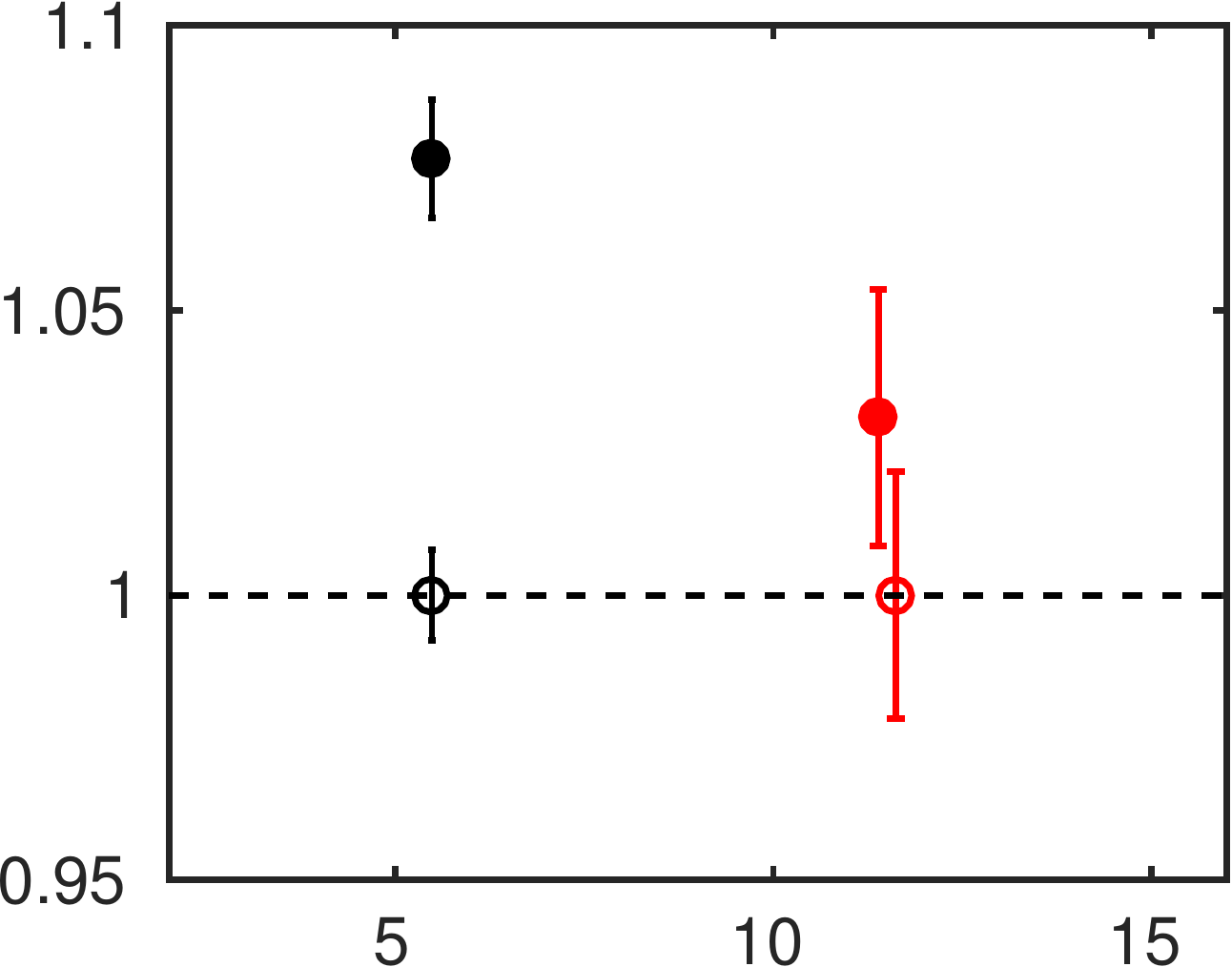}
      \\
      \centerline{$D/\eta$}
   \end{minipage}
   \caption{%
     $(a)$ 
     Time evolution of the standard deviation of the normalized volume
     of the cells of the three-dimensional Vorono\"i tesselation with
     the particle centers as ``sites''. 
     Line styles correspond to: 
     {\color{black}\solidthick},~case~D5; 
     {\color{red}\solidthick},~case~D11,  
     The dashed lines indicate the average value for a sequence of
     random distributions of non-overlapping spheres computed for the
     same number of particles and domain sizes as in the respective DNS. 
     $(b)$ 
     The time-average of the standard-deviation of the Vorono\"i cell
     volumes (normalized by the corresponding value for a random
     particle arrangement), plotted as a function of the particle
     diameter (normalized by the Kolmogorov length scale). The DNS
     data is indicated by circular symbols
     ({\color{black}$\bullet$},~case~D5; 
     {\color{red}$\bullet$},~case~D11), with the errorbars
     corresponding to the standard-deviation in time. 
     The errorbars attached to the random data (open circles) indicate
     the standard-deviation over the respective random sequence. 
     Note that the red open circle has been shifted slightly in the
     horizontal direction for clarity. 
   }
   \label{fig-voronoi-cell-vols}
\end{figure}

Vorono\"i tesselation is a computationally efficient means for the
analysis of the spatial structure of the dispersed phase
\citep{monchaux:10b,monchaux:12}. In the context of particulate flow,
one typically takes the particle centers as the sites for which a 
three-dimensional partition of space into cells is performed such that
each point in a given cell is located closer to the cell's site than
to any other site. The inverse of the volume of a Vorono\"i cell is
then a measure of the local concentration: a smaller cell corresponds
to a denser particle arrangement and vice-versa. Particle positions
drawn at random (from a random Poisson process) yield Vorono\"i cell
volumes which (when normalized with the mean cell volume) exhibit a
universal p.d.f.\ given by a Gamma distribution with parameters
determined numerically by 
\citet{ferenc:07}. 
When considering
finite-size particles, this result is no longer valid, since particles
should not overlap. Therefore, the parameters of the random
distribution need to be computed specifically for each given solid
volume fraction and box-size.  
It turns out that the data for the Vorono\"i cell volumes of the real
particle phase is also Gamma-distributed, however with a different
variance: those cases which exhibit ``clustering'' feature a larger
variance than the random data \citep[cf.][]{monchaux:12,uhlmann:14a}. 
This suggests that the variance of the normalized Vorono\"i cell
volumes can be taken as a single indicator of particle ``clustering'', 
i.e.\ of a particle arrangement which is statistically distinct from
that obtained by a random Poisson process in the sense that it
features more closely located particles and more voids than the
latter. 
Figure~\ref{fig-voronoi-cell-vols}$(a)$ shows the time evolution of the
variance of Vorono\"i cell volumes in the two present cases. 
  Also marked are the averages of the random data computed from a
  large ensemble for the present solid volume fraction and the two
  box-sizes. 
It can be seen that the variance in both cases D5 and D11
is clearly above the random value. It is also obvious that the same
quantity fluctuates more in time in case 
D11 which features 
an eight times 
smaller number of particles. 
In figure~\ref{fig-voronoi-cell-vols}$(b)$ we report the time-average
of the instantaneous variance of Vorono\"i cell volumes, normalized by
their respective average random value, along with their temporal
standard-deviation (in form of error-bars). It is clearly visible that
the particle phase in case D5 exhibits stronger clustering than in
case D11 with values of the variance which are 7.7\% and 3.1\% larger
than the random value, respectively. 
It should be noted that this clustering behavior can be qualified as 
relatively weak. For comparison, in the clustering case of
\cite{uhlmann:14a} the variance was approximately 50\% larger than the
random value. 

\begin{figure}%
    \begin{minipage}{2.5ex}
      \rotatebox{90}{$g(r)$}
    \end{minipage}
    \begin{minipage}{0.45\linewidth}
      \centerline{$(a)$}
      \includegraphics[width=\linewidth]
      {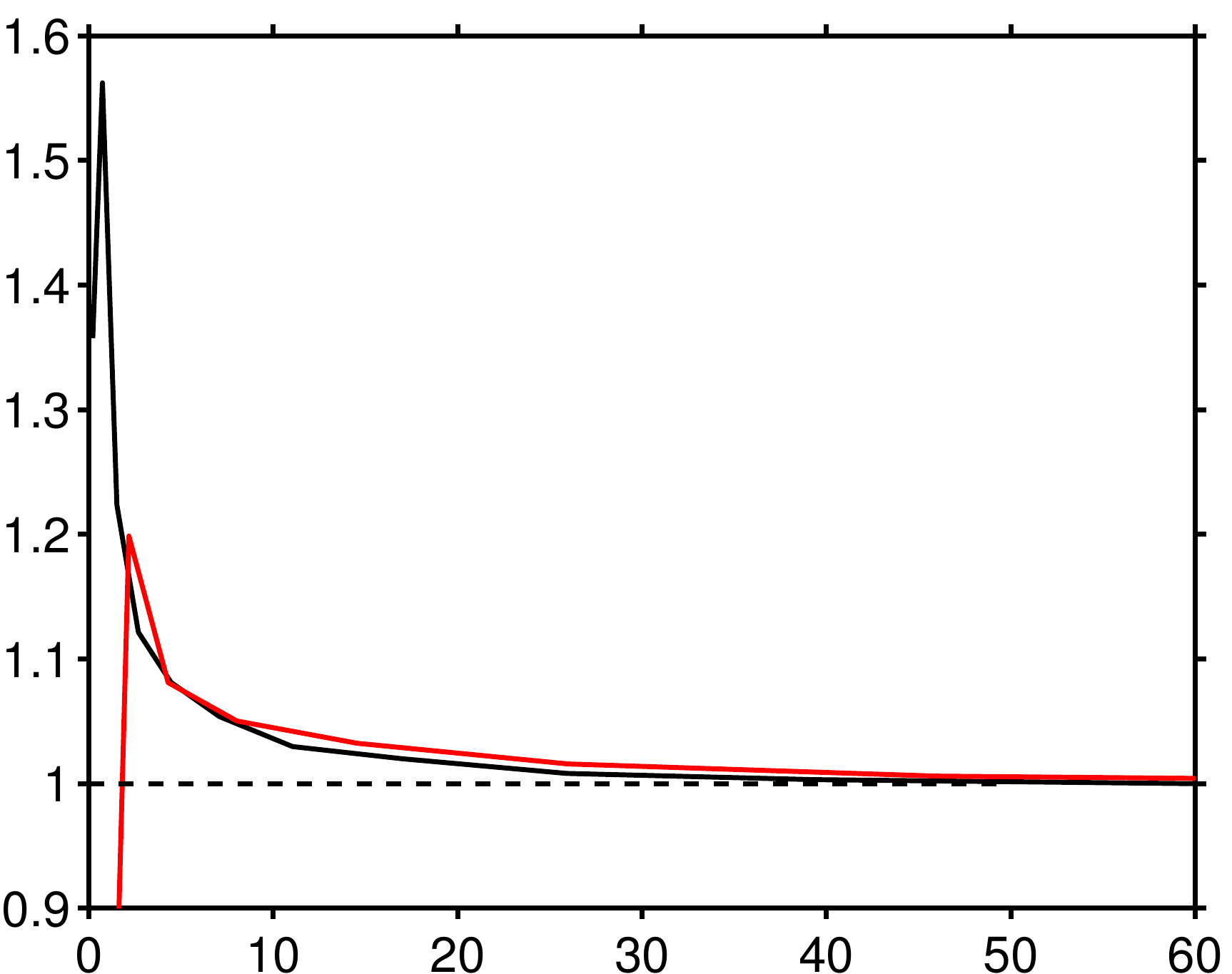}
      \\
      \centerline{$(r-D)/\eta$}
    \end{minipage}
    \hfill
    \begin{minipage}{1.5ex}
      \rotatebox{90}{$g(r)-1$}
    \end{minipage}
    \begin{minipage}{0.45\linewidth}
      \centerline{$(b)$}
      \includegraphics[width=\linewidth]
      {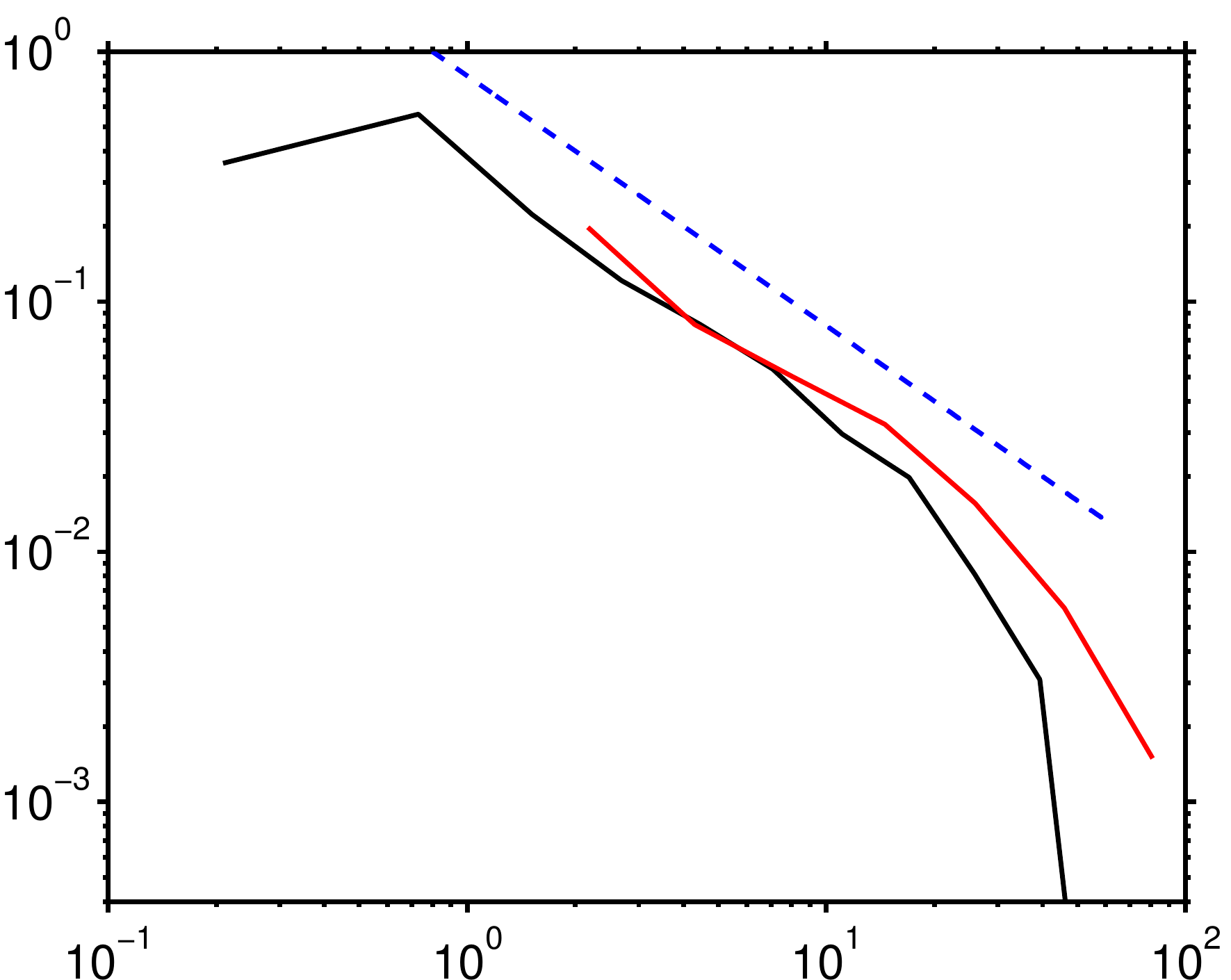}
      \\
      \centerline{$(r-D)/\eta$}
    \end{minipage}
    \caption{%
        $(a)$ 
        Normalized radial distribution function $g(r)$ 
        of particle positions in case D5 ({\color{black}\solidthick})
        and D11 ({\color{red}\solidthick}) in the statistically
        stationary regime.   
        $(b)$ 
        The same data as in $(a)$, but showing the excess value
        $g(r)-1$ in double-logarithmic scaling.  
        The blue dashed line ({\color{blue}\dashed}) corresponds to a 
        $-1$ power-law.  
    }
    \label{fig-radial-dist-function-RDF}
  \end{figure}

  An alternative way of characterizing a spatial distribution of a set
  of points which is often employed in the context of particulate flow
  is the radial distribution function (RDF), i.e.\ the probability of
  finding a second particle at a given distance of a test particle. 
  Figure~\ref{fig-radial-dist-function-RDF} shows the normalized RDF
  for both cases D5 and D11 in linear and logarithmic scaling. 
  This quantity confirms that the clustering is more pronounced in
  case D5 than D11. The decay with distance $r$ from the test particle
  is found to follow a power-law with an approximate scaling as
  $r^{-1}$ for $r\lesssim20\eta$. 

\begin{figure}%
  \centering
   \begin{minipage}{2.5ex}
     \rotatebox{90}{pdf}
   \end{minipage}
   \begin{minipage}{0.55\linewidth}
      \includegraphics[width=\linewidth]
      {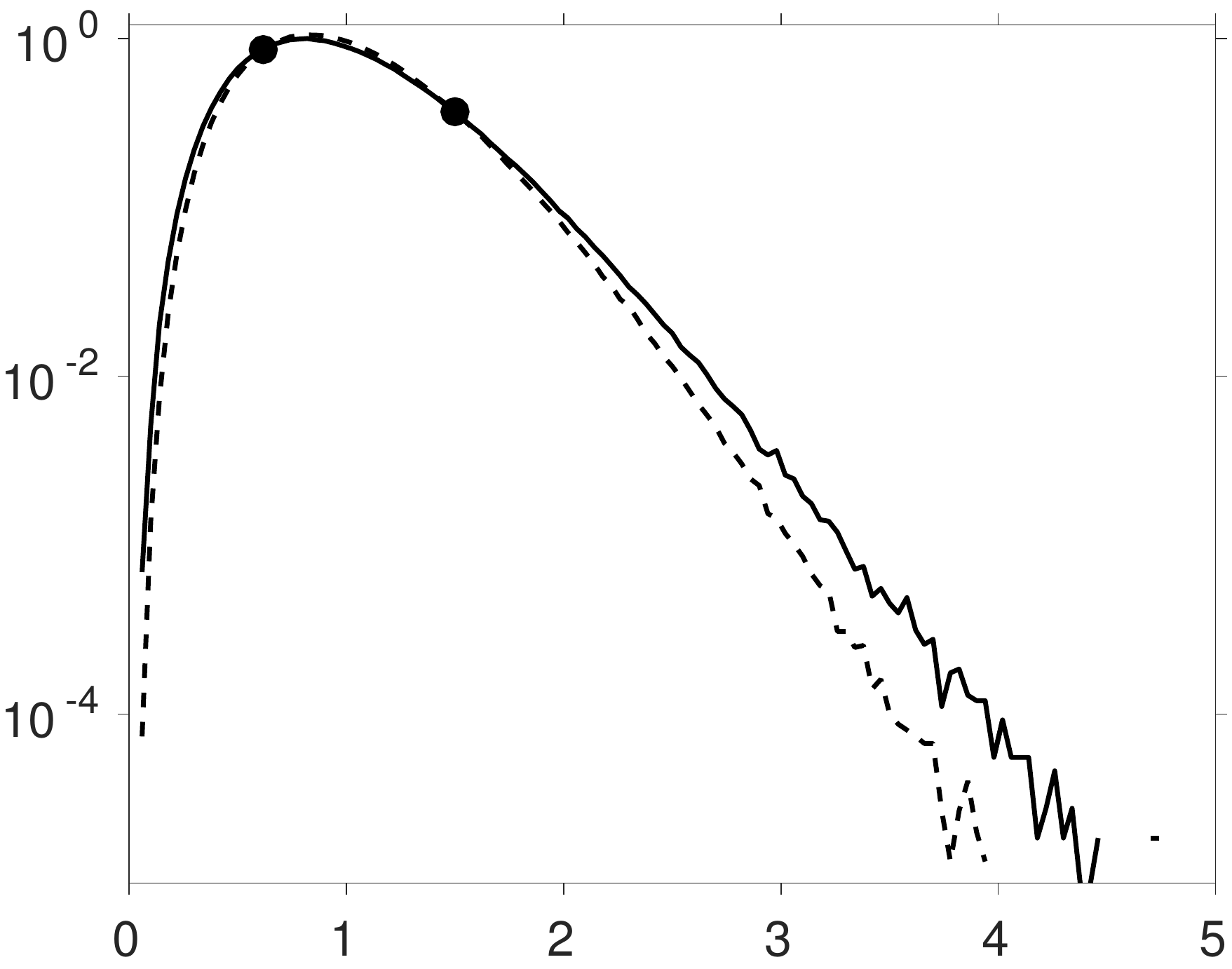}
      \\
      \centerline{$V^{(i)}/\langle V^{(i)}\rangle$}
   \end{minipage}
   \caption{%
     P.d.f.\ of the Vorono\"i cell volumes, sampled over the entire
     statistically stationary interval in case D5
     ({\color{black}\solidthick}), as compared with the distribution
     obtained from a random Poisson process
     ({\color{black}\dashed}), 
       sampled over a large ensemble. 
     The cross-over points between the two curves are marked with
     solid circles ({\color{black}$\bullet$}).  
   }
   \label{fig-voronoi-cell-vol-pdf}
\end{figure}

\begin{figure}%
   \begin{minipage}{2.5ex}
     \rotatebox{90}{pdf}
   \end{minipage}
   \begin{minipage}{0.45\linewidth}
     \centerline{$(a)$}
      \includegraphics[width=\linewidth]
      {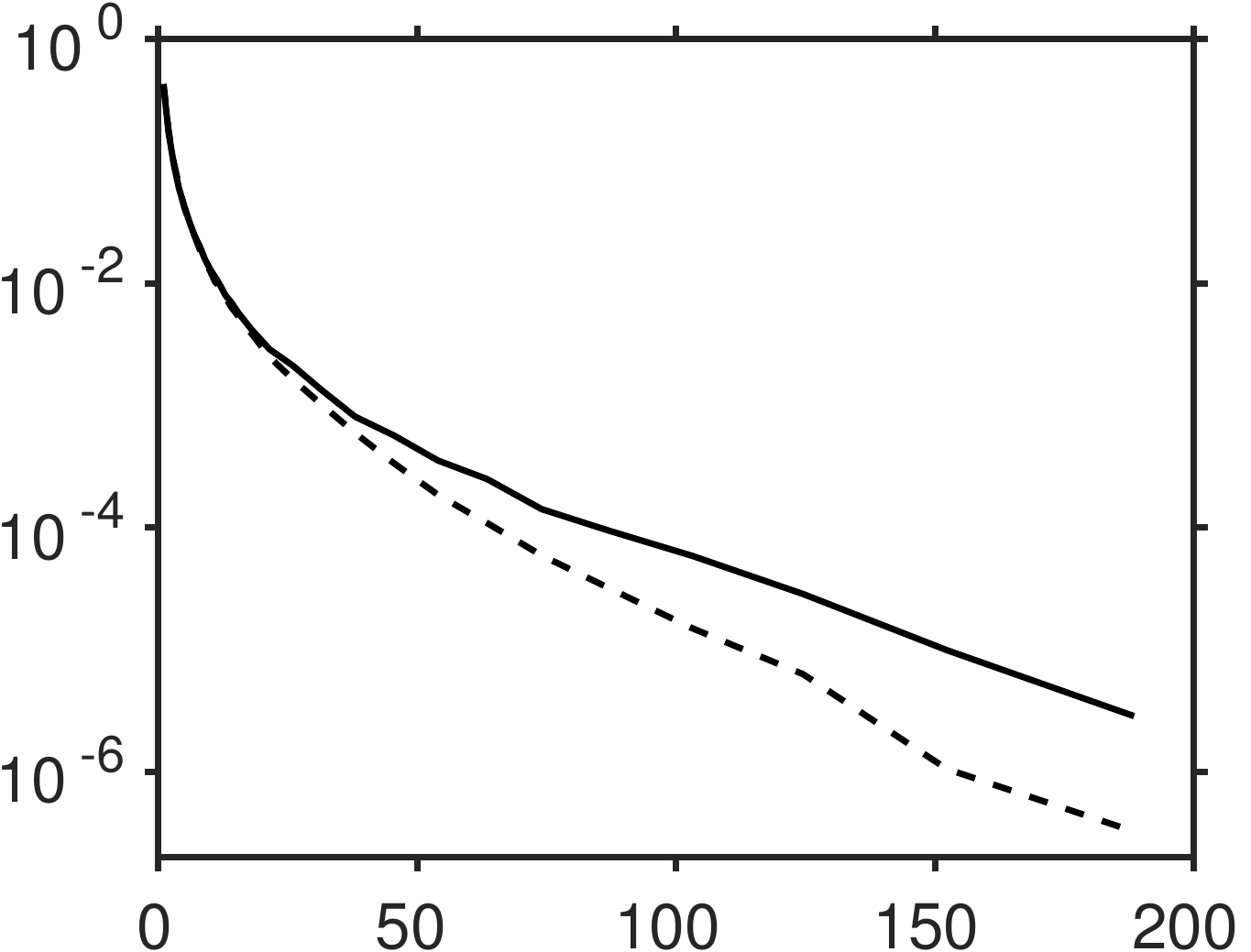}
      \\
      \centerline{number of member cells per cluster}
   \end{minipage}
   \hfill
   \begin{minipage}{1.5ex}
     \rotatebox{90}{pdf}
   \end{minipage}
   \begin{minipage}{0.45\linewidth}
     \centerline{$(b)$}
     \includegraphics[width=\linewidth]
     {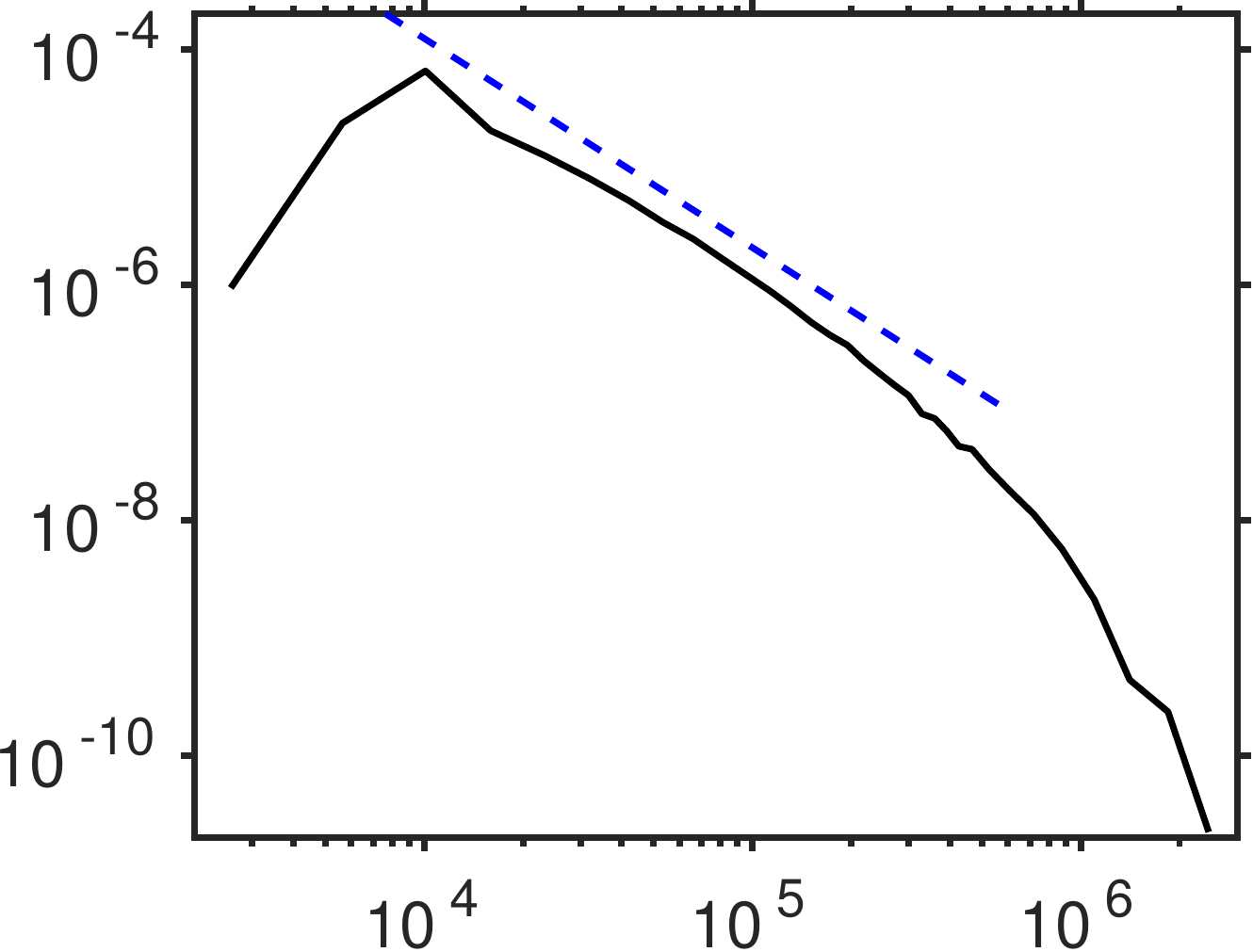}
     \\
     \centerline{cluster volume/$\eta^3$}
   \end{minipage}
   \caption{%
         $(a)$ 
         P.d.f.\ of the size distribution of particle clusters expressed
         in terms of the number of member cells per cluster. 
         The black solid line ({\color{black}\solidthick}) corresponds to
         case D5;  
         the dashed line ({\color{black}\dashed}) shows the distribution
         for a corresponding random particle arangement. 
         $(b)$ 
         P.d.f.\ of the distribution of the volume of particle clusters.  
         The blue dashed line ({\color{blue}\dashed}) corresponds to a 
         $-16/9$ power-law.  
   }
   \label{fig-voronoi-cell-clust-pdf}
\end{figure}

The Vorono\"i data can be utilized to define a host of additional
diagnostic quantities \citep{monchaux:10b}. One of them is the
objective definition of a cluster. As can be seen in
figure~\ref{fig-voronoi-cell-vol-pdf}, the fact that the variance in
the DNS data is larger than the corresponding random particle
arrangement leads to a cross-over between the two p.d.f.s. The two
cross-over points (marked in the figure) can then be used as 
objective thresholds for defining clusters and void areas. 
Here it is found that the lower cross-over point measures $0.62$ times
the average Vorno\"i cell volume in case D5. Note that this threshold
is equivalent to 124 times the particle volume $V_p$ (as compared to 
$200V_p$ for the mean volume), which is still far from a dense
particle arrangement. 
\cite{monchaux:10b} propose that all those particles
belong to a cluster whose associated cell has a Vorono\"i cell volume
smaller than the one given by the lower cross-over point. This
information can e.g.\ be used to track the residence time of particles
in a given cluster \citep{uhlmann:14a}. 
Here we attempt to extract information on the cluster size in terms of
the number of members per cluster. For this purpose a cluster is
constructed in the spirit of \citep{monchaux:10b} by first marking 
all Vorono\"i cells according to the lower volume threshold, then
connecting those which share at least one vertex of their respective
Vorono\"i cells. For each instant in time this leads to a set of
clusters with a number of individual cells (each associated to a
single particle) as members.  
Figure~\ref{fig-voronoi-cell-clust-pdf}$(a)$ shows the p.d.f.\ of the
cluster size computed over the statistically stationary time
interval. In comparison to the data for an artificial particle
arrangement drawn from a random Poisson process the DNS data-set
clearly exhibits a significantly higher number of very large
clusters, i.e.\ much more marked tails in the p.d.f. 
  Additionally, we have computed the volume of each detected cluster
  as the sum of the members' Vorono\"i
  cells, and figure~\ref{fig-voronoi-cell-clust-pdf}$(b)$ shows the
  corresponding p.d.f.\ in log-log scaling. It can be seen that the
  cluster volume distribution features a peak 
  at approximately $10^4\eta^3$, an approximate
  power-law decay over a range of nearly two decades, and an
  exponential decay for the largest volumes. 
  The most probable cluster size corresponds to a linear dimension 
  of approximately $20\eta$ which is somewhat larger than, but
  comparable to the experimental results of \cite{aliseda:02} and
  \cite{obligado:14} for sub-Kolmogorov size particles, both
  determined from planar measurements.  
  \cite{goto:06} have proposed a model for the statistical
  distribution of the area of the voids in a point-particle
  arrangement in forced two-dimensional turbulence. Their model is
  based upon the assumptions of a fully-developed Kolmogorov 
  inertial range and upon 
  a one-to-one correspondence between the ``holes'' in the particle
  field and the size of the fluid eddies from which particles are
  ejected; this leads to a $-5/3$ power-law over the range of scales
  affecting the particle motion. 
  The model predictions have recently been confirmed by
  \cite{sumbekova:16} based upon their laboratory measurements of
  small-particle data in active-grid turbulence in a
  wind-tunnel. These latter authors have also observed that the same
  power-law scaling as the one for the voids applies to the areas
  occupied by the clusters themselves.  
  When applying the model assumptions of \cite{goto:06} to the volume
  of the structures in a three-dimensional flow (instead of the area)
  one obtains an exponent of $-16/9$, which is not too far from what
  is presently observed, as can bee seen in
  figure~\ref{fig-voronoi-cell-clust-pdf}$(b)$.  
  However, since the flow Reynolds number is not very large here, the
  extent of the inertial range is rather limited. 

\begin{figure}%
   \begin{minipage}{0.46\linewidth}
     \centerline{$(a)$}
      \includegraphics[width=\linewidth]
      {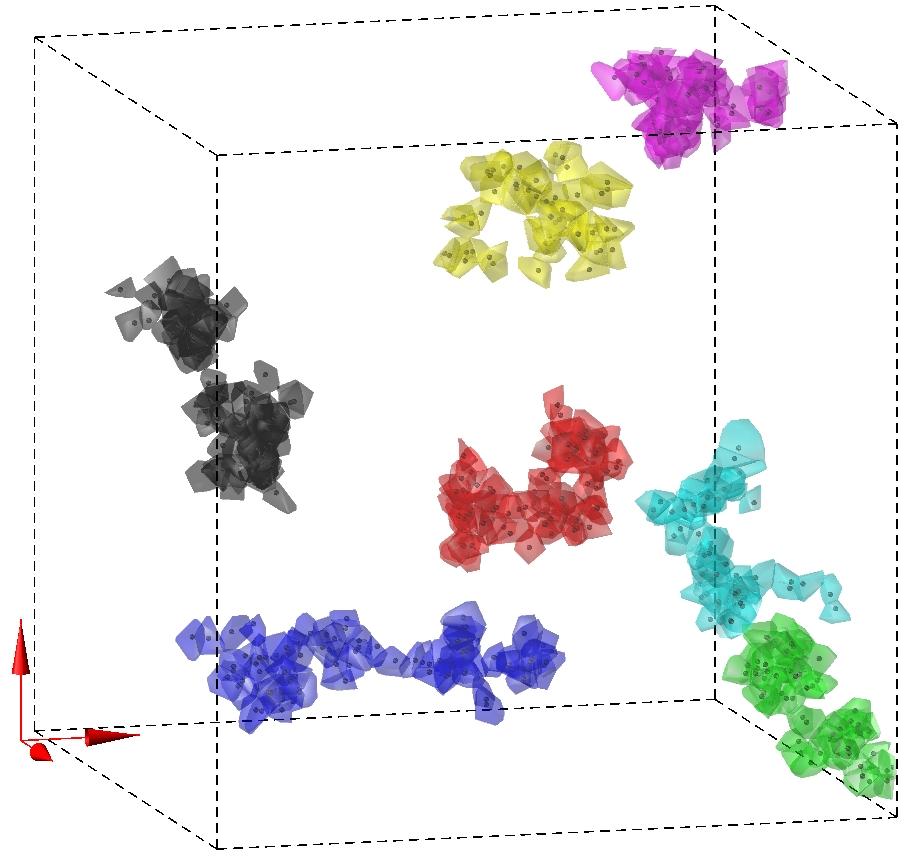}
   \end{minipage}
   \hfill
   \begin{minipage}{0.46\linewidth}
     \centerline{$(b)$}
      \includegraphics[width=\linewidth]
      {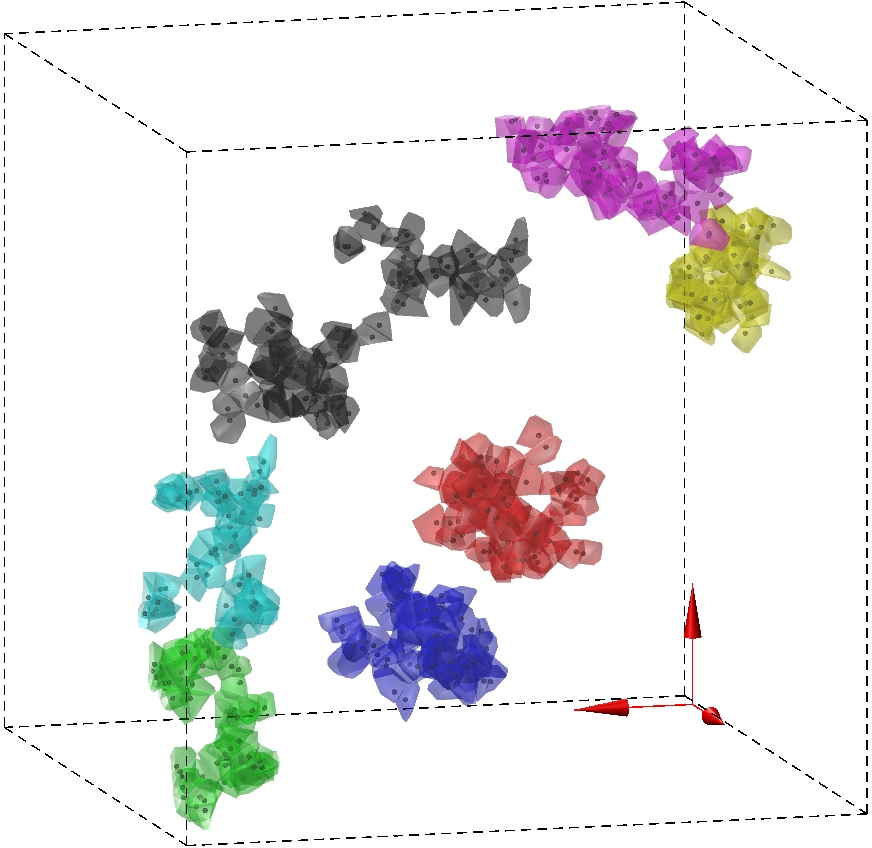}
   \end{minipage}
   \caption{%
     The 7 largest clusters in some snapshot of case D5, with the
     Vorono\"i cell faces colored uniformly per cluster. 
     The color code is in descending order (the number of member cells
     is given in parenthesis): 
     black (90), blue (80), red (77), magenta (76), green (64), cyan
     (59) and yellow (56). 
     The particles inside those cells are colored in black, while all 
     remaining particles %
     have been omitted from the figure. 
     Graphs $(a)$ and $b)$ show the same data from two different
     perspectives.  
   }
   \label{fig-voronoi-cell-clust-3dvisu}
\end{figure}

A visualization of the largest clusters in a snapshot of simulation D5 
is shown in figure~\ref{fig-voronoi-cell-clust-3dvisu}. It can be
observed that the large clusters feature an irregular shape which can
neither be described as filamentary nor as compact. A more detailed
investigation of the geometrical properties of the clusters, however,
has not been attempted at the current time. 

\subsection{Preferential particle locations}
\label{sec-results-preferential}
The preceding section 
(in particular figure~\ref{fig-voronoi-cell-vols}$b$) has shown that
the solid phase is not randomly distributed in space; instead it
features a mild level of clustering. 
In the absence of gravity and considering the low value of the global
solid volume fraction in the present simulations, it is clear that
this effect is not a consequence of collective particle dynamics, but
that it is directly induced by the turbulent background flow.  
The question is then: which mechanism is responsible for the formation
of the statistically significant excess of clusters in the case of
suspended finite-size particles? 
In the following we will consider %
two of the principal mechanisms 
which have been proposed in the context of sub-Kolmogorov particles,
as discussed in the introduction, i.e.\ 
the centrifugal effect \citep{maxey:87} 
and the sweep-stick mechanism \citep{goto:08}. 

\begin{figure}%
  \centering
  \begin{minipage}{0.45\linewidth}
    \centerline{$(a)$}
    \includegraphics[width=\linewidth]
    %
    {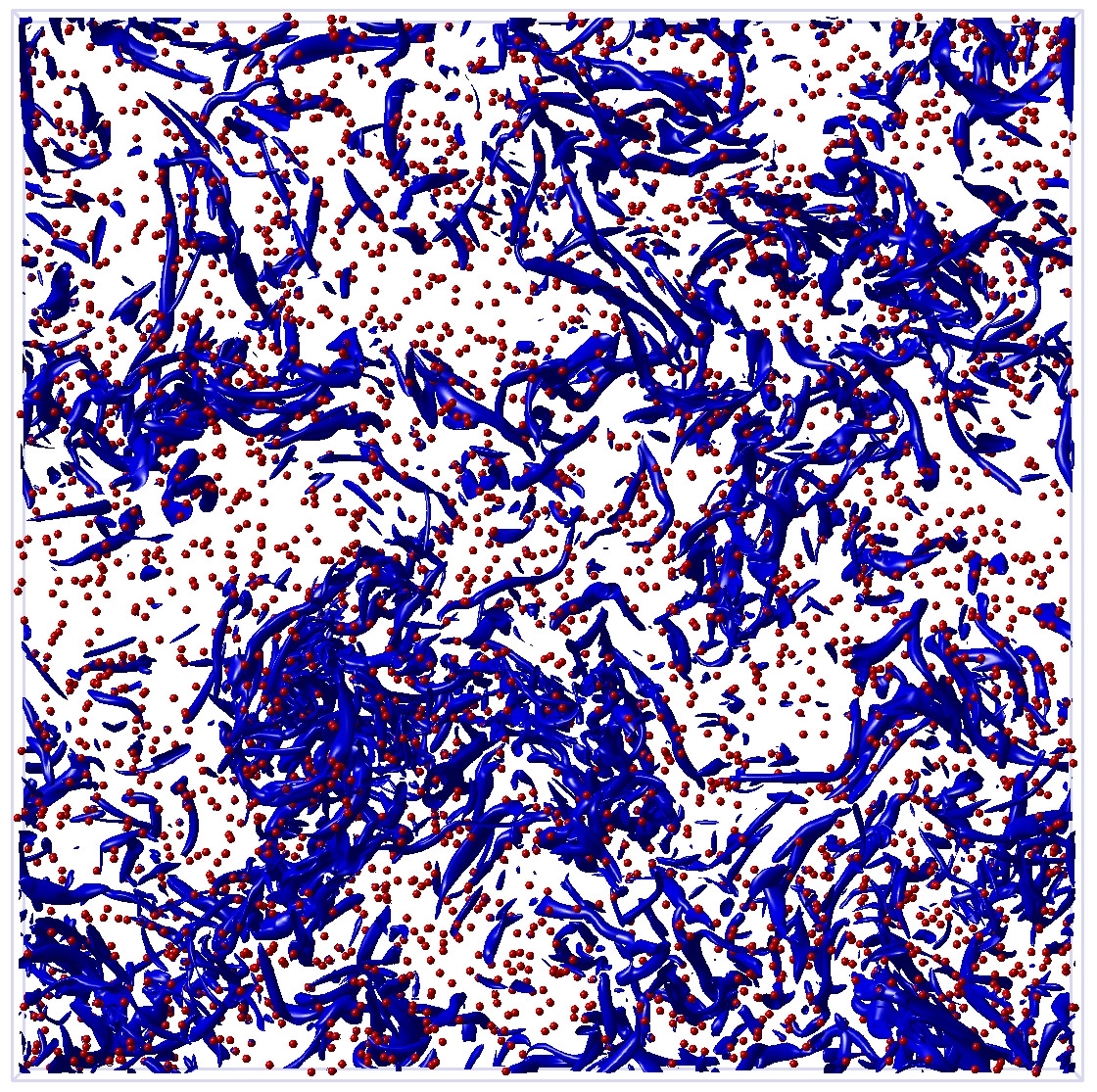}
   \end{minipage}
   \hspace*{2ex}
   \begin{minipage}{0.45\linewidth}
     \centerline{$(b)$}
     \includegraphics[width=\linewidth]
     %
     {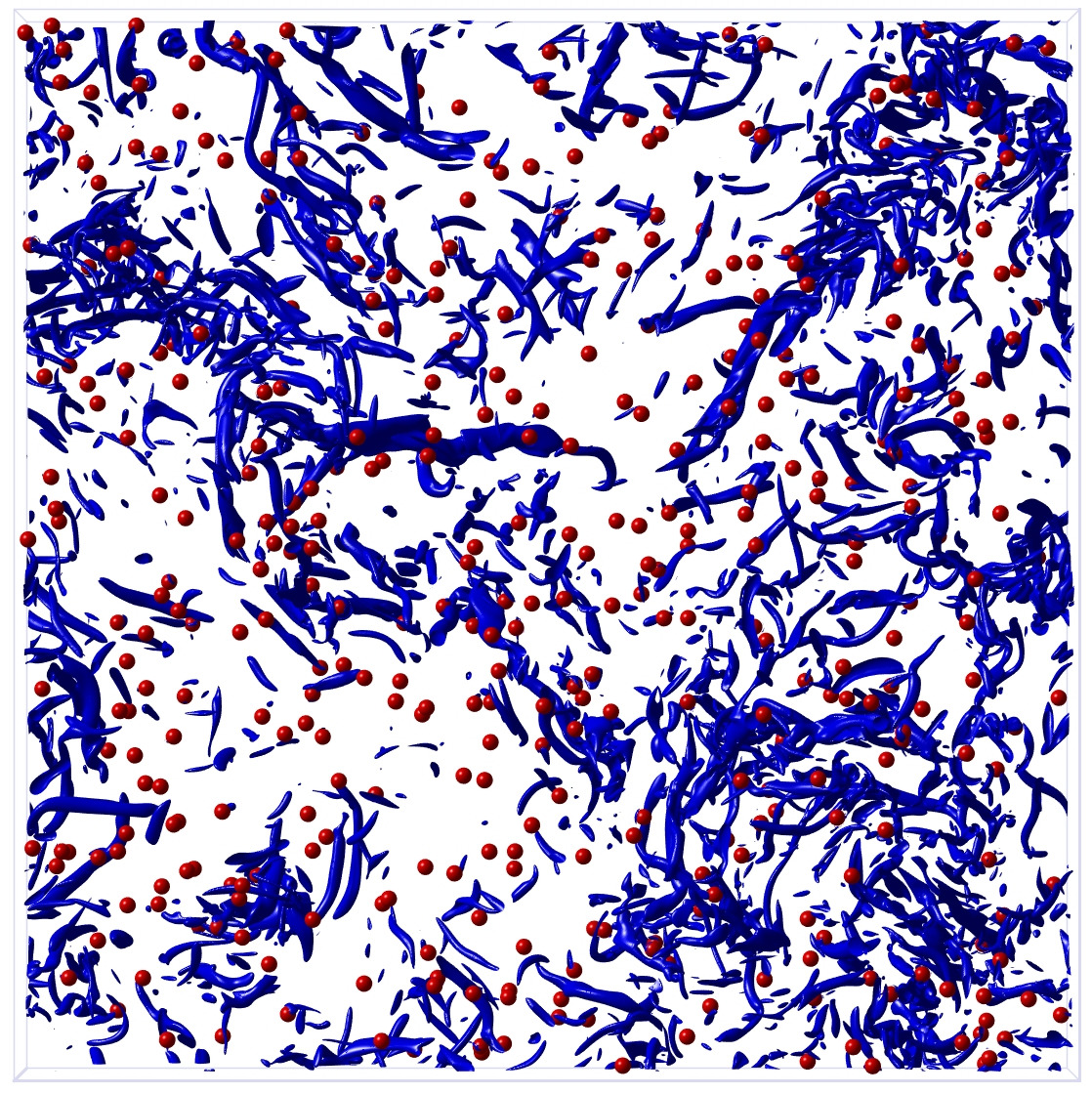}
   \end{minipage}
   \caption{%
     Snapshot of the flow field and the particle positions in case: 
     $(a)$ D5; 
     $(b)$ D11. 
     Intense vortical structures are educed with the q-criterion of
     \cite{hunt:88}, and locations with a value of $1.5$ times the
     standard-deviation of $q$ are shown as blue colored surfaces.  
     Only a slice with thickness equal to one-eighth of the domain
     size (equivalent to approximately $90$ Kolmogorov units) is shown
     for clarity; in the other two directions the entire domain is
     shown.    
   }
   \label{fig-worm-3dvisu}
\end{figure}

\begin{figure}%
  \centering
  \begin{minipage}{0.45\linewidth}
    \centerline{$(a)$}
    \includegraphics[width=\linewidth]
    {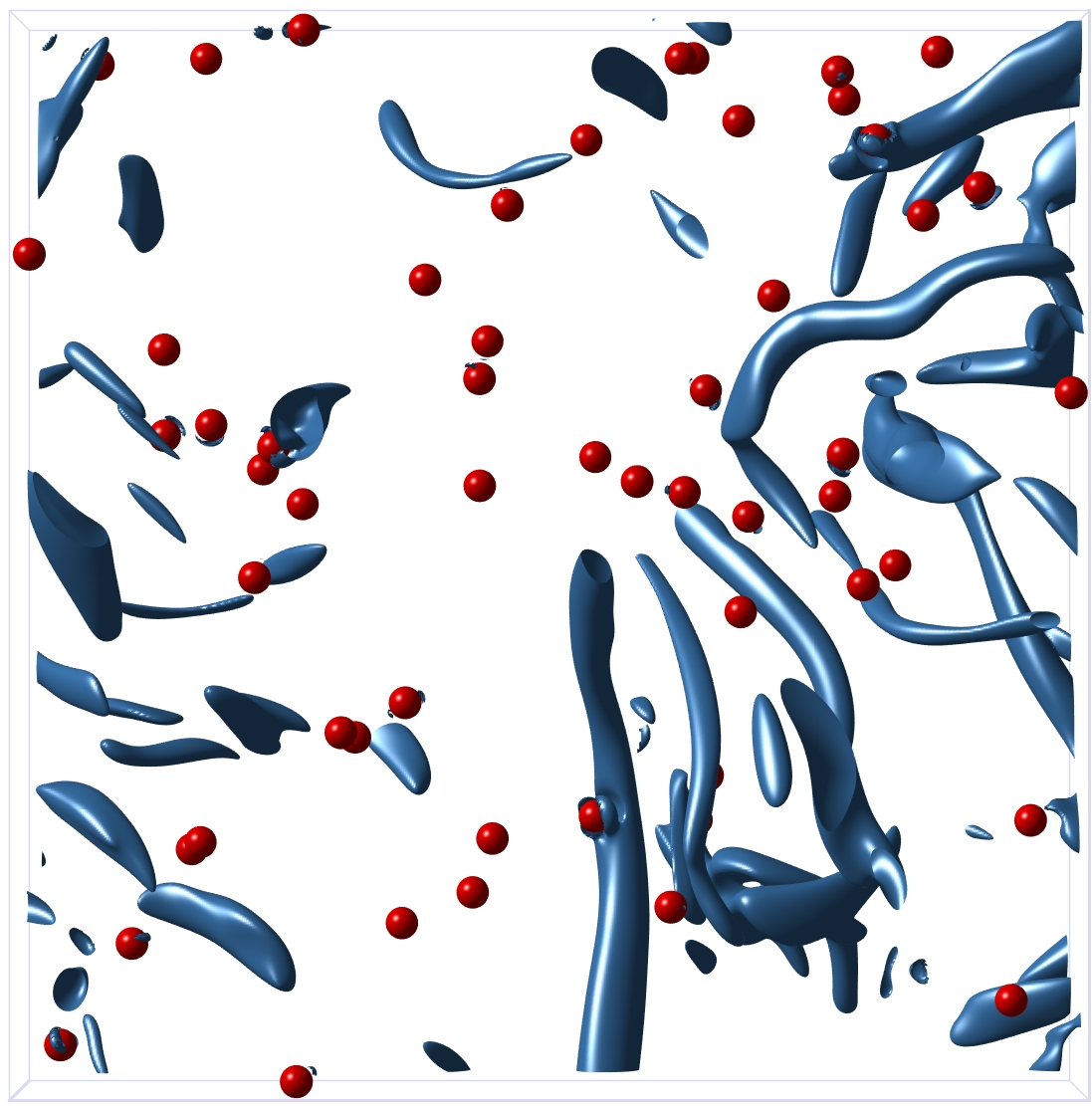}
   \end{minipage}
   \hspace*{2ex}
   \begin{minipage}{0.45\linewidth}
     \centerline{$(b)$}
     \includegraphics[width=\linewidth]
     {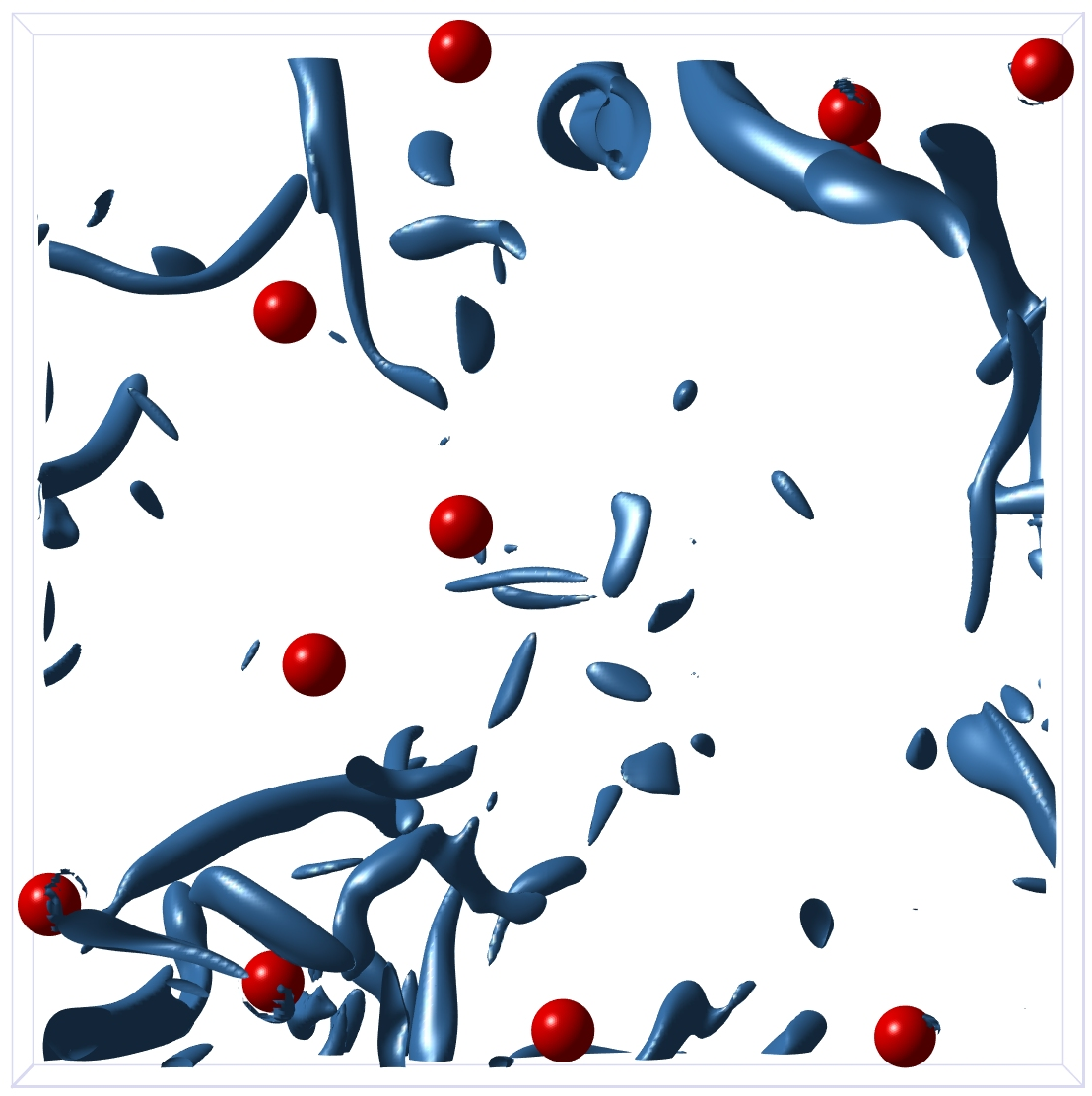}
   \end{minipage}
   \caption{%
     The same data as in figure~\ref{fig-worm-3dvisu}, but showing a
     close-up with side-length equal to $180\eta$ and a reduced
     depth of $45\eta$.
   }
   \label{fig-worm-3dvisu-zoom}
\end{figure}

\begin{figure}%
  \centering
   \begin{minipage}{2.5ex}
     \rotatebox{90}
     {pdf}
   \end{minipage}
   \begin{minipage}{0.55\linewidth}
      \includegraphics[width=\linewidth]
      {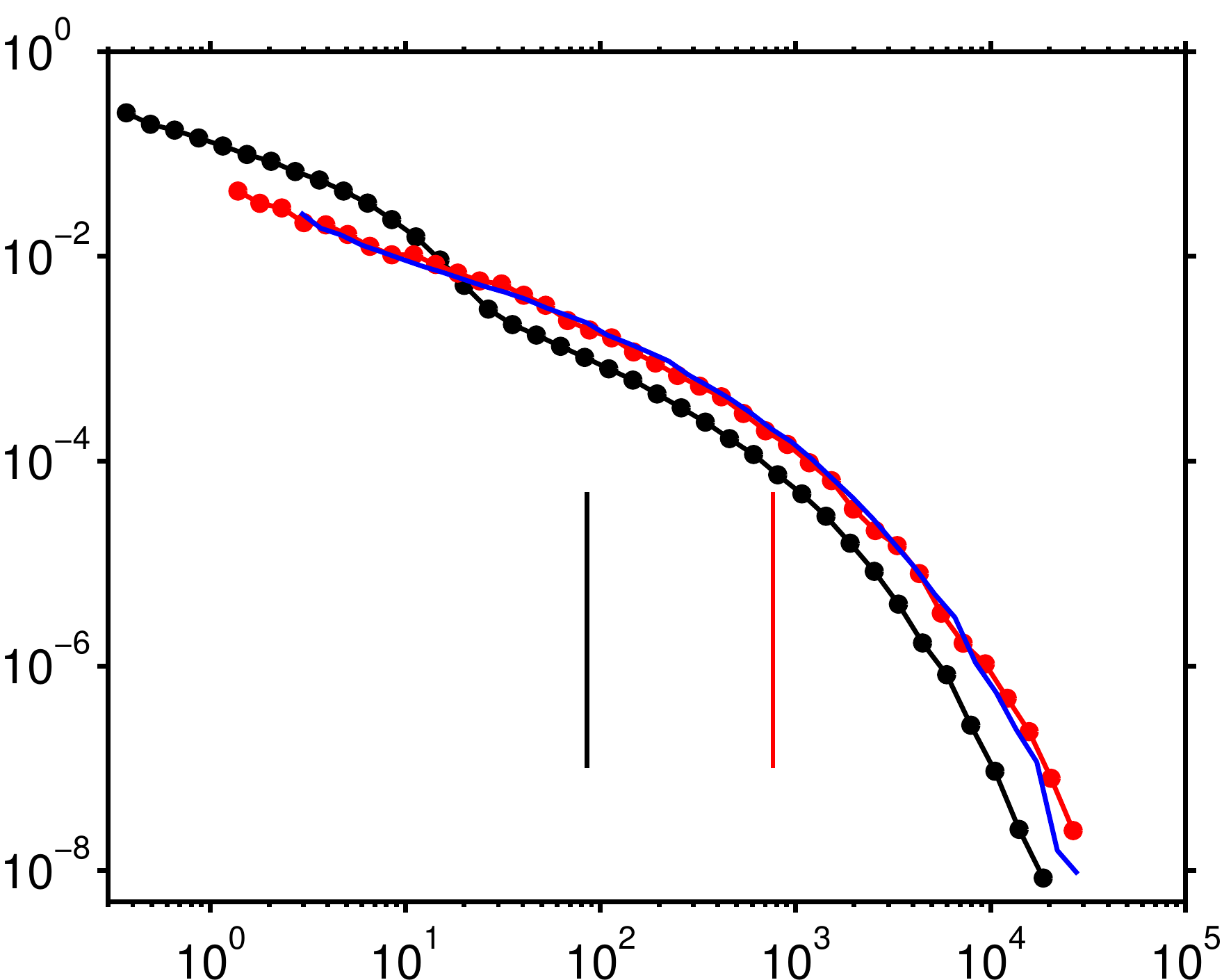}
      \hspace*{-.6\linewidth}\raisebox{.11\linewidth}{$V_p$(D5)}
      \hspace*{+0\linewidth}\raisebox{.11\linewidth}{\color{red}$V_p$(D11)}
      \\
      \centerline{$V/\eta^3$}
   \end{minipage}
   \caption{%
     The p.d.f.\ of the volume of the vortical structures, educed with
     the criterion of \cite{hunt:88} (using a threshold equal to $1.5$ 
     times the standard deviation), 
     and normalized in Kolmogorov length scales. 
     Line styles correspond to: 
     {\color{black}\solidthickbullet},~case~D5; 
     {\color{red}\solidthickbullet},~case~D11; 
     {\color{blue}\solidthick}, single-phase data. 
     The vertical lines indicate the particle volume of cases D5 and
     D11. 
     Note that the different lower ends of the three curves correspond
     to the specific values of $\Delta x^3$ in each case. 
   }
   \label{fig-worm-volume-pdf}
\end{figure}

Let us first consider the centrifugal effect. 
It hinges upon the rotational property of the fluid motion and the
conditions that: 
(a) the particle inertia is sufficiently large such that
particle paths deviate from fluid particle paths,
and 
(b) the excess inertia still does not place the particles in the
ballistic regime. 
  In the context of the simplest point-particle assumption the argument
  then boils down to a comparison of time scales, viz.\ considering
  the value of the Stokes number. 
  Now in the case of finite-size particles 
  it is doubtful that the Stokes number alone is a sufficient
  parameter to predict clustering, as has been demonstrated by
  \cite{fiabane:12}.  
  On the other hand, an analysis of the role of coherent vortices 
  upon the motion of particles with a diameter
  comparable to or larger than the Kolmogorov length has to our
  knowledge previously not been undertaken.  
As a first step 
in this direction 
we educe the coherent vortical structures with the aid
of the ``q-criterion'' of \cite{hunt:88}. 
Denoting by ``$q$'' the second invariant of the velocity gradient tensor 
($q=u_{i,i}^2/2-u_{i,j}u_{j,i}/2$), 
we define an intense vortical region as one where a given threshold
value of this quantity is exceeded, viz.\ 
$q>q_{thresh}=1.5\,\sigma_q$ (where $\sigma_q$ is the standard-deviation of $q$). 
Figure~\ref{fig-worm-3dvisu} gives an impression of the typical
worm-like structures which are educed with this method in conjunction
with the relative locations of particles, while a close-up of the same
data is  provided in figure~\ref{fig-worm-3dvisu-zoom}. 
Globally the visualization looks similar to the
single-phase counterpart (not shown), featuring elongated vortices
with a diameter of several Kolmogorov lengths, as previously observed
by many authors \citep[e.g.][]{she:90,jimenez:93,moisy:04}. 
Close inspection of the present images reveals, however, that a
relatively large fraction of the particles is accompanied by small
vortices in their vicinity. 
This is expected in the present cases, since the particle Reynolds
number based upon the instantaneous relative velocity reaches
substantial values (cf.\ figure~\ref{fig-rep-shell-avg}) such that the
formation of small transient wakes occurs frequently. 
In order to gain quantitative insight into the geometry of the educed
vortices, we have computed the volume enclosed by each individual
`worm'. 
After determining a triangulation of the surface defined by
$q=q_{thresh}$, we compute the enclosed volume from a 
surface integral (making use of the divergence theorem); this
procedure is repeated for a number of flow fields in order to
accumulate significant statistics.  
The resulting probability distribution of the `worm' volumes is shown
in figure~\ref{fig-worm-volume-pdf}. 
It can be seen that the distribution in case D11 is essentially
identical to the single-phase data. 
Case D5 on the other hand features a larger number of small-volume
vortical structures than the single-phase data, the rise being
observed for volumes smaller than approximately one-half of the
particle volume $V_p=D^3\pi/6$.  
Due to the normalization of the p.d.f., the curve corresponding to
case D5 is then slightly shifted downward (as compared to single-phase
data) at larger volumes. 
In absolute terms, however, the occurrence of vortical structures with
a volume comparable to and larger than the particle volume are also
essentially the same in case D5 as in the single-phase data-set. 
This discussion shows that the presence of the finite-size particles
in both present simulations only affects the number density of
small structures with a volume smaller than the particle volume. The
fact that this feature is much more pronounced in case D5 can be
explained with the larger number density of particles (by a factor of
8) as compared to case D11. 
Note that \cite{moisy:04} have performed an analysis of the volume
distribution of intense vortices in single-phase homogeneous-isotropic
turbulence. Although their definition of a `worm' is slightly
different (i.e.\ based upon the magnitude of the vorticity field),
we observe a good qualitative agreement when choosing a similar
objective threshold value.  

\begin{figure}%
  \centering
   \begin{minipage}{2.5ex}
     \rotatebox{90}
     {pdf}
   \end{minipage}
   \begin{minipage}{0.55\linewidth}
      \includegraphics[width=\linewidth]
      {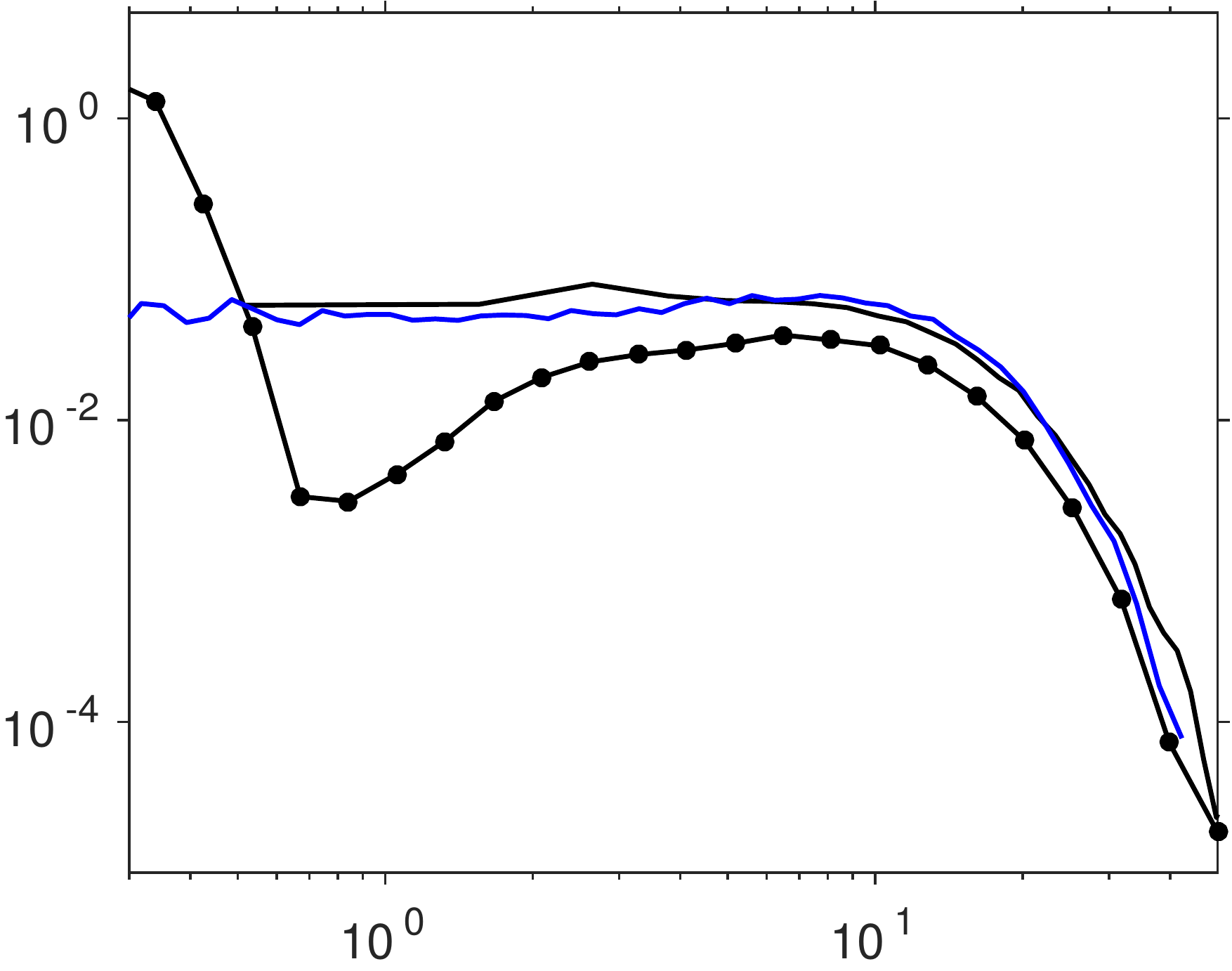}
      \\
      \centerline{$d_{CS}^{(i)}/\eta$}
   \end{minipage}
   \caption{%
     P.d.f.\ of the distance from the particles to the nearest point
     on any vortical structure \citep[as educed with the
     criterion of][]{hunt:88}. Note that the distance $d_{CS}^{(i)}$
     (defined in the text) is measured from the particle surface.
     Line styles correspond to: 
     {\color{black}\solidthickbullet},~case~D5; 
     {\color{black}\solidthick},~case~D5 and removing coherent
     structures with a volume inferior to $V_p$; 
     {\color{blue}\solidthick}, distance from randomly chosen points
     to the coherent structures in case~D5. .  
   }
   \label{fig-dist-to-CS-pdf}
\end{figure}

Next we proceed to the characterization of the particle locations with 
respect to the coherent vortices educed by the method presented in the
previous paragraph. 
Here we again use the information provided by the surface
triangulation of the `worms'. 
More specifically, at a given instant in time we determine for the
$i$th particle the smallest distance $d_{CS}^{(i)}$ from the particle
surface to any vertex point of the triangulation of any `worm'
surface. The principal quantity of interest is the distribution of the
set of so-defined minimum distances $d_{CS}^{(i)}$, but in this
fashion we also obtain the identification of the nearest coherent
vortex, which gives us access to all its properties. 
This additional information will be used below in order to further
refine the analysis.  
Figure~\ref{fig-dist-to-CS-pdf} shows the resulting p.d.f.\ of the
distance $d_{CS}^{(i)}$ in case D5, accumulated over the set of
particles and over a number of snapshots.  
As a reference the graph also includes the data corresponding to a set
of points which were distributed in the flow fields of 
case~D5 by means of a random Poisson process.   
Note that the random-point data, which features an
essentially flat distribution that decays exponentially for large 
distances ($d_{CS}^{(i)}\gtrsim10\eta$), solely reflects the geometry
of the `worms' themselves. 
By comparison, the finite-size particle DNS data in case D5 exhibits a
distinct distribution, namely featuring a marked dip at
distances around the Kolmogorov scale and a dominant peak when the 
distance goes to zero. This result clearly corresponds to a
high probability of a vortical structure being practically attached to
a particle. 
The fact that the particles in case D5 induce a
substantial amount of small-scale vortical structures 
(cf.\ figure~\ref{fig-worm-volume-pdf}) 
suggests that the difference between the random data and the DNS
results in figure~\ref{fig-dist-to-CS-pdf} is actually due to the
attached wake structures and not a manifestation of a significant
preferential location with respect to coherent structures of the
background turbulent flow.  
In order to remove this bias, we have first eliminated from each
snapshot those `worms' that have an enclosed volume smaller than the
particle volume. 
The resulting distribution of the distance to the nearest coherent
flow structure is also included in figure~\ref{fig-dist-to-CS-pdf}. 
It can be seen that filtering the flow field in this manner indeed has 
the effect that (to within statistical uncertainty) the p.d.f.\ for
the minimum distances measured from the real particles becomes
equivalent to the one measured from random points. 
Therefore, we can conclude from this geometrical analysis 
that the particles in case D5 do not preferentially accumulate with 
respect to the intense vortical structures of the turbulent background
flow. 
  We now turn to the analysis of the relation between the fluid
  acceleration field $\mathbf{a}_f$ and the particle locations 
  motivated by the results of \cite{goto:08} in the context of
  sub-Kolmogorov particles. 
In the framework of the simplest one-way coupled point-particle model  
(with the hydrodynamic force given by Stokes drag) \cite{goto:08}
argue that the particle phase potentially accumulates at points which
satisfy the following criterion:
\begin{equation}\label{equ-def-sticky-points}
  \mathbf{e}_1\cdot\mathbf{a}_f=0
  \,,
  \qquad
  \mbox{and}
  \qquad
  \lambda_1>0
  \,.
\end{equation}
In (\ref{equ-def-sticky-points}) the symbol $\lambda_1$ denotes the
largest eigenvalue of the symmetric part of the acceleration gradient
tensor $\mathbf{A}\equiv\nabla\mathbf{a}_f+(\nabla\mathbf{a}_f)^T$, and 
$\mathbf{e}_1$ is the associated eigenvector (normalized to unit
length). 
Recall that the argument relies on the asymptotic result (valid for
this point-particle model in the limit of small Stokes number)
obtained by \cite{maxey:87} which states that the slip velocity is
proportional to the fluid acceleration, viz.\  
\begin{equation}\label{equ-maxey-asymptote}
  \mathbf{u}_p-\mathbf{u}_f\approx-\tau_p\,\mathbf{a}_f
  \,.
\end{equation}
Taking the divergence of (\ref{equ-maxey-asymptote}) then yields the
result that the particle `field' should converge wherever the fluid
acceleration field diverges (regions where $\nabla\cdot\mathbf{a}_f >
0$).  
Note that the divergence of the fluid acceleration is directly linked
to the Laplacian of pressure and to the $q$-criterion of
\cite{hunt:88}, viz.\ 
$\nabla\cdot\mathbf{a}_f=-\nabla^2 p/\rho=-2\,q$ 
\citep[][p.76]{jeong:95}. 
In order to properly account for the multi-dimensional nature of the
acceleration field, 
\cite{goto:08} propose to selectively consider the most expansive
direction $\mathbf{e}_1$ of the fluid acceleration $\mathbf{a}_f$ 
instead of directly working with zero-acceleration points. 
Hence they suggest to investigate the points defined in
(\ref{equ-def-sticky-points}), to which we will henceforth simply
refer as `sticky points'. 
Since this criterion focuses the attention on the first principal axis
of the tensor $\mathbf{A}$, it is not equivalent to an analysis of
points with positive divergence of acceleration, and consequently it
is not equivalent to simply analyzing points with negative values of
$q$ \citep[i.e.\ strain-dominated locations in the sense
of][]{hunt:88}. 
The 
DNS data of \cite{goto:08} strongly supports this
argument with a large accumulation of point-particles in the vicinity
of the sticky points. 

\begin{figure}%
  \centering
  \begin{minipage}{0.46\linewidth}
    \centerline{$(a)$}
    \includegraphics[width=\linewidth]
    {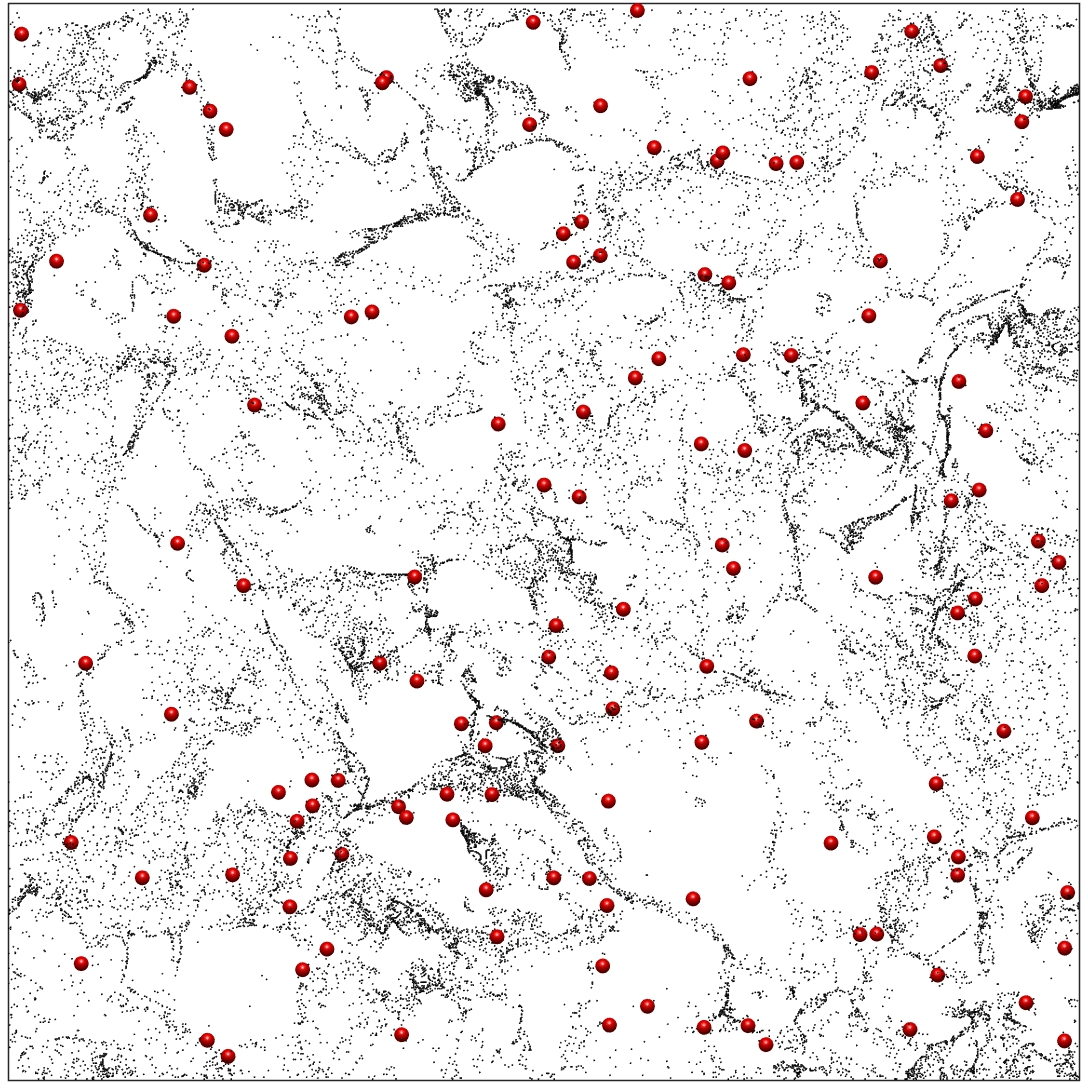}
  \end{minipage}
  \hfill
  \begin{minipage}{0.46\linewidth}
    \centerline{$(b)$}
    \includegraphics[width=\linewidth]
    {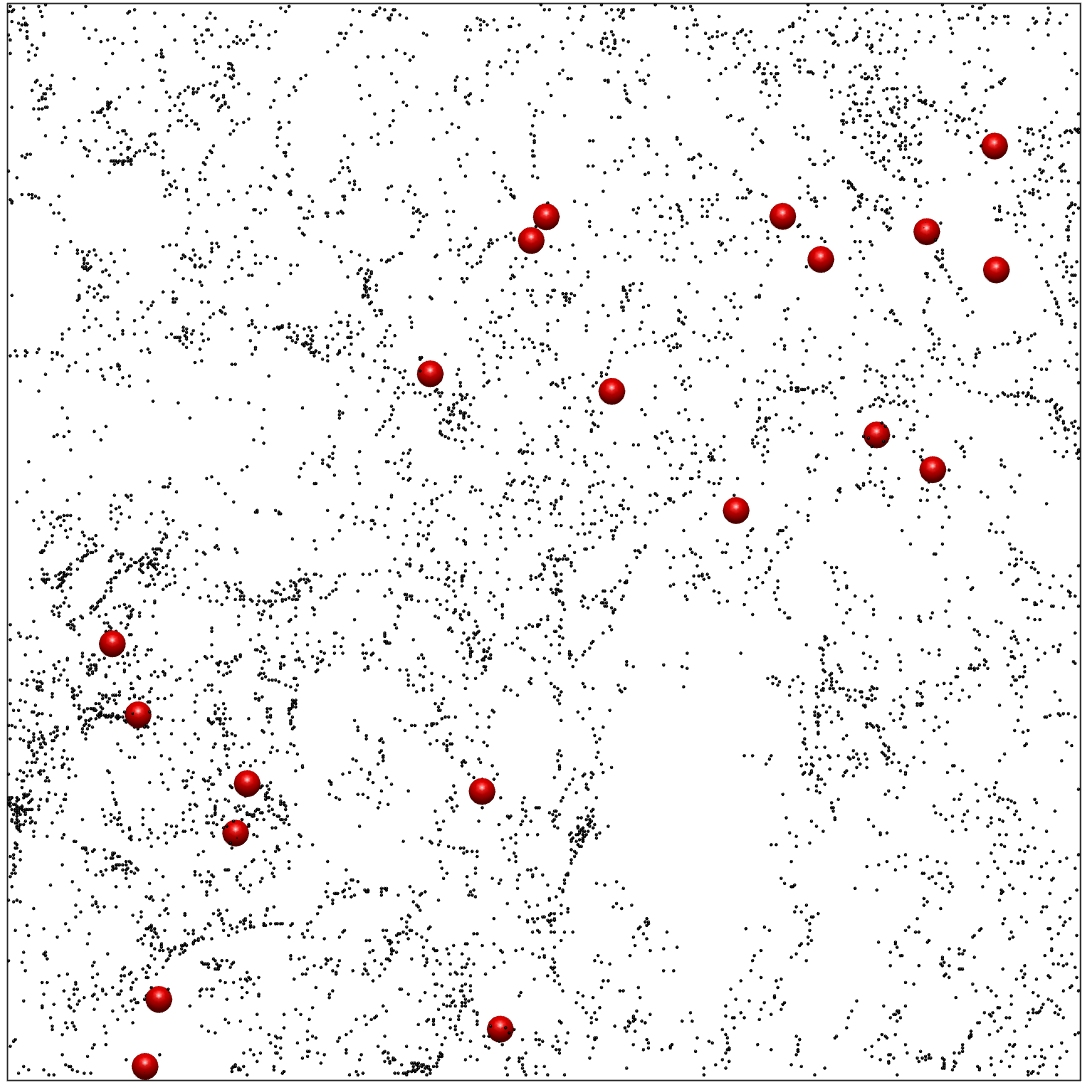}
  \end{minipage}
  \caption{%
    Instantaneous location of particles (red spheres) and the
    points of the fluid velocity field (grey dots) which satisfy the
    criteria of (\ref{equ-def-sticky-points}). 
    The data is shown in a slice of thickness equal to
    $15\eta$ (direction into the page), the length in the other two 
    directions being $380\eta$: 
    $(a)$ case D5; 
    $(b)$ case D11.
   }
   \label{fig-voronoi-particles-sticky-points-slice}
\end{figure}

\begin{figure}%
  \centering
   \begin{minipage}{2.5ex}
     \rotatebox{90}
     {$n_S(r)/n_{SG}$}
   \end{minipage}
   \begin{minipage}{0.45\linewidth}
      \includegraphics[width=\linewidth]
      {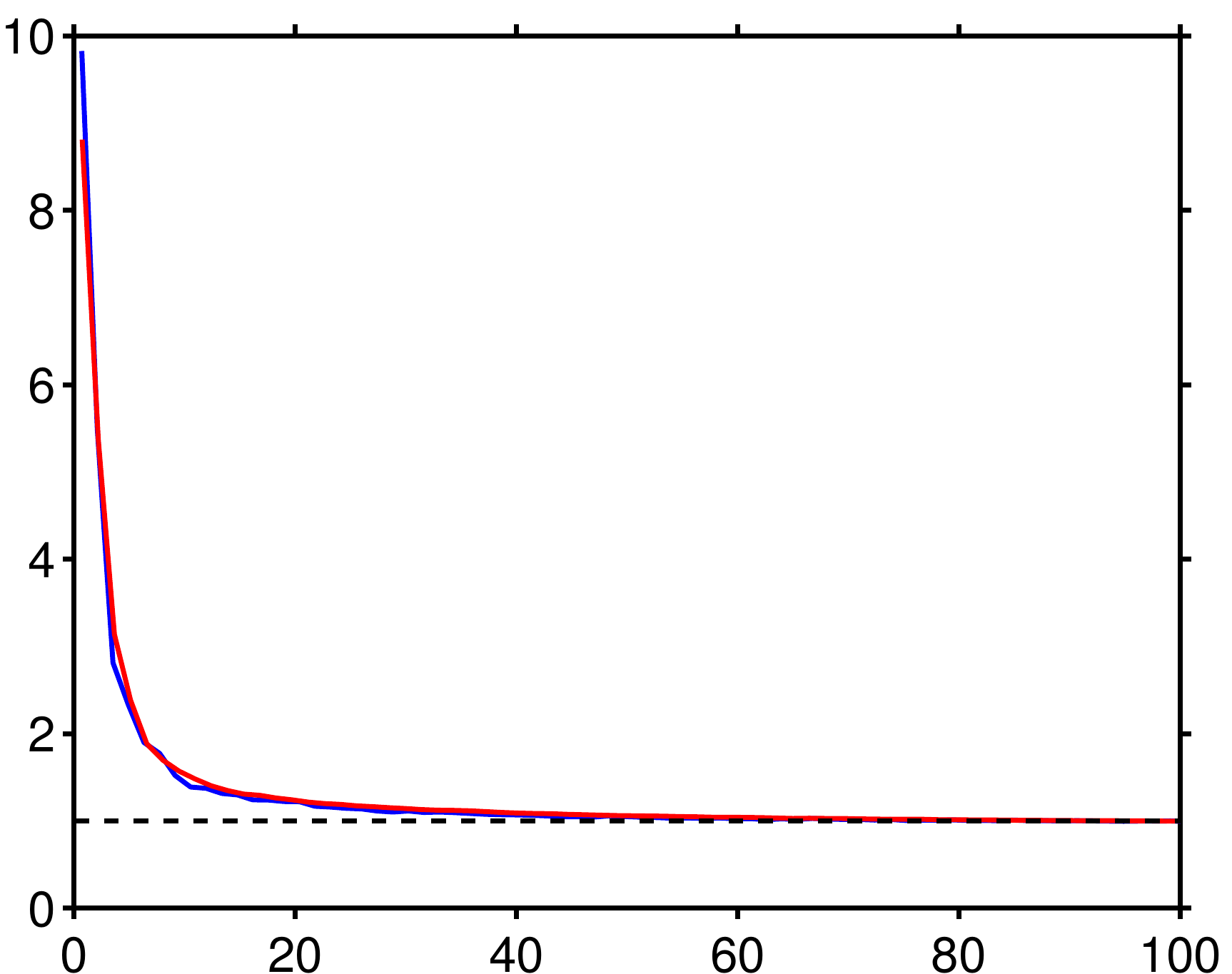}
      \\
      \centerline{$r/\eta$}
   \end{minipage}
   \caption{%
     Radial distribution function of 'sticky points' as
     defined in (\ref{equ-def-sticky-points}). 
     The number density $n_S(r)$ is normalized with the global value
     $n_{SG}$. 
     The lines correspond to: 
     {\color{red}\solidthick} case D11; 
     {\color{blue}\solidthick} single-phase flow.
   }
   \label{fig-stickyPoints-radDist}
\end{figure}

\begin{figure}%
   \begin{minipage}{2.5ex}
     \rotatebox{90}
     {$n_{PS}(r)/n_{PSG}$}
   \end{minipage}
   \begin{minipage}{0.45\linewidth}
     \centerline{$(a)$}
      \includegraphics[width=\linewidth]
      {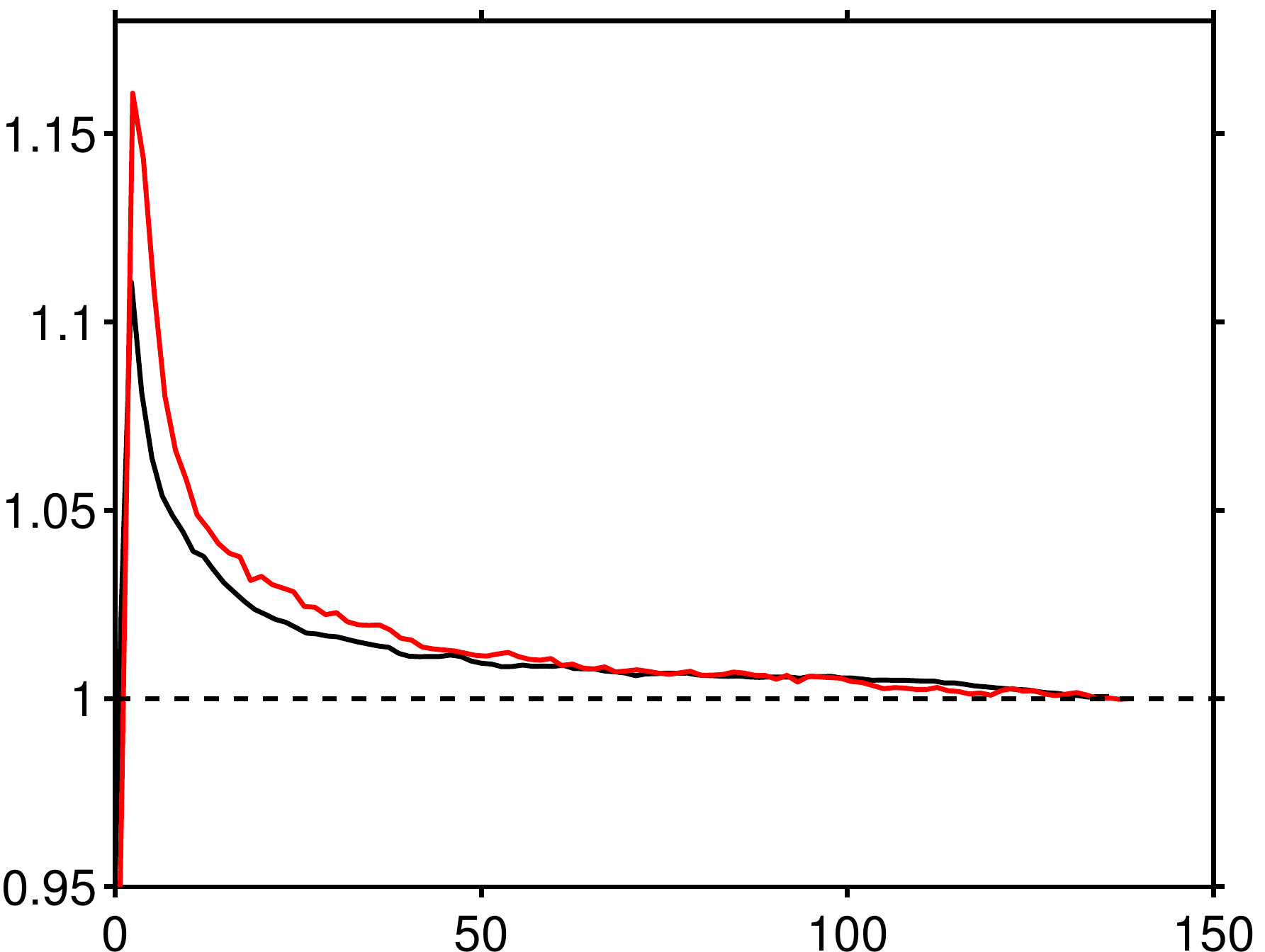}
      \\
      \centerline{$r/\eta$}
   \end{minipage}
   \begin{minipage}{2.5ex}
     \rotatebox{90}
     {$n_{PS}(r)/n_{PSG}$}
   \end{minipage}
   \begin{minipage}{0.45\linewidth}
     \centerline{$(b)$}
      \includegraphics[width=\linewidth]
      {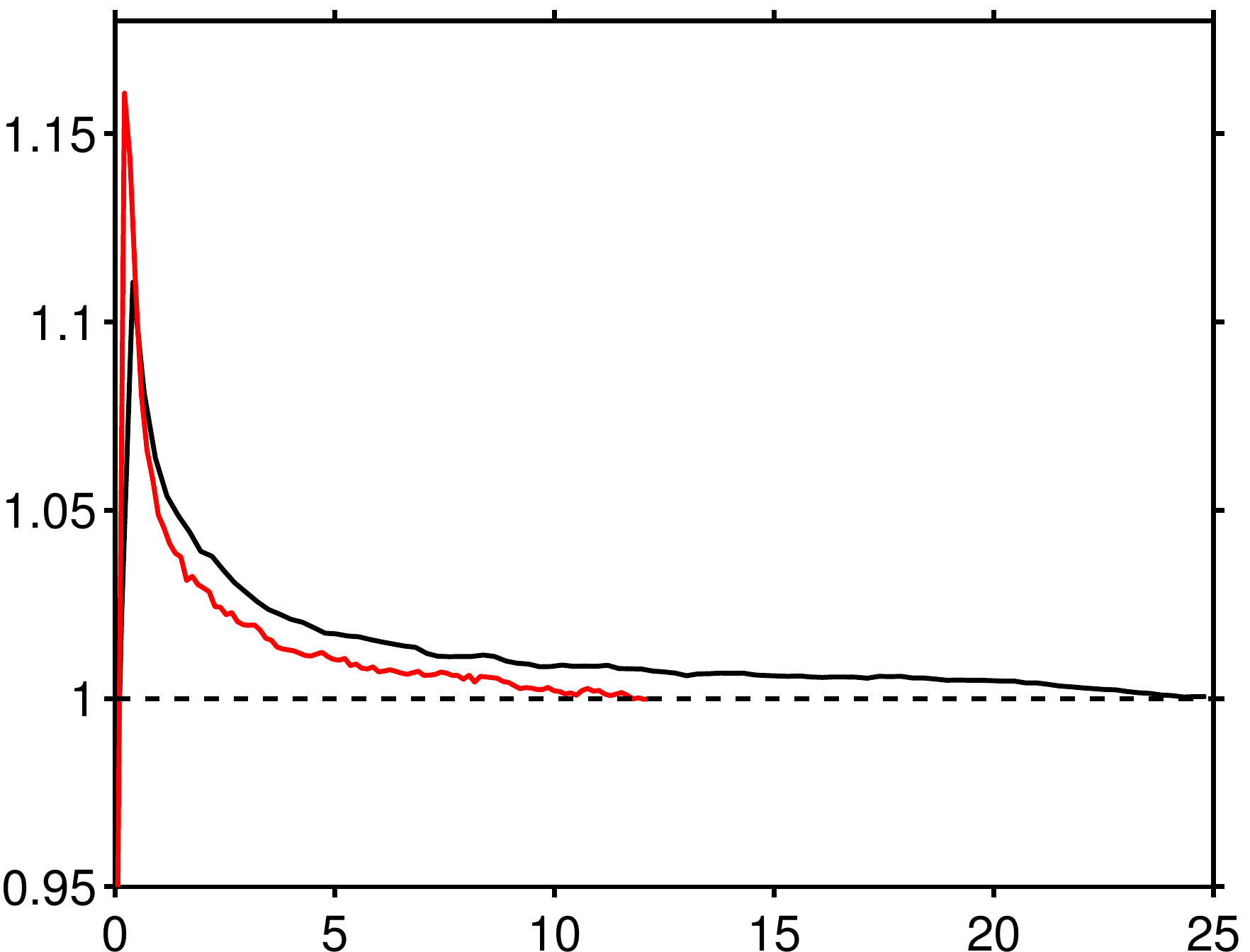}
      \\
      \centerline{$r/D$}
   \end{minipage}
   \caption{%
     Particle-conditioned, relative number density of 'sticky
     points' 
     (as defined in \ref{equ-def-sticky-points}). 
     The number density $n_{PS}(r)$ is normalized with the global value
     $n_{PSG}$. 
     In  $(a)$ the distance is scaled with the Kolmogorov length
     scale $\eta$, in 
     $(b)$ it is scaled with the particle diameter $D$. 
     Line styles: 
     {\color{black}\solidthick} case D5; 
     {\color{red}\solidthick} case D11.
   }
   \label{fig-stickyPoints-partCondDist}
\end{figure}

\begin{figure}%
   \begin{minipage}{2.5ex}
     \rotatebox{90}
     {$n_{PS}(r)/n_{PSG}$}
   \end{minipage}
   \begin{minipage}{0.45\linewidth}
     \centerline{$(a)$}
     \centerline{$r/D$}
      \includegraphics[width=\linewidth]
      {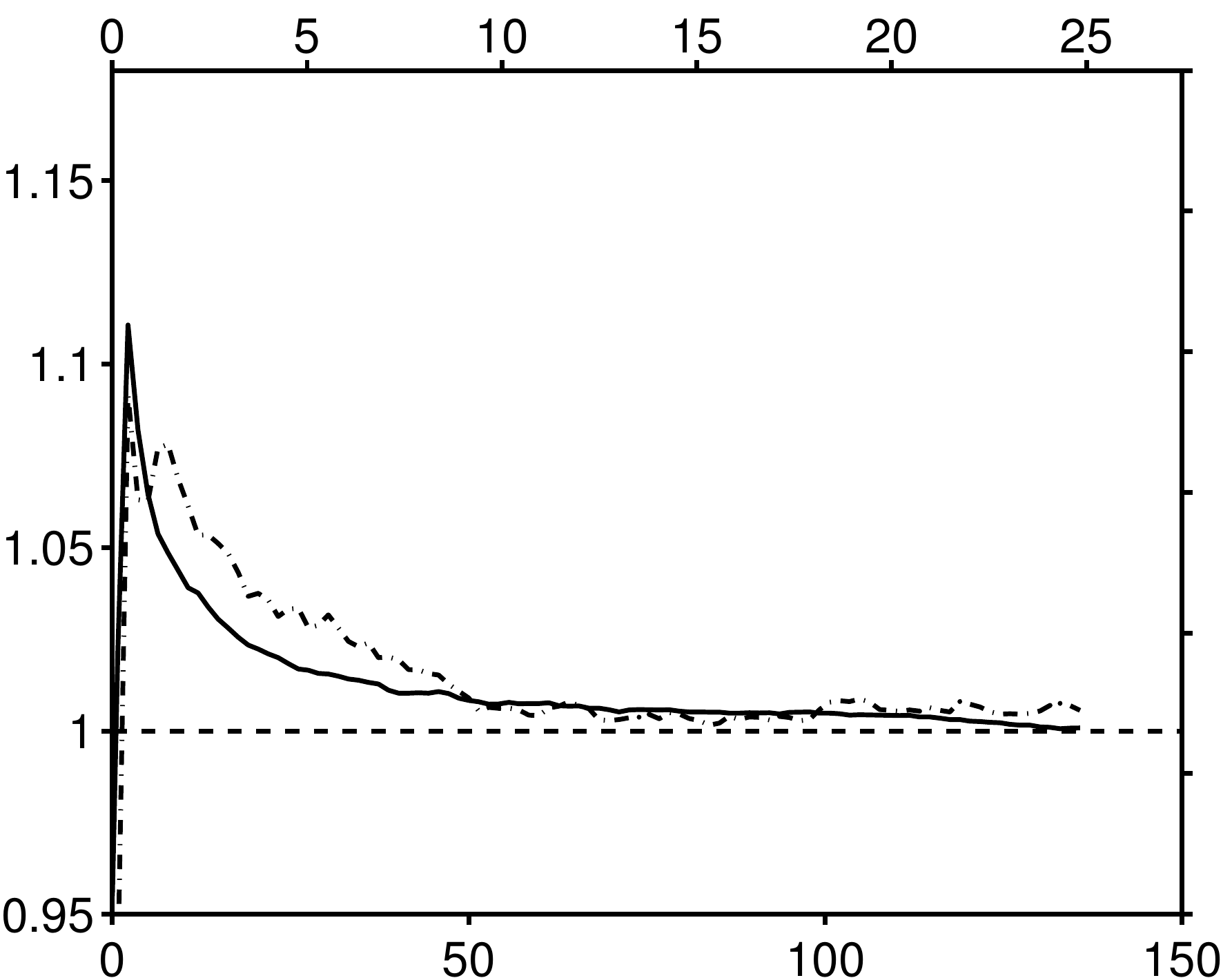}
      \\
      \centerline{$r/\eta$}
   \end{minipage}
   \begin{minipage}{2.5ex}
     \rotatebox{90}
     {$n_{PS}(r)/n_{PSG}$}
   \end{minipage}
   \begin{minipage}{0.45\linewidth}
     \centerline{$(b)$}
     \centerline{$r/D$}
      \includegraphics[width=\linewidth]
      {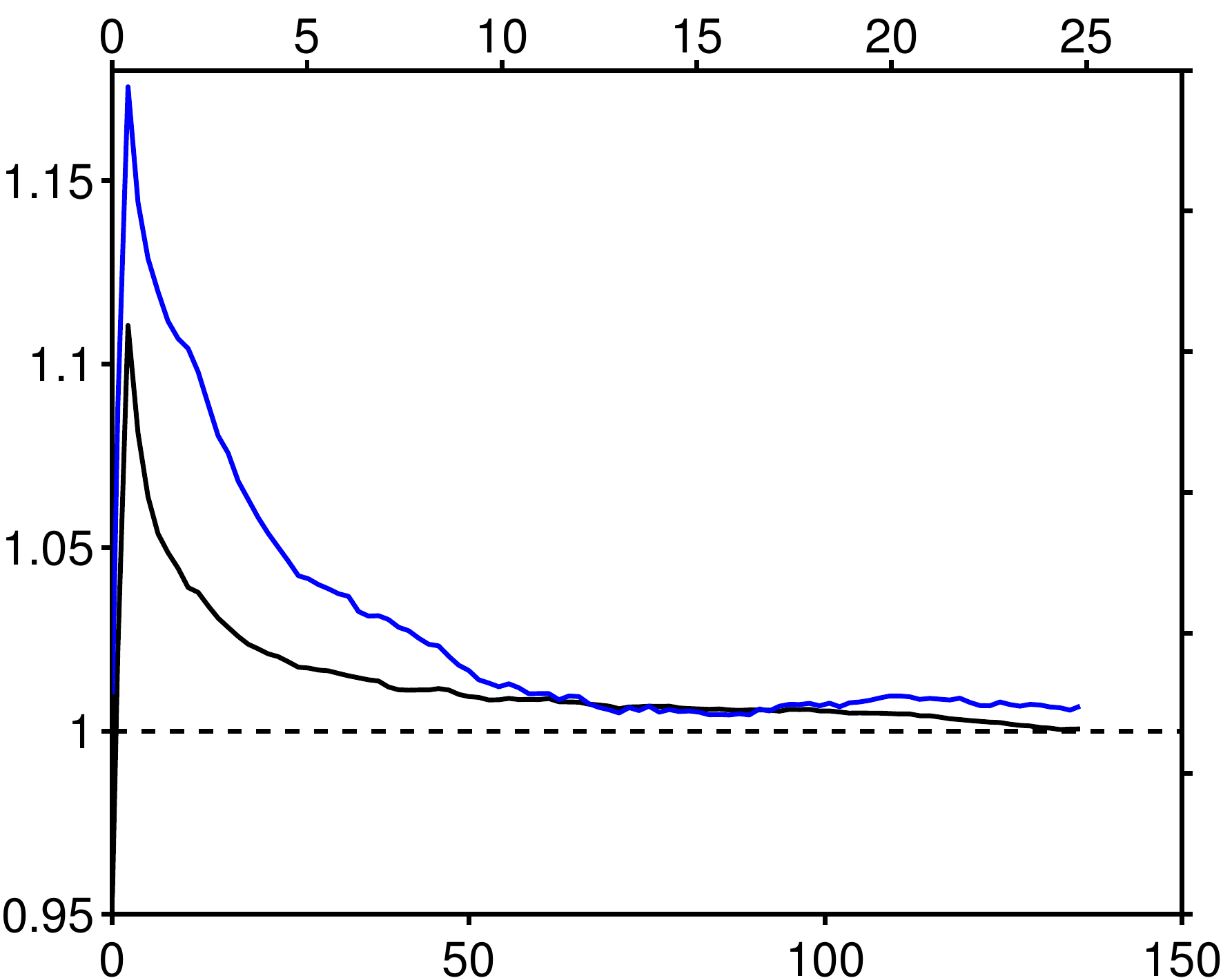}
      \\
      \centerline{$r/\eta$}
   \end{minipage}
   \caption{%
       $(a)$
       Comparison of the particle-conditioned radial distribution
       function in case D5 according to different criteria: 
       {\color{black}\solidthick}, criterion
       (\ref{equ-def-sticky-points}); 
       {\color{black}\chndot}, zero-acceleration points. 
       $(b)$
       Particle-conditioned, relative number density of 'sticky
       points' (as in figure~\ref{fig-stickyPoints-partCondDist}), 
       showing the full set of samples versus only those samples
       corresponding to clustering particles (having Vorono\"i cell
       volumes smaller than the lower cross-over point in
       figure~\ref{fig-voronoi-cell-vol-pdf}).  
       Line styles: 
       {\color{black}\solidthick} case D5; 
       {\color{blue}\solidthick} only clustering particles of case D5. 
   }
   \label{fig-stickyPoints-partCondDist-clustCond}
\end{figure}

The implications of the sweep-stick mechanism for finite-size
particles have not been explored in detail \citep[cf.\ the final
discussion by][]{qureshi:08}. 
As a first step in this direction we characterize the structure of the
set of points which verify the criterion (\ref{equ-def-sticky-points})
in our simulations. 
Note that in practice we have not determined the zeros of the
projected acceleration component from a minimization. Instead we have
identified grid points at which the value falls below a very small
threshold value 
($\mathbf{e}_1\cdot\mathbf{a}_f\leq
a_{P,thresh}=4\cdot10^{-4}\,\sigma_{a,P}$, where $\sigma_{a,P}$ is 
the standard-deviation of $\mathbf{e}_1\cdot\mathbf{a}_f$).  
We have checked the sensitivity of the following analysis with
respect to the choice of the threshold by varying the value of 
$a_{P,thresh}$ by a factor of four with no visible differences. 
Figure~\ref{fig-voronoi-particles-sticky-points-slice} provides
an instantaneous visualization of the set of sticky points in both
present particulate flow cases.  
It can be observed that their spatial distribution is clearly
inhomogeneous, featuring a succession of denser filaments, dilute
regions and voids. 
A quantitative measure of the level of clustering in these sets is
provided by the radial distribution function, which is computed from a
number of flow fields and shown in
figure~\ref{fig-stickyPoints-radDist}.  
A highly excessive probability of finding sticky points in
the immediate vicinity of a given sticky point is observed, which
reaches values of the order of ten times the global probability. The
decay with distance is found to follow an exponential approach to
unity for $r/\eta\gtrsim20$. 
Figure~\ref{fig-stickyPoints-radDist} also includes the data for
single-phase flow, which practically collapses with the curve for the
present case D11. 
This very good agreement between the curves in particulate and
single-phase flow implies that the particles do not
significantly alter the spatial structure of the set of sticky
points, with the consequence that we can consider the latter ones as a
quantity pertaining principally to the turbulent background flow. 
Now we are in a position to characterize the relation between the
particle locations and the set of sticky points. 
For this purpose we have computed the particle-conditioned
radial-distribution function of sticky points, i.e.\ the probability
of finding a sticky point at a given distance of a particle. 
The result is shown in figure~\ref{fig-stickyPoints-partCondDist},
where a normalization of the distance in both Kolmogorov units and
particle diameter is presented. 
A statistically significant excess probability is indeed observed in
both cases D5 and D11. However, the maximum amplitude of this 
probability (which occurs at small distances from the test particle)
only amounts to roughly $1.1-1.15$ times the global probability. 
This moderate level of statistical correlation between the positions of
the particles and those of the sticky points is consistent with the
relatively modest amount of clustering exhibited by the former ones
(cf.\ \S~\ref{sec-results-clustering}). 
Despite its relatively weak amplitude, the present data clearly links
the particle locations to the distribution of the sticky points
proposed by \cite{goto:08}. 
  The work of \cite{coleman:09} suggests that point-particles
  (under Stokes drag) in homogeneous-isotropic turbulence accumulate
  preferentially in the vicinity of zero-acceleration points. They
  conclude that locations identified with this simpler criterion are
  more strongly correlated with particle clusters than those defined
  through the projection upon the most expansive eigenvector
  (i.e.\ the sticky points given by \ref{equ-def-sticky-points}). 
  We have tested this alternative criterion by repeating the previous
  analysis with respect to zero-acceleration points. Again, the set of
  'zero-acceleration' points was determined by applying a small
  threshold ($0.05$ times the standard deviation of the fluid
  acceleration), whose influence on the results has been found to be
  weak.  
  Figure~\ref{fig-stickyPoints-partCondDist-clustCond}$(a)$ shows a
  comparison between the particle-conditioned radial distribution
  functions of the zero-acceleration points and the previous one
  pertaining to the sticky points defined by
  (\ref{equ-def-sticky-points}) in case D5. 
  It can be seen that both criteria lead to very similar results, with
  a comparable excess probability of roughly 10\% in the direct
  vicinity of the particles. 
  It can therefore be concluded that the two criteria, 
  (a) zero fluid
  acceleration, and 
  (b) zero fluid acceleration projected upon the most
  expansive direction of the acceleration gradient tensor, feature a 
  comparable statistical correlation with the positions of the present
  finite-size particles. 

It is also of interest to analyze the influence of the particle size
upon the particle-conditioned radial distribution function of sticky
points.  
Figure~\ref{fig-stickyPoints-partCondDist} shows that scaling the
distance in terms of the particle diameter is not appropriate. On the
other hand, it can be observed that a normalization in Kolmogorov
units leads to a reasonable collapse of the curves for larger
distances ($r/\eta\gtrsim50$).
This can be taken as an indication that the level of particle 
clustering does not influence the far-field decay of the 
particle-conditioned radial distribution function too much, 
where the particles essentially `see' the same sticky point
distribution. 
Note that in case D11 
at small distances from the particle somewhat larger values of the
radial distribution function are recorded than in case D5, although
the former case exhibits less intense clustering.  
Since this effect is restricted to the very near-field of the flow 
around the particle ($r/D\lesssim0.55$, as obtained from the
cross-over coordinate in
figure~\ref{fig-stickyPoints-partCondDist}$b$), it is presumably
caused by the difference in the transient wake flows the particles
generate as a response to the two respective particle Reynolds number
distributions (cf.\ figure~\ref{fig-rep-shell-avg}). 

In order to test whether the correlation between sticky-point
locations and particle positions is really relevant for particle
clustering, we have additionally computed the radial distribution
function for those particles only which have a Vorono\"i cell volume
smaller than the lower cross-over point in
figure~\ref{fig-voronoi-cell-vol-pdf}. 
The effect of this additional condition upon particle clustering is
shown in figure~\ref{fig-stickyPoints-partCondDist-clustCond}$(b)$ for case
D5. It can be seen that the restriction to clustering particles leads 
to significantly larger probabilities of finding a sticky point in the
vicinity of the test particle, in particular for distances smaller
than approximately $10D$.  
This additional piece of evidence strongly suggests that there is
indeed a causal relationship between the location of sticky points
and the tendency of particles to exhibit locally increased levels of
concentration. 

\begin{figure}%
  \centering
   \begin{minipage}{2.5ex}
     \rotatebox{90}
     {pdf}
   \end{minipage}
   \begin{minipage}{0.55\linewidth}
      \includegraphics[width=\linewidth]
      {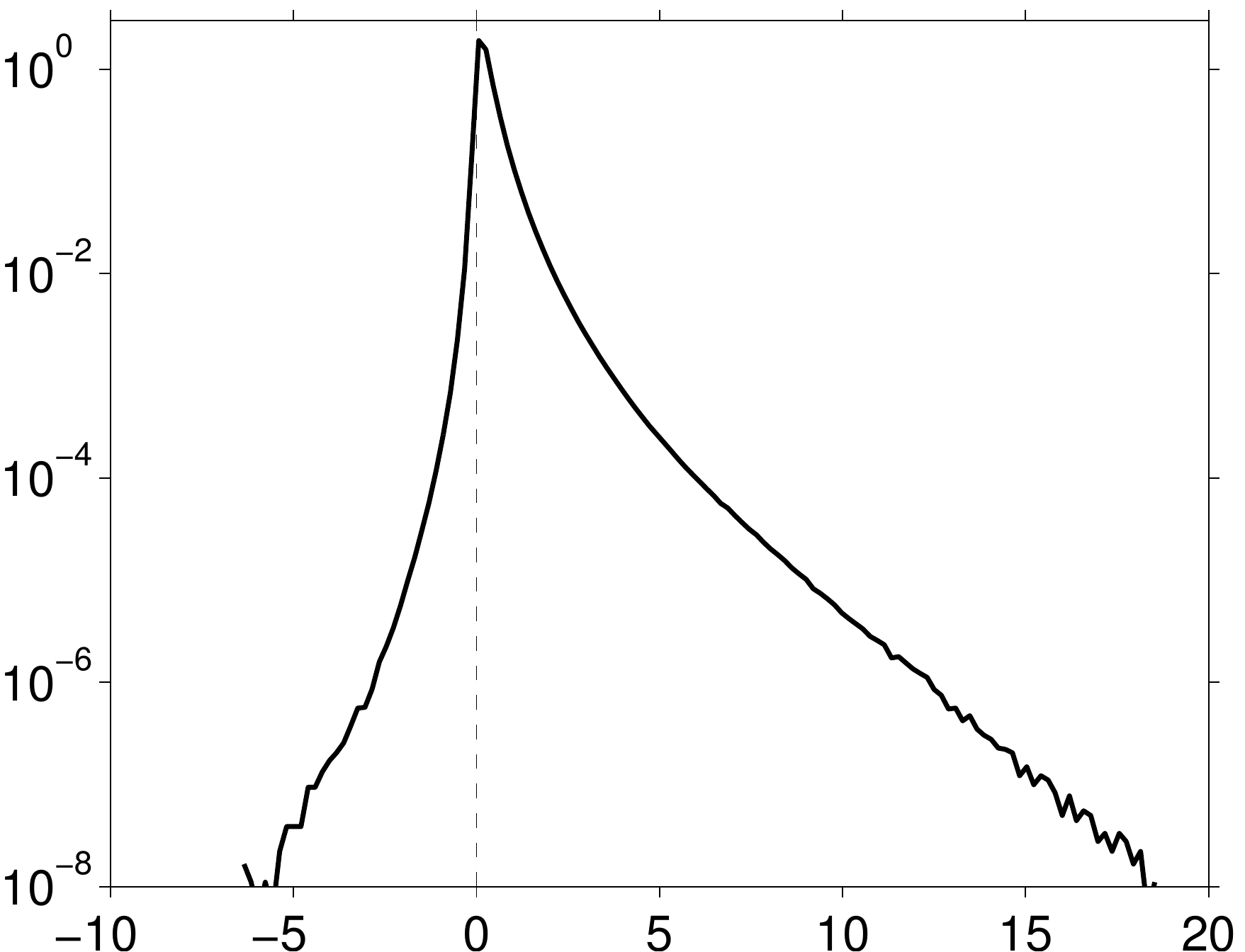}
      \\
      \centerline{$\lambda_1\,\tau_\eta^2$}
   \end{minipage}
   \caption{%
     P.d.f.\ of the largest eigenvalue $\lambda_1$ of the symmetric part of
     the fluid acceleration gradient tensor,
     $\nabla\mathbf{a}_f+(\nabla\mathbf{a}_f)^T$, in single-phase
     turbulence case S5. 
   }
   \label{fig-eig1-accel-gradient-pdf}
\end{figure}

  Finally, let us consider the characteristic time scales of the
  proposed sweep-stick mechanism.  
  \cite{goto:08} have noted that the values of the largest eigenvalue
  $\lambda_1$ of the symmetric part of the acceleration gradient tensor
  are of the order of $\tau_\eta^{-2}$ ($\tau_\eta$ being the Kolmogorov
  time scale).  
  Figure~\ref{fig-eig1-accel-gradient-pdf} shows the probability
  distribution of $\lambda_1$ in the single-phase case S5. 
  Since we are only concerned with positive values, 
  we have defined the set of points where $\lambda_1>0$, and denote this 
  quantity $\lambda_1^+$, for which the statistical moments have been
  determined. 
  It turns out that in the present single-phase case S5 the mean and
  standard-deviation of $\lambda_1^+$ 
  measure $0.34/\tau_\eta^2$ and $0.52/\tau_\eta^2$, respectively. 
  Consequently, following \cite{goto:08} one can estimate a
  characteristic time scale of the compressive effect of the fluid
  acceleration field as $\tau_{af}=\lambda_1^{-1/2}$ which indeed turns
  out to be of the order of the Kolmogorov time scale $\tau_\eta$ (e.g.\
  when computing $\tau_{af}$ from the sum of the mean plus one standard
  deviation of $\lambda_1^+$).  
  This result suggests that the particle Stokes number which would be
  most susceptible to clustering through the sweep-stick mechanism is of
  the order of unity when defined with the Kolmogorov time scale, i.e.\ 
  $St_\eta=\tau_p/\tau_\eta$. 
  However, as can be seen in figure~\ref{fig-eig1-accel-gradient-pdf}, 
  the probability distribution of $\lambda_1^+$ is very broad, and, 
  consequently, it corresponds to a broad range of time scales,
  including ones which are much larger than the Kolmogorov time. 

  \cite{sumbekova:16} have recently argued that the characteristic
  time-scale of the carrier flow in view of preferential accumulation
  of particles in the vicinity of zero-acceleration points
  ($|\mathbf{a}_f|=0$) is expected to be of the order of the
  Lagrangian correlation time of the magnitude of fluid acceleration. 
  This time scale in turn has been shown to scale with the integral
  scale (as opposed to the correlation time of individual fluid
  acceleration components which scale with the Kolmogorov time scale),
  cf.\ \cite{mordant:04}. 
  This consideration then suggests that from the point of view of
  time-scales the particles can be expected to exhibit some degree of
  preferential accumulation as long as their time scale does not
  exceed the time scale of the energetic eddies. This latter condition
  is presently fulfilled, as $\tau_p/T$ is below unity in
  both cases D5 and D11. 

  Another way to estimate a relevant fluid time-scale for particle
  clustering is to consider a length scale deduced from the actual
  observed cluster-volume distribution. The peak occurrence in
  figure~\ref{fig-voronoi-cell-clust-pdf}$(b)$ corresponds to a length 
  scale of approximately $20\eta$. If we assume classical inertial-range
  scaling, the fluid time-scale commensurate with the cluster size is
  then of the order of $20^{2/3}\tau_\eta$, which means that the Stokes
  number which we can form with this quantity is roughly one order of
  magnitude smaller than the Stokes number based upon the Kolmogorov
  scale in the present case D5. 
\section{Summary and discussion}  
\label{sec-conclusion}
We have performed interface-resolved DNS of forced
homogeneous-isotropic turbulence of a dilute suspension of spherical,
solid particles in the absence of gravity. 
The Taylor micro-scale Reynolds number was in the range
$Re_\lambda=115-142$, while the particle size measured $5.5$ and
$11.4$ times the Kolmogorov scale in our two simulations (denoted as
``D5'' and ``D11'', respectively). 
The particle density was set to $1.5$ times the value of the fluid
density in both cases, such that the effect of inertia can be studied
independently of gravity. 
In the present setup we are therefore able to investigate: 
the effect of particle inertia, 
the effect of finite particle size and 
collective effects. 
Excluded are: effects of mean relative velocity (e.g.\ particle
settling), statistical anisotropy in general, and spatial
inhomogeneity. 
Despite the low solid volume fraction of one half percent, the chosen
computational domain size accommodates a number of particles which is
sufficiently large to allow for a meaningful analysis of the particle
clustering behavior. Likewise, the observation interval in the
statistically stationary regime of the two-phase flow system is
sufficiently long in order to permit us to draw conclusions about a
broad number of statistical quantities of interest. 

In our analysis we have first considered the basic Eulerian and
Lagrangian statistics.
Our present results confirm earlier observations \citep{chouippe:15a}  
that solid particles at the present concentration level have only a
mild influence upon the global flow properties. In particular the
kinetic energy budget is very little affected by the addition of the 
particles. 
On the other hand, the particle and fluid inertia lead to a relative
motion with average particle Reynolds numbers measuring approximately
8 and 32 in these two cases, while the distribution of that quantity
is very broad (with exponential tail). 
This in turn means that the local flow in the vicinity of the
particles is obviously modified by the solid inclusions, which is
detectable e.g.\ as an increase in the number density of intense 
vortices which are formed in the transient wakes. 
As a consequence, the interpretation of the results needs to take into
account the two-way coupled character of the fluid-particle
interaction carefully, despite the small global influence of the
particles upon the fluid flow. 

The particle acceleration in the present simulations exhibits a
normalized probability distribution which is consistent with the
empirical fit of \cite{qureshi:07} if the samples corresponding to
particle-particle contact are eliminated from our data. This result
once more confirms the universality of the log-normal shape of the
p.d.f.\ of the acceleration which has previously been shown to
represent a wide range of experimental and numerical data-sets  
\citep{qureshi:08,volk:08a,homann:10,YeoClimentMaxey:10,villalba:12,chouippe:15a}.  
With this in mind, knowledge of the variance of particle acceleration
(which is parameter dependent) then completely determines the
distribution of the acceleration.  
Since our simulations only span a modest range of particle diameters,
we are not able to resolve the discrepancy in acceleration variance
data which exists between the various experimental data-sets on the
one hand (suggesting a scaling of approximately $D^{-2/3}$ whenever the
particle size exceeds a few multiples of the Kolmogorov length) and
the numerical data of \cite{homann:10} and \cite{cisse:15} on the
other hand (which is reasonably well represented by a $D^{-4/3}$ law).  
Concerning the effect of solid-fluid density ratio, 
our results are consistent with the ones of \cite{YeoClimentMaxey:10},
who have provided the only previous data-set addressing this effect 
in the absence of gravity in the
present flow configuration, 
albeit at lower Reynolds number ($Re_\lambda\approx60$) and larger solid
volume fraction ($\Phi_s\approx0.06$). 
The fact that the present particle acceleration data falls within the 
cloud of previous DNS data (involving both density-matched
single-particle cases and inert particle collectives) suggests that
the density ratio has only a limited (damping) effect. 

We have also found that the Lagrangian auto-correlation function of
particle acceleration (i.e.\ of the hydrodynamic force acting upon the
particles) decays within a few multiples of the Kolmogorov time, as
already observed by several authors in the context of
neutrally-buoyant particles \citep{homann:10,volk:11}. 
The integral time scale of this quantity (by integration up to the
first zero-crossing) roughly follows a linear law as a function of the
particle diameter, with the same slope as in the two above-mentioned
studies. This means that the Lagrangian auto-correlation of particle
acceleration is not sensitive to the value of the solid-fluid density
ratio, 
at least up to the present value of $\rho_p/\rho_s=1.5$. 

Particle clustering has been quantified with the aid of Vorono\"i
tesselation. The standard deviation of the volume of the Vorono\"i
cells is increased by 7.7\% (3.1\%) in case D5 (D11) with respect to a
non-overlapping set of particles distributed by a random Poisson
process. 
This is a comparatively mild but significant level of
clustering. 
We have further computed the objective clustering threshold proposed
by \cite{monchaux:10b}, and used it for the purpose of cluster
identification in case D5. The probability distribution of the number
of members per cluster is found to be much broader than in the random
case, such that in the DNS data there is a significantly higher 
probability to encounter very large clusters. 
  Furthermore, we observe that the distribution of the volume
  occupied by each particle cluster (as measured by the sum of their 
  associated Vorono\"i cells) exhibits a clear peak and a power-law
  decay consistent with the self-similar inertial-range model proposed
  by \cite{goto:06}.  
We have explored the possibility that particle clustering might be
caused by similar effects as documented in the point-particle
literature, namely the centrifugal mechanism \citep{maxey:87} and
the sweep-stick mechanism \citep{goto:08}. 
Concerning the centrifugal effect, 
we have focused upon the role of the intense vortical structures
(i.e.\ `worms') which have been educed 
by means of the $q$-criterion of \cite{hunt:88}. 
We have first analyzed the volume enclosed by surfaces where the value
of $q$ is equal to 1.5 times its standard deviation. It is found that
the present particles do not significantly alter the worm volume
statistics, except for the addition of small vortices (with enclosed
volume smaller than the volume of an individual particle) which can be 
linked to the transient wake flow. 
We have then conducted an analysis of the distances from each particle
to the nearest point on any surface defined by the above threshold. 
We have shown in case D5 that the p.d.f.\ of these distances from
particle to coherent vortex is similar to the one in the case of
randomly chosen points, except for small distances (of the order of
the Kolmogorov length), where the two clearly differ. 
However, by eliminating all vortices with enclosed volume smaller than
the particle volume, the difference with respect to randomly chosen
points disappears. 
From this analysis we conclude that the positions of the particles in
case D5 do not significantly correlate with the presence of strong
vortices. 
  This result then suggests that the centrifugal mechanism 
  is not responsible for the clustering of particles in the present
  cases. 
Next we have turned to the analysis of the relation between the
acceleration field and the particle positions, as implied by 
the sweep-stick mechanism proposed by
\cite{goto:08}. 
These authors define `sticky points' as those points, 
where the largest eigenvalue of the symmetric part of the acceleration
gradient tensor is positive, and where at the same time the projection
of the acceleration upon the direction of the largest eigenvalue
vanishes. 
  Note that only in the framework of the simplest point-particle model
  (one-way coupled, with Stokes drag as the only hydrodynamic force)
  in the limit of small Stokes numbers it can be rigorously argued
  that the particle ``field'' will cluster at these points. 
Here we have computed the radial distribution function for the highly
clustered set of sticky points, and it turns out that the addition of
particles (in case D11) does not have any significant effect. 
We have then computed the particle-conditioned radial distribution
function of the sticky points. Our results exhibit a small but
significant increase in the probability of finding a sticky point in
the vicinity of both present particulate flow data-sets. 
This enhanced probability further increases when only those particles
are sampled which are members of a cluster \citep[i.e.\ which have an
associated Vorono\"i cell with a small volume in the sense of the 
objective definition of][]{monchaux:10b}. 
The evidence in the present study therefore strongly suggests that a
direct link between these sticky points and the preferential locations
of the particles exists. 
This is the first time that such a relation has been established in
the context of finite-size particles, 
and it appears indeed surprising that a mechanism which has been
proposed in the point-particle context under very restrictive
assumptions should be relevant to these finite-size particles moving
at finite particle Reynolds numbers. 
However, it should be kept in mind that both intensities are
comparatively weak, the one of the observed clustering and the one of
the preferential particle positioning with respect to the sticky
points.  

Many questions regarding the 
relevance of the 
sweep-stick mechanism for finite-size
particles still remain. 
First, the properties of the flow in the vicinity of the sticky points
is not clear, and a detailed analysis in terms of coherent structures
is outstanding. 
For this reason, the precise mechanism by which particles are
attracted to these locations is not known. 
As a consequence, it is also unclear how effective the sweep-stick
mechanism is at parameter points other than those for which detailed
data is available. 
Second, while we have been able to find evidence in favor of the
sticking part of the sweep-stick mechanism, the sweeping part has
proved to be elusive. In particular, we did not detect a statistically
significant correlation between the local, instantaneous relative
velocity and the fluid acceleration in the vicinity of the particles 
(figure omitted). 
  This means that the finite-size particles in the vicinity of
  zero acceleration points do not follow the ``local'' fluid velocity,
  and an adequate description in the framework of a point-particle
  approach would require a much more complex model of the force balance
  (cf.\ \citealp{calzavarini:09} for a discussion of the relevance of
  Fax\'en corrections,  
  and \citealp{daitche:15} for 
  the effect of the Basset history force), 
  as well as it would probably require to take into account local
  two-way coupling effects.    
  Statistical models for the description of clustering in
  sub-Kolmogorov particles 
  \citep{zaichik:07,bragg:14a,gustavsson:16} 
  point to the importance of the history of the velocity gradient
  tensor experienced by the particles. 
  In the future it would be interesting to explore the role of this
  non-local effect in the case of finite-size particles which are
  fully coupled to the fluid flow. For this purpose, an extensive
  analysis based upon time-resolved flow data along the particle
  trajectories will be necessary, which is indeed an extremely
  data-intensive task. 
In the present study we have mainly focused on the influence of the
size of the particles as the density ratio was kept constant at the
value of $1.5$. 
Further investigations 
over a broader range of density ratios should be conducted in order to
obtain more insight into the impact of particle inertia not only on
global quantities, but also on local interaction mechanisms between
the particles and the surrounding fluid. 
\vspace*{1ex}
This work was supported by the German Research Foundation (DFG) under
project UH~242/1-2. 
The simulations were partially performed at LRZ M\"unchen (under
grant pr83la) and at SCC Karlsruhe (project DNSPARTHIT).      
The computer resources, technical expertise and assistance 
provided by these centers are thankfully acknowledged. 
In particular we would like to express our gratitude to the generous
storage space provided by SCC in their ``Large Scale Data Facility''
(LSDF).
Thanks is also due to M.\ Bourgoin and R.\ Volk for many 
stimulating discussions.   
        The anonymous referees have significantly contributed to the
        improvement of the manuscript by suggesting the following
        aspects: 
        analysis of cluster volume distribution, 
        relevance of the scaling of the Lagrangian
        auto-correlation of the fluid acceleration magnitude, 
        additional particle cluster quantification by means of the
        radial distribution function.  
\bibliographystyle{model2-names}
\addcontentsline{toc}{section}{References}
\appendix
\begin{table}
  \setlength{\tabcolsep}{4pt}
  \centering
  \begin{tabular}{lcccc}
    case & 
    $D/\eta$ &
    $Re_\lambda$ &
    $\Phi_s$ & 
    $\rho_p/\rho_f$ 
    \\[.5ex]  
    D5 &
    $5.5$   &
      $116.8$
    &%
    $0.005$ &%
    $1.5$ 
    \\
    D11 &
    $11.4$  & 
    $141.6$ &
    $0.005$ &%
    $1.5$  
    \\[1ex]
    S5  &--&$119.0$&0&--\\
    S11 &--&$142.8$&0&--\\
  \end{tabular}
  \caption{%
    The main physical parameters in the present simulations: 
    length scale ratio $D/\eta$, 
    Taylor-scale Reynolds number $Re_\lambda=\lambda u_{rms}/\nu$, 
    global solid volume fraction $\Phi_s$, 
    density ratio $\rho_p/\rho_f$. 
    Cases S5 and S11 are the corresponding single-phase
    simulations with the same turbulence-forcing parameters as D5 and
    D11, respectively, cf.\ table~\ref{tab-turbforce-params}.
  }
  \label{tab-phys-params}
\end{table}
\begin{table}
  \setlength{\tabcolsep}{4pt}
  \centering
  \begin{tabular}{lccc}
    case & 
    $\kappa_f/\kappa_{1}$ 
    & $T_L\nu/{\cal L}_x^2$ 
    & $\varepsilon^{*}{\cal L}_x^4/\nu^3$
    \\[.5ex]  
    D5, S5&
    $3.61$&
    $5.37\cdot10^{-5}$&
    $1.31\cdot10^{9}$
    \\
    D11, S11&
    $2.50$&
    $5.94\cdot10^{-5}$&
    $3.52\cdot10^{9}$
  \end{tabular}
  \caption{%
    Imposed parameters related to the turbulence forcing scheme of 
    \cite{eswaran:88} in the notation of \cite{chouippe:15a}: 
    the forcing cut-off wavenumber $\kappa_f$, normalized by the
    smallest discrete wavenumber $\kappa_{1}$; 
    the characteristic time of the random forcing, $T_L$; 
    the dissipation-rate parameter $\varepsilon^{*}$. 
  } 
  \label{tab-turbforce-params}
\end{table}
\begin{table}
  \setlength{\tabcolsep}{4pt}
  \centering
  \begin{tabular}{lcccccccc}
    case &
    $N_p$ &
    $N_x$ &
    $L/{\cal L}_x$ & 
    $\eta/\Delta x$ & 
    ${\cal L}_x/D$ & 
    $D/\Delta x$ & 
    $T_{obs}/T$ &
    $T_{obs}/\tau_p$
    \\[.5ex] 
    D5 &
    $20026$ &
    $2048$ &
      $0.43$
    &
    $2.89$&
    $128$ &
    $16$  &
      $24.98$
    &
      $460.64$
    \\
    D11 &
    $2504$ &
    $1024$ &
    $0.56$ &
    $1.41$ &
    $64$ &
    $16$ &
    $26.40$ &
    $135.64$
  \end{tabular}
  \caption{The principal numerical parameters pertaining to the
    present simulations: 
    number of particles $N_p$, 
    number of Eulerian grid nodes per linear dimension $N_x$, 
    ratio between large-eddy length scale and box-sitze $L/{\cal L}_x$, 
    ratio between Kolmogorov scales and grid width $\eta/\Delta x$, 
    ratio between box-size and particle diameter ${\cal L}_x/D$, 
    particle resolution $D/\Delta x$, 
    observation time normalized with large-eddy time scale $T_{obs}/T$
    and normalized with particle time-scale $T_{obs}/\tau_p$.
  } 
  \label{tab-num-params}
\end{table}
\begin{figure}%
  \centering
  \begin{minipage}{2ex}
    \rotatebox{90}
    {p.d.f.}
  \end{minipage}
  \begin{minipage}{0.45\linewidth}
    \includegraphics[width=\linewidth]
    {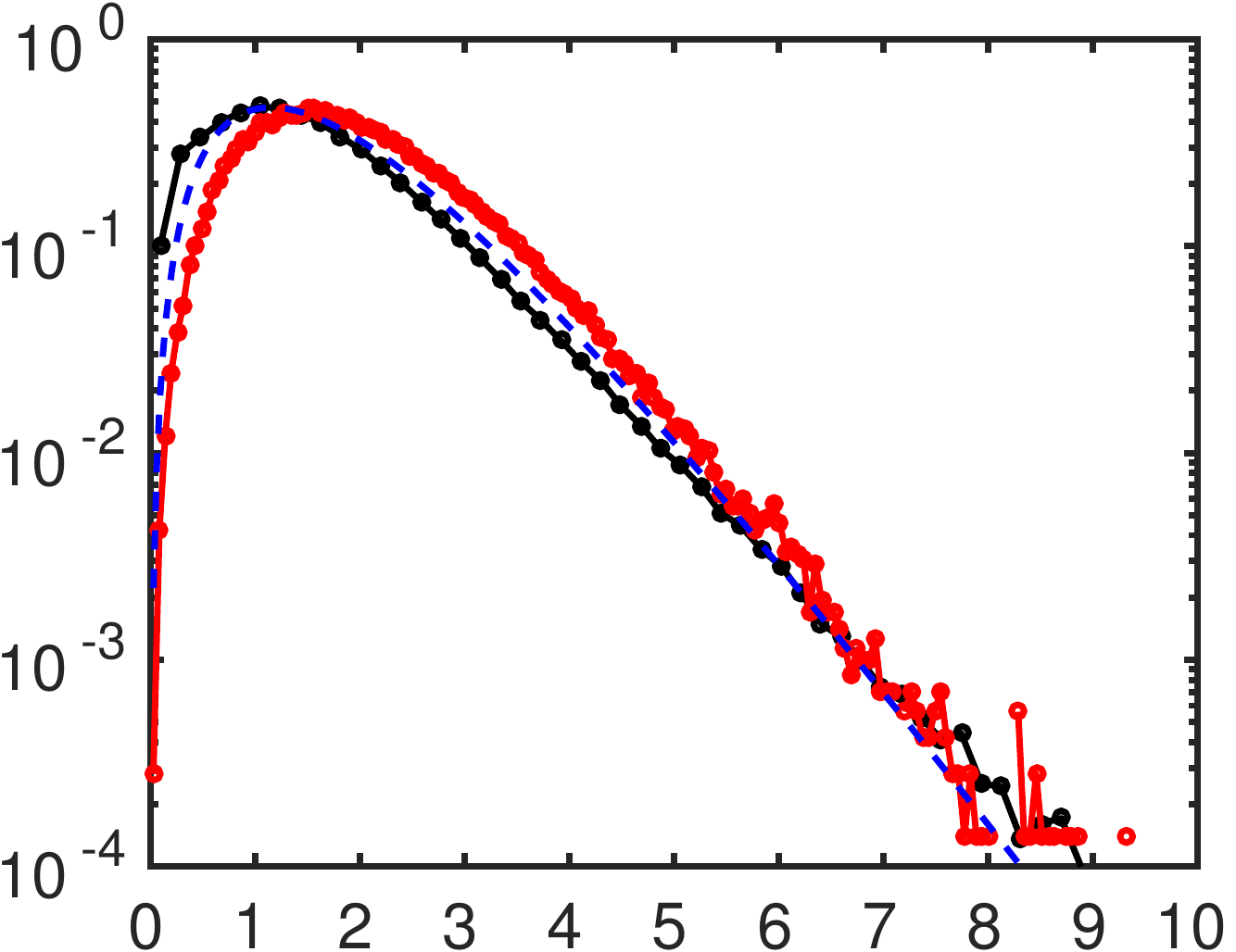}
    \\
    \centerline{$Re_p/\sigma(Re_p)$}
  \end{minipage}
  \hspace*{3ex}
  \begin{minipage}{0.35\linewidth}
    \setlength{\tabcolsep}{4pt}
      \begin{tabular}{lcc}
        case&
        \multicolumn{1}{c}{$\langle Re_p\rangle$}&
        \multicolumn{1}{c}{$\langle Re_p^\prime
          Re_p^\prime\rangle^{1/2}$} 
        \\[.5ex]
        D5&
        $7.60$&
        $4.84$
        \\
        D11&
        $32.17$&
        $16.24$
      \end{tabular}
  \end{minipage}
  \caption{%
    Normalized p.d.f.\ of the particle Reynolds number computed
    from the sphere-averaged relative velocity $\mathbf{u}_{pr}$ as
    defined in the text. 
    Line styles correspond to: 
    {\color{black}\solidthick},~case~D5; 
    {\color{red}\solidthick},~case~D11.   
    The blue dashed line indicates a Gamma distribution 
    with shape parameter $k=3$ which for a quantity $x$ with variance
    unity reads:  
    $f(x)=(3^{3/2}/2)x^2\exp(-x\sqrt{3})$. 
    Mean and standard-deviation of the particle Reynolds number are
    listed next to the graph.  
  }
  \label{fig-rep-shell-avg}
\end{figure}
\begin{figure}%
  \centering
   \begin{minipage}{2.5ex}
     \rotatebox{90}
     {$\langle a_{p,\alpha}^\prime a_{p,\alpha}^\prime\rangle/
       \langle a_{f,\alpha}^\prime a_{f,\alpha}^\prime\rangle$}
   \end{minipage}
   \begin{minipage}{0.7\linewidth}
       \includegraphics[width=\linewidth]
       {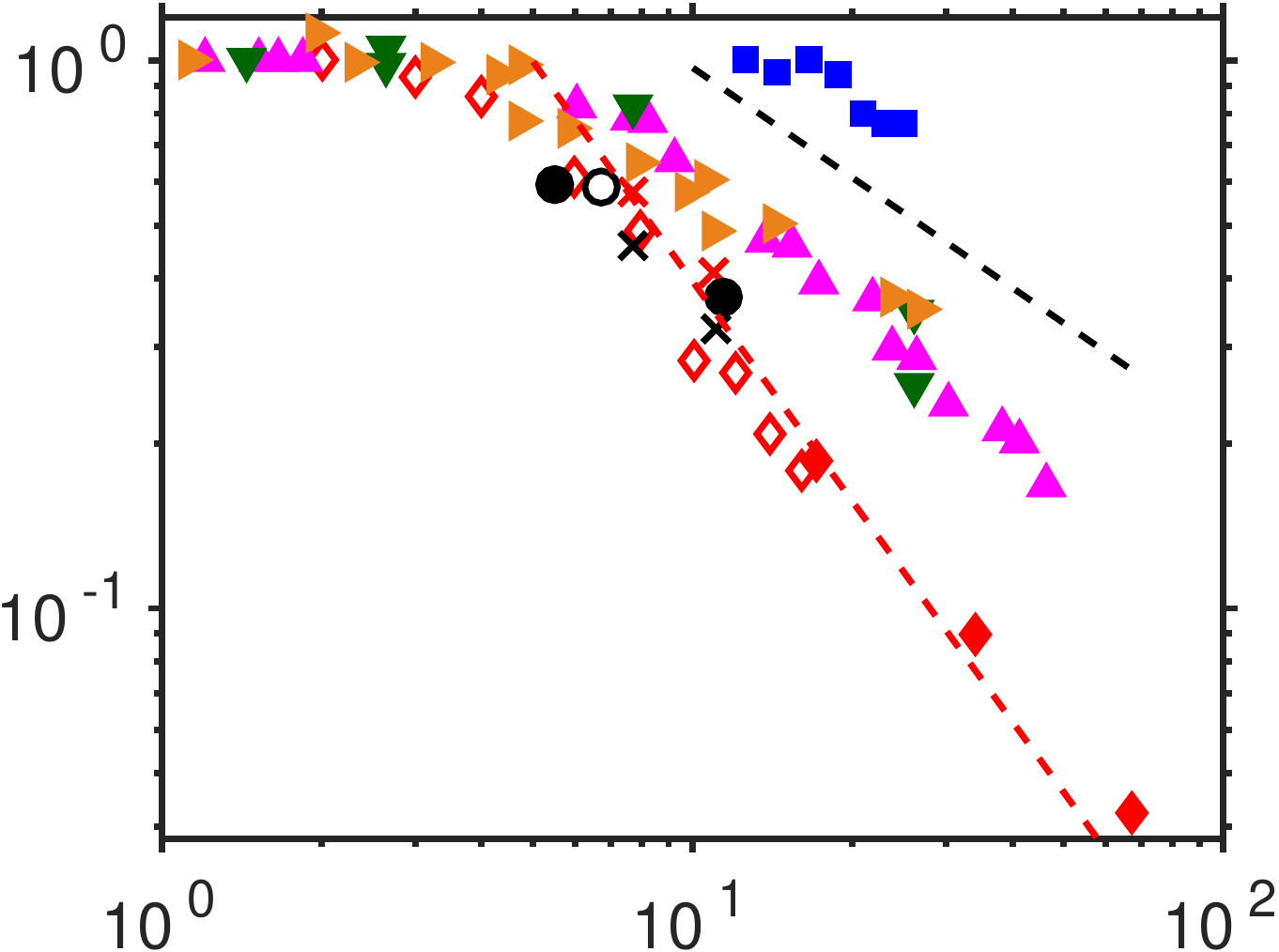}
      \\
      \centerline{$D/\eta$}
   \end{minipage}
   \hspace*{1ex}
   \begin{minipage}{0.2\linewidth}
     \setlength{\tabcolsep}{4pt}
     \begin{tabular}{lc}
       case&
       $%
       \displaystyle
       \frac{\langle a_{p,\alpha}^\prime a_{p,\alpha}^\prime\rangle}
       {\langle a_{f,\alpha}^\prime a_{f,\alpha}^\prime\rangle}
       $
       \\[3ex]
       D5&
         $0.59$ %
       \\
       D11&
       $0.37$ %
     \end{tabular}
   \end{minipage}
   \caption{%
       Variance of the single-component particle acceleration normalized
       by the variance 
       of fluid particles, plotted as function of the particle
       diameter. 
     The present data is represented by the filled circular symbols
     ($\bullet$); additionally, numerical values are listed next to the graph. 
     The open circle ({\color{black}$\boldsymbol{\circ}$}) indicates
     the case ``A-G0'' from \cite{chouippe:15a}. 
     The dark green triangles
     ({\color{darkgreen}$\mathsmaller{\blacktriangledown}$}) mark the
     experimental results of 
     \cite{voth:02} which were obtained in a
     von K\'arm\'an flow  at different Reynolds
     numbers ($Re_\lambda=140\ldots970$). 
     The orange colored triangles 
     ({\color{niceorange}$\mathsmaller{\blacktriangleright}$}) 
     indicate the more recent measurements in the same flow,
     carried out at Reynolds numbers $Re_\lambda=400\ldots800$, 
     as reported by \cite{brown:09}. 
     The blue squares ({\color{blue}\solidsquare}) 
     correspond to the wind-tunnel data of \cite{qureshi:07} for
     neutrally-buoyant particles at $Re_\lambda=160$.  
     The magenta-colored triangles
     ({\color{magenta}$\mathsmaller{\blacktriangle}$}) are
     experimental measurements in a von K\'arm\'an flow by
     \cite{volk:11}, performed at different Reynolds numbers
     ($Re_\lambda=590\ldots1050$).  
     Red diamonds represent the simulation data of \cite{homann:10}
     ({\color{red}$\mathsmaller{\lozenge}$}) 
     and those of \cite{cisse:15} ({\color{red}\soliddiamond}). 
     The numerical data for density-matched particles (cases ``N1'',
     ``N2'') of \cite{YeoClimentMaxey:10} are indicated by the symbol
     ``{\color{red}$\boldsymbol{\mathsmaller{\times}}$}'', and their
     particles with density ratio $\rho_p/\rho_f=1.4$ (cases ``S1'',
     ``S2'') are shown as  ``{\color{black}$\boldsymbol{\mathsmaller{\times}}$}''.
     The guiding lines are proportional to $(D/\eta)^{-2/3}$
     ({\color{black}\dashed}) 
     and to $(D/\eta)^{-4/3}$ ({\color{red}\dashed}). 
   }
   \label{fig-part-accel-variance}
\end{figure}
\clearpage
\end{document}